\documentclass[
 twocolumn,
]{ceurart}
\usepackage[nolist]{acronym}
\usepackage{xspace}
\usepackage[round, comma, authoryear]{natbib}
\usepackage{booktabs,colortbl}
\usepackage{pdflscape}
\usepackage{amsmath}
\newcommand{\fontTables}{\small}
\usepackage{rotating}
\usepackage{graphicx}
\usepackage{subcaption}
\usepackage{multirow}
\usepackage{tabularx}
\usepackage{longtable}
\usepackage{calc}
\usepackage{float}
\usepackage{pbox}
\usepackage{mathtools}
\newtheorem{definition}{Definition}

\usepackage{paralist}
\usepackage{pgfplots}\pgfplotsset{compat=1.9}
\usepackage{pgfplotstable}
\usepackage[nolist]{acronym}
\usepackage[ruled,linesnumbered]{algorithm2e}

\usepackage{listings}
\usepackage[prependcaption]{todonotes} 
\usepackage{xargs} 
\newcommandx{\pablo}[1]{\todo[linecolor=green,backgroundcolor=green!25,bordercolor=green,inline]{Pablo: #1}}
\newcommandx{\linus}[1]{\todo[linecolor=orange,backgroundcolor=orange,bordercolor=black,inline]{Linus: #1}}
\newcommandx{\alejandro}[1]{\todo[linecolor=blue,backgroundcolor=blue!25,bordercolor=black,inline]{Alejandro: #1}}

\newcommand{\Popularity}{Pop\xspace}
\newcommand{\Rnd}{Rnd\xspace}
\newcommand{\IB}{IB\xspace}
\newcommand{\UB}{UB\xspace}
\newcommand{\HKV}{HKV\xspace}
\newcommand{\PopGeoNN}{PopGeoNN\xspace}
\newcommand{\GeoBPRMF}{GeoBPRMF\xspace}
\newcommand{\IRENMF}{IRENMF\xspace}
\newcommand{\BPRMF}{BPRMF\xspace}
\newcommand{\RankGeoFM}{RankGeoFM\xspace}

\newcommand{\NDCG}{nDCG\xspace}
\newcommand{\NDCGAux}{nDCG}

\newcommand{\checkin}{check-in\xspace}
\newcommand{\checkins}{check-ins\xspace}

\newcommand{\LBSNsAbbr}{\acp{LBSN}\xspace}

\begin{document}

\copyrightyear{2024}
\copyrightclause{Copyright for this paper by its authors.
  Use permitted under Creative Commons License Attribution 4.0
  International (CC BY 4.0).}

\conference{Author Accepted Version of the article published in Information Technology \& Tourism \url{https://doi.org/10.1007/s40558-024-00304-0}}

\title{Understanding the Influence of Data Characteristics on the Performance of Point-of-Interest Recommendation Algorithms}

\author[1]{Linus W. Dietz}[%
orcid=0000-0001-6747-3898,
email=linus.dietz@kcl.ac.uk,
]
\address[1]{King's College London, UK}

\author[2,3]{Pablo S\'anchez}[%
orcid=0000-0003-1792-1706,
email=psperez@icai.comillas.edu
]
\address[2]{Instituto de Investigación Tecnológica (IIT), Universidad Pontificia Comillas, Spain}

\author[3]{Alejandro Bellogín}[%
orcid=0000-0001-6368-2510,
email=alejandro.bellogin@uam.es
]
\address[3]{Universidad Aut\'onoma de Madrid, Spain}

\begin{abstract}
	Point-of-interest (POI) recommendations are essential for travelers and the e-tourism business. They assist in decision-making regarding what venues to visit and where to dine and stay. While it is known that traditional recommendation algorithms’ performance depends on data characteristics like sparsity, popularity bias, and preference distributions, the impact of these data characteristics has not been systematically studied in the POI recommendation domain. To fill this gap, we extend a previously proposed explanatory framework by introducing new explanatory variables specifically relevant to POI recommendation. At its core, the framework relies on having subsamples with different data characteristics to compute a regression model, which reveals the dependencies between data characteristics and performance metrics of recommendation models. To obtain these subsamples, we subdivide a POI recommendation data set on New York City and measure the effect of these characteristics on different classical POI recommendation algorithms in terms of accuracy, novelty, and item exposure. Our findings confirm the crucial role of key data features like density, popularity bias, and the distribution of check-ins in POI recommendation. Additionally, we identify the significance of novel factors, such as user mobility and the duration of user activity. In summary, our work presents a generic method to quantify the influence of data characteristics on recommendation performance. The results not only show why certain POI recommendation algorithms excel in specific recommendation problems derived from a LBSN check-in data set in New York City, but also offer practical insights into which data characteristics need to be addressed to achieve better recommendation performance.
\end{abstract}


\maketitle

\begin{acronym}

\acro{UI}{user interface}
\acro{LBSN}{location-based social network}
\acro{RS}{recommender system}
\acro{POI}{point of interest}
\acrodefplural{POI}{points of interest}
\acro{EV}{explanatory variable}
\acro{URM}{user-rating matrix}
\acrodefplural{URM}{user-rating matrices}
\acro{UCM}{user-check-in matrix}
\acrodefplural{UCM}{user-check-in matrices}
\acro{VIF}{variance inflation factor}
\end{acronym}

\section{Introduction}
\label{s:Introduction}

Understanding which recommendation algorithm is most effective for a specific data set is crucial, as it has been widely acknowledged that no single recommender can achieve optimal performance in all scenarios~\citep{Im2007,Anelli2022}.
Apart from the performance variation of the algorithms across different data sets, it should be taken into account that the quality of any recommender can be evaluated through a wide array of different dimensions~\citep{DBLP:reference/sp/GunawardanaSY22}.
While accuracy may be the primary concern when recommending items that a user might actually consume, it is equally important to prioritize recommending different items for the users (diversity), items that may be unfamiliar to users and, hence, surprise them (novelty), or ensure that our recommendations are not biased towards individual users or items (fairness)~\citep{DBLP:reference/sp/CastellsHV22, Ekstrand2022}. However,
designing models that perform well in all these dimensions is challenging, as they need to deal with, for example, the accuracy-diversity trade-off~\citep{DBLP:journals/ipm/IsufiPH21}.
Whereas such analysis of accuracy versus novelty, diversity, and other dimensions has been conducted in traditional recommendation scenarios like movies or books, within the %
point-of-interest recommendation domain, the problem has not been analyzed in such detail, although some researchers have started to examine these dimensions in this context~\citep{DBLP:journals/jitt/MassimoR21,Sanchez2022a}.

Moreover, while most existing studies have primarily focused on evaluating the quality of recommendations based on the accuracy of recommended venues through offline experimentation metrics (i.e., using ranking accuracy metrics like Precision or Recall), there remains a lack of consensus on the other crucial aspects of evaluation methodology, such as data sets, data filtering, data partitioning, and other evaluation metrics~\citep{Sanchez2022}.
In addition, it is important to consider that users can be grouped based on simple touristically relevant information, such as their origin or the categories of visited POIs, all of which can correlate with their preferences.
It has been shown that recommendation performance may fluctuate substantially depending on the user group a user belongs to, especially between local users and visiting tourists \citep{Sanchez2022a}.

\cite{DBLP:journals/ipm/DeldjooBN21} proposed a method to analyze the impact of different data characteristics on the accuracy and fairness of matrix factorization algorithms in the movie and book recommendation domains.
As opposed to such classical recommendation problems, %
\ac{POI} recommendations are influenced in a way larger degree by further factors, such as seasonality, geographical influences of the venues to be recommended, and the type of users of such system~\citep{Sanchez2022}. Hence, in this paper, we investigate the success factors of classical and \ac{POI} recommendation algorithms through the lens of data characteristics %
present in the data set used as input by the recommendation models.
To do so, we incorporate the most relevant  influences in a framework derived from~\citep{DBLP:journals/ipm/DeldjooBN21} to analyze the effect they have on the performance of a set of recommenders.

\subsection{Using Data Characteristics  to Explain Recommender Performance}
When proposing a new recommendation model, researchers often begin with intuitions or anecdotal evidence, until they ultimately obtain empirical evidence through experimentation to validate the model effectiveness.
However, relying solely on intuitions or anecdotal evidence to justify the quality of recommendations is insufficient. This approach may explain why some models obtain excellent results on some data sets while, at the same time, they perform poorly in others, as recent efforts on reproducibility have demonstrated~\citep{DBLP:conf/recsys/DacremaCJ19,DBLP:conf/recsys/SaidB14}. %

It should be noted that some studies have analyzed the effect of different aspects in the recommendations, like the data partitioning \citep{DBLP:conf/recsys/MengMMO20} and the hyperparameters of the models \citep{DBLP:conf/recsys/AnelliNSPR19}.
The experiments presented in this paper are based %
on the approach  by
the works of \cite{DBLP:journals/ipm/DeldjooBN21, DBLP:conf/sigir/DeldjooNSM20}, where the authors defined a set of explanatory variables that model the characteristics of the data set (e.g., ratings per user, per item, population bias, etc.).
On the basis of those works, herein, we use a similar methodology as proposed by \cite{DBLP:journals/ipm/DeldjooBN21}, but adapt the framework towards \ac{POI} recommendation  by incorporating additional variables that capture the unique dynamics of the POI recommendation domain.
The core idea is that recommendation performance is influenced by quantifiable patterns in the data, which result in easier or more difficult recommendation problems. For example, a high density of the user-rating matrix, i.e., where many users have already rated a large portion of items, generally provides recommendation models with ample signal to compute suitable recommendations. Hence, many such data characteristics can potentially influence performance. In this paper, we study not only the impact of these data characteristics on ranking quality, but we also analyze the effect on the recommendations in terms of both novelty and the  amount of exposure each item receives across all users. %
These aspects are important as they help identifying recommenders that may be amplifying unfairness in the exposure of items. For example, if a model recommends popular POIs significantly more frequently than they appear in the test set, it can lead to low novelty values and greater disparities in item exposure. In e-tourism, this is a key consideration as it can decide which businesses thrive.

Despite the commonalities with the approach presented in \citep{DBLP:journals/ipm/DeldjooBN21}, we further develop several aspects of the general method, which requires generating a sufficiently large number of recommendation data sets with varying data characteristics to compute a regression model.
To achieve this, we start from a widely-used \checkin data set based on the location-based social network Foursquare and generate domain-driven subsamples; that is, considering characteristics of special importance in both traditional and POI recommendation.
The subsamples are created using filter-like rules targeting the interaction density, popularity of venues, seasonality and origin of users, e.g., locals visitors of a travel destination.
This is in contrast with the original approach which used a constraint-based \emph{random} sampling method to derive subsamples. By subdividing the set of all \ac{POI} visits in the data set based on the aforementioned \emph{domain-specific} rules, we can better steer the subsampling process and simultaneously obtain interpretable subsamples, which can be used to understand inherent attributes and characteristics of the POI recommendation domain. As an outcome, we obtain individual subsamples of the data set with varying data characteristics.

This variability in the data characteristics in the individual subsamples is important as, in the next step, we independently perform recommendation experiments with each subsample and record the outcome variables in terms of accuracy, novelty, and item exposure.
Herein, it is important to note that these subsamples are synthetic simulations of recommendation data sets.
We compute regression models using the data characteristics of the individual subsamples as independent variables and the performance metrics as dependent variables. In other terms, we quantify data characteristics using \emph{explanatory} variables to explain the %
performance changes of the recommenders in terms of ranking accuracy, novelty, and item exposure
using the regression model.
Through the quantification of the statistical significance of the explanatory variables within the regression model, we ensure that the determined influences are robust and not spurious effects.
To capture all potential influences on the recommendation outcome in the \ac{POI} recommendation domain, we further extend the variables proposed for classical recommendation domains \citep{DBLP:journals/ipm/DeldjooBN21, DBLP:conf/sigir/DeldjooNSM20,Adomavicius2012} with spatio-temporal features.
An analysis of the statistical significance of the coefficients in the regression model reveals which data characteristics are needed to explain the recommenders' performance.

\subsection{Overview of Contributions}

While the core concepts of this paper are based on previous approaches of~\cite{DBLP:journals/ipm/DeldjooBN21} and \cite{Adomavicius2012}, we go beyond of simply adapting an established method to the domain of \ac{POI} recommendations.
We make the following conceptual and methodological contributions:

\begin{enumerate}
	\item We extend the corpus of explanatory variables for analyzing the effect of different data characteristics, including geographic aspects,
	in a varied set of state-of-the-art POI recommendation algorithms (\autoref{subsec:evs}). This analysis considers three complementary evaluation dimensions: ranking accuracy, novelty, and item exposure.
		\item We introduce a domain-specific subsampling algorithm for POI recommendation (\autoref{sec:subsampling}). This algorithm ensures that the simulated data sets are grounded in the domain instead of random subsampling, as done in previous works.
	\item We perform a comprehensive analysis of a set of recommendation algorithms by considering $144$ different simulations (\autoref{sec:exp_setup}).
	Each simulation corresponds to a subsampled recommendation data set of
	a specific city within the Foursquare \checkin data set.
	In this way, we conduct an analysis of different samples with disparate characteristics to detect which explanatory variables help us to better explain the performance of the recommenders.
\end{enumerate}

\subsection{Impact on e-Tourism}

This research can have significant implications for the tourism industry.
\ac{POI} recommendations play a crucial role in shaping tourists' experiences and guiding their choices of which places to visit; thus, they are a key factor in decision-making.
We offer valuable insights that can enhance the tourism industry's ability to provide more tailored and satisfying experiences to travelers by determining which recommendation algorithms should be used in which kind of specific recommendation scenarios. %
The analysis of the beyond-accuracy metrics, i.e., novelty and item exposure, offers valuable perspectives that cater to the user needs and local businesses.
Novelty is especially relevant to local users, as it fosters the exploration of their city, whereas item exposure is a precondition to ensure that the flow of visitors is dissipated on many venues, contributing to provider fairness \citep{Deldjoo2023Fairness}.
By harnessing data characteristics such as density, popularity bias, and user activity duration, %
the proposed methods can be used to select algorithms that better align with the business goals of destination management stakeholders.
Thus, understanding how user behavior varies in different parts of a destination can enable businesses to tailor their offerings and marketing efforts more effectively.
Our research underscores the value of data-driven decision-making in the tourism sector.
By leveraging the insights gained from this study, both providers of \ac{POI} recommendation platforms and the tourism industry can enhance their ability to provide personalized and engaging experiences to users.

\section{Background %
}

\subsection{Point-of-Interest Recommendation}
\label{ss:POIRecommendation}
The \ac{POI} recommendation problem is typically defined as suggesting  %
venues in a city or particular region that the target user has not previously visited \citep{DBLP:journals/aim/MassimoR22}. %
These venues have a specific location, typically expressed as latitude and longitude, and might be varied in nature, including museums, parks, restaurants, or bars, among others. As in the traditional recommendation scenario, the objective is to maximize the number of relevant items (in this case, venues) that are being recommended to the user. %
Formally, as pointed in a recent survey by~\cite{Sanchez2022}, the POI recommendation problem can be formulated as follows:
\begin{equation}
\label{eq:rec_problem}
	p^*(u) = \arg\max_{p \in \mathcal{P}} g(u,p,\Phi)
\end{equation}
\noindent where $\mathcal{P}$
denotes all POIs available in the region, $p^*$ is the optimal venue that maximizes the utility for user $u$ among all POIs\footnote{Even though we use the symbol $\mathcal{P}$ to refer to the POIs, in line with the standard notation from the traditional recommendation problem, we shall use the letter $I$ to refer to the items of the system, i.e., the POIs.} in $\mathcal{P}$, as defined by the utility function g, and $\Phi$ represents the set of influences, also referred to as contextual information in some works~\citep{DBLP:reference/sp/AdomaviciusBTU22}.
This contextual information should be considered to perform meaningful POI recommendations \citep{DBLP:conf/sigir/ManotumruksaMO18}.

Temporal, sequential, social, categorical, and, most importantly, geographical information are normally exploited in most POI recommendation approaches \citep{DBLP:conf/sigir/LiCLPK15, DBLP:conf/recsys/GriesnerAN15, DBLP:conf/sigir/ZhangC15,DBLP:conf/cikm/LiuWSM14}.
In order to perform \ac{POI} recommendations, researchers often use the information available in \acp{LBSN}.
Foursquare, Gowalla, or Yelp are examples of this type of application where users are allowed to register the \checkins they perform at the different venues they visit. Data sets extracted from such \acp{LBSN} are invaluable for understanding the visiting behavior of users. %
Information about friendship links, along with the geographical coordinates of the venues, their categories, and the timestamps of the check-ins, can be used to model the aforementioned influences and generate potentially interesting recommendations to the users who are new in a specific geographical region, sometimes even requiring completely different approaches, such as reinforcement learning \citep{DBLP:conf/um/Massimo023}.

It is important to note that \LBSNsAbbr typically contain \checkins from different cities around the world.
However, as this type of recommendation is affected by all the aforementioned influences (especially the geographical information), many researchers perform recommendations considering each city/region as independent data sets
\citep{DBLP:conf/cikm/LiuWSM14,DBLP:conf/sigir/LiCLPK15}. %
This strategy is not only practical from an experimental point of view, but it is also quite reasonable since when a user is in a particular city, she will be interested in visiting venues belonging to that region and not from distant cities.
Furthermore, each city may exhibit a unique distribution of POIs to visit, along with distinct cultural characteristics and specific urban planning.
In fact, this is one of the reasons why POI recommendation is closely related to the tourism industry \citep{DBLP:conf/www/WangYCHWZH20, DBLP:journals/urban/SantosAMGM19}, since tourists, when they arrive to a new city, are usually interested in visiting the most relevant venues of that specific city and immerse themselves in the local culture. Besides, we need to consider that tourism is the base of many economies, such as some countries in southern Europe \citep{https://doi.org/10.1002/jtr.646}, due to the large number of stakeholders involved, including tourists, venue owners, and local residents. %

\subsection{Specific Considerations of the POI Recommendation Domain}
\label{ss:Considerations}
When addressing the POI recommendation problem, it is necessary to regard several domain-specific considerations and problems.
Some of them include:
\begin{description}
	\item[\textbf{Sparsity:}] The ratio between the stored preferences in the data set and all the possible interactions between the users and the venues is extremely low.
While the densities of \acp{LBSN} data sets from Foursquare and Gowalla are approximately 0.0034\%, the density of the Netflix and Movielens20M data sets, typically used in classical recommendation, are around 1.77\%, hence showing much higher values of sparsity.
\item[\textbf{Additional influences:}] As discussed in Section~\ref{ss:POIRecommendation}, POI recommendation is influenced by geographical aspects, social connections, and temporal information.
Due to the high data sparsity, it is crucial to leverage additional information to enhance the performance of the algorithms. Among all the information sources, geographical influence plays the most important role since users often prefer to visit nearby POIs in accordance to Tobler's Law: \textit{``[...]Everything is related to everything else, but near things are more related than distant things''} \citep{Tobler1970Law}. However, temporal influence can also provide valuable insights, such as the duration of users' visits and their movement patterns between POIs. Therefore, exploiting information suitable to the respective use cases is essential for the success of the recommendations.
\item[\textbf{Implicit information:}] Traditionally, classical recommender systems model user-item interactions using ratings. However, in POI recommendation data sets, we typically lack of explicit ratings and we only have timestamps of user visits. Moreover, users may check in multiple times at the same POI, which classical recommendation systems typically do not account for~\citep{DBLP:reference/sp/NikolakopoulosNDK22}. In POI recommendation, repeated check-ins to the same venue can serve as implicit information, refining the model of a user's preference similar to explicit ratings. As normally we do not have explicit information to create a user-item matrix, these repeated check-ins provide valuable implicit feedback. POI recommender systems capture latent user preferences using frequency matrices, providing better recommendations.
\item[\textbf{Popularity bias:}]		Popularity bias is a well-studied problem in the recommender systems domain that occurs when popular items are recommended more frequently than less popular ones, regardless of whether they actually match the interests of the target user~\citep{DBLP:conf/flairs/AbdollahpouriBM19}. %
The effect of popularity bias is evident in multiple layers within the context of POI recommendation.
Firstly, at the city level, an analysis of the original Foursquare data set reveals that out of the 415 cities worldwide, the top 1\% of the most popular cities (based on the highest number of check-ins) represent the 20\% of the total \checkins in the data set.
However, when considering the top 2\% of the most popular cities, this percentage increases to 28\% of the total \checkins.
	Additionally, within each specific city, we can observe the impact of popularity bias on individual POIs. Taking New York City and Tokyo as examples, two extensively studied cities in the Foursquare data set \citep{Sanchez2022}, we find that in New York City, the top 1\% of the most popular venues are responsible for 27\% of all \checkins, while the top 2\% of venues account for 36\% of the total \checkins. Similarly, in Tokyo, the top 1\% of venues comprise 48\% of all \checkins, and the top 2\% represent 57\% of the \checkins in the city.

	\end{description}

\subsection{Offline Evaluation in Point-of-Interest Recommendation}
\label{ss:Offline_evaluation}
In offline evaluation of POI recommendation methods, most works follow the same protocols used in the traditional recommendation scenario.
The data set is split into a training and test set with, occasionally, an additional validation set being used for model parameter tuning. All models are trained on this data, and for each user in the test set, a top-N list of recommendations is generated based on predicted user satisfaction \citep{DBLP:conf/recsys/CremonesiKT10}.

As in classical Machine Learning, subsets are often generated through random splits or cross-validation from the original data set~\citep{DBLP:journals/tist/ChengYKL16, DBLP:journals/tii/WangCWCLG20}. However, recently, temporal dimension has been considered in these splits~\citep{DBLP:conf/aaai/ZhaoZLXLZSZ19,DBLP:journals/fgcs/HuangMLS20}. Currently, two main types of temporal splits are common: per user, where the $n$ oldest interactions for each user are used for training, and the rest for validation and testing, and a global split, where interactions before a specific timestamp are used for training and the rest, for validation and test~\citep{Sanchez2022}. The latter is more natural, mimicking production-scale recommender system evaluation and avoiding data leakage to the test set~\citep{Ji2022}.

Different metrics are used to evaluate recommendation algorithms, and most of them focus on measuring recommendation accuracy. This is determined by the overlap between recommended and actually visited venues in the test set. Greater overlap implies better recommendation quality.
However, there is a recognition in the community that solely focusing on items in the ground truth may overlook other user-centric evaluation dimensions such as novelty (recommending non-popular items), diversity  (recommending items that are different), and serendipity (recommending items that are novel and not easy to discover)~\citep{DBLP:reference/sp/CastellsHV22}. In the POI recommendation domain, which is influenced by categorical, geographical, and social factors, additional metrics can be used. For instance, category-level accuracy metric~\citep{DBLP:conf/ksem/ZhaoLLSC15} and the error in geographical distance between recommended and visited POIs~\citep{DBLP:journals/tkdd/YinCCHZ15} are used for measuring additional dimensions.

Another aspect that has received considerable attention in recent years in
the recommender systems community, as in other areas of Artificial Intelligence and Machine Learning, is trying to understand the inherent biases learned by these systems and how they get reinforced by the recommendations.
Thus,
many researchers have focused on analyzing potential biases that may be found in either the
data sets %
or the recommendations produced by the models.
These biases vary widely, ranging from gender \citep{DBLP:journals/umuai/EkstrandK21, DBLP:journals/ipm/MelchiorreRPBLS21} to popularity bias \citep{DBLP:conf/flairs/AbdollahpouriBM19, DBLP:conf/sigir/CanamaresC17}.
Moreover, whereas in the classical recommendation scenario it is currently established that analyzing different types of biases is important, the POI recommendation domain appears to lack comprehensive exploration in this regard. To the best of our knowledge, only a limited number of studies have analyzed this aspect.
For example,
\cite{Sanchez2022a} observed biases in the recommendations provided to different groups of tourists and locals; %
and \cite{DBLP:conf/gis/WeydemannSW19} studied three types of fairness in this domain, i.e.,  fairness regarding the popularity of the venues, fairness with respect to the nationalities of the users, and an assessment if the recommendations are aligned with the category distribution observed in previous visits.

In light of the complexities and influences in the POI recommendation domain, there is little known about the impact of data characteristics on the recommendation performance. We address this gap using an explanatory framework, which was adapted to this domain from \cite{DBLP:journals/ipm/DeldjooBN21}.

\section{An Explanatory Framework for POI Recommendation}
\label{sec:framework}

The overall goal of this work is to understand which factors influence the performance of different recommendation models in the \ac{POI} recommendation domain in terms of different evaluation dimensions such as ranking accuracy, novelty, and item exposure.
Similar to previous approaches addressing this challenge \citep{Adomavicius2012,DBLP:journals/ipm/DeldjooBN21}, we use data characteristics to describe subsamples of a recommendation data set and use a regression model to capture the impact of each feature on the recommendation outcome.
The idea behind analyzing different subsamples of the same data set is that the regression analysis would reveal the influence of variations in data characteristics on dependent variables.
The main difference to previous studies is that our analysis is regarding \ac{POI} recommendation, which enables us to define further explanatory variables capturing the geographical influences of users visiting venues in a city.
Moreover, we are able to use a domain-driven subsampling approach, which yields additional insights into the performance of recommenders. %

In this section, we describe the explanatory framework, which is a regression model applied to a series of %
data characteristics for capturing the interactions of users with the respective venues in a city. %
These data characteristics are computed for each subsample independently using a domain-driven approach outlined in the subsequent \autoref{sec:subsampling}.

\subsection{Regression Model}
\label{subsec:regression_theory}

Given all subsamples, we aim to model the relationship between the data characteristics and the recommendation performance of each individual recommendation algorithm.
This allows us to test hypotheses regarding which explanatory variables are able to describe the variations in the dependent variables in a statistically significant way.
\autoref{eq:regression} shows the regression model, which is the core of the explanatory analysis.

\begin{equation}
	\label{eq:regression}
	y^r = \epsilon^r + \theta_0^r + \sum_{ev \in EV} \theta_{ev}^r x_{ev}
\end{equation}
where $\epsilon$ is the error term (residuals), $\theta_0$ is the intercept, i.e., the mean value of the dependant variable when the rest of the independent variables are zero,
$\theta_{ev}$ is the regression coefficient of the respective explanatory variable $ev$ (among the set of variables $EV$), $x_{ev}$ represents the value of the explanatory variable in the current training example, and $y$ is the value of the dependent variable according to the recommendation models.
Since some of these values will depend on a specific recommendation model $r$, the notation shows this with a superscript.
In particular, this means that, for a specific recommender system $r$, the value of the dependent variables will be potentially different for each
$r$%
. Then, the performance would be modeled upon the set of explanatory variables $x_{ev}$,
that will depend on the characteristics of the data set.
Based on this, the regression model will produce coefficients %
$\theta^r_{ev}$
specifically tailored for this particular recommender, as it will consider the explanatory variables and the dependent variables at the same time.
When using the EVs within the regression model, we apply min-max normalization to obtain coefficients that are directly comparable.

\subsection{Explanatory Variables}
\label{subsec:evs}

We define 32 \acp{EV} which serve as independent variables in the regression analysis.
Unlike the approaches of \cite{Adomavicius2012} and \cite{DBLP:journals/ipm/DeldjooBN21}, we do not have a \acl{URM}, but a \acl{UCM}, since the interaction between the users and \acp{POI} is a visit and not a rating.
While the \ac{UCM} is conceptually very similar to a \ac{URM}, the \ac{UCM} is established based on unique visits, i.e., the cell values of the \ac{UCM}  are $1$ if a user has visited a venue; otherwise, it is $0$.
Multiple visits of one user to the same venue also result in a value of $1$.
Despite this small conceptual difference, most \acp{EV} proposed and used by \cite{Adomavicius2012} and \cite{DBLP:journals/ipm/DeldjooBN21} can also be computed for a \ac{UCM}.
Since there is a significant geographic influence in \ac{POI} recommendation \citep{DBLP:conf/sigir/LiCLPK15}, we also propose some further \acp{EV} that capture such geographic information about the visited venues.
Thus, the \acp{EV} we use can be categorized in the following four categories:

\begin{enumerate}
	\item \acp{EV} that describe the structure of the \ac{UCM}.
	\item \acp{EV} that describe the \checkin
	distribution of the \ac{UCM}.
	\item \acp{EV} that are based on item and user properties in the \ac{UCM}.
	\item \acp{EV} that capture the underlying user activity and mobility.
\end{enumerate}

\subsubsection{EVs Based on the Structure of the UCM}

These \acp{EV} capture the general structure of the \ac{UCM} and are well established to describe properties of recommendation data sets.
Thus, we keep the discussion around them succinct.

\begin{definition}{(SpaceSize)}
	\label{def:spacesize}
	Given a \ac{UCM}, SpaceSize is defined as:
	\begin{equation}
		ev_1 = SpaceSize(UCM) = |U| \cdot |I|
	\end{equation}
\end{definition}

We use the SpaceSize instead of its components, the number of users, $|U|$, and the number of items $|I|$ since it reduces the number of variables. As pointed out in Section~\ref{ss:POIRecommendation}, although we are dealing with POIs, denoted in that section as $\mathcal{P}$, we use the letter $I$ to refer to items in general.
\begin{definition}{(Shape)}
	\label{def:shape}
	Given a \ac{UCM}, Shape is defined as:
	\begin{equation}
		ev_2 = Shape(UCM) =  \frac{|U|}{|I|}
	\end{equation}
\end{definition}

The ratio between the number of users and the number of items can be an initial indicator of whether user-based collaborative filtering or item-based collaborative filtering approaches might be more successful \citep{DBLP:reference/sp/NikolakopoulosNDK22}.

\begin{definition}{(Density)}
	\label{def:denisty}
	Given a \ac{UCM}, Density is defined as:
	\begin{equation}
		ev_3 = Density(UCM) =  \frac{|C|}{|U|\times|I|}
	\end{equation}
\end{definition}

Density, or its inverse, sparsity, is a commonly reported metric to give an estimation of the recommendation difficulty for collaborative recommendation algorithms. Here we use $C$ to refer to all the check-ins performed by the users and registered in a data set.
Generally, the higher the density, the more signal is available for the algorithm to compute fitting recommendations.
Density typically varies a lot depending on the data set and the domain, as mentioned in Section~\ref{ss:Considerations}.

\begin{definition}{($Cp_u,Cp_i$)}
	\label{def:rpu}
	Given a \ac{UCM}, \checkins per user ($Cp_u$) and \checkins per item ($Cp_i$) are defined as:
	\begin{equation}
		ev_4 = Cp_u(UCM) =  \frac{|C|}{|U|}
	\end{equation}
	\begin{equation}
		ev_5 = Cp_i(UCM) =  \frac{|C|}{|I|}
	\end{equation}
\end{definition}

The number of interactions per user/item is also a simple but effective measure to put the recommendation quality into perspective.
If the number of interactions is remarkably small for a user or an item, it can be regarded as \emph{``cold''}, indicating that there is not sufficient information to compute meaningful recommendations.
Given the high sparsity of the data sets in the POI recommendation domain, it is very common to impose a minimum number of interactions for both items and users, i.e., enforcing a $k$-core, cf.\,\autoref{subsec:subsampling_variables}.
This is done to avoid evaluating cold-start recommendations.

\subsubsection{EVs Based on the Check-in Distribution of the UCM}

Naturally, some users are more active than others, and not all items get the same attention.
In the \ac{POI} recommendation domain, this is a very natural phenomenon since major highlights inherently attract more visits.
This is a major challenge with respect to several dimensions of the recommendation algorithms, including accuracy, novelty, and fairness,
as there is a delicate trade-off between recommending popular \acp{POI} and items in the long-tail \citep{Rahmani2022}.

\begin{definition}{(Gini$_I$, Gini$_U$)}
	\label{def:gini}

	Given a \ac{UCM}, $|C_i|$ and $|C_u|$  be the number of \checkins associated with item $i$ and user $u$ and the users/items are sorted according to $C_i$ and $C_u$, respectively; then Gini$_I$ and Gini$_U$ are defined as follows~\citep{DBLP:journals/ipm/DeldjooBN21}:

	\begin{equation}
		ev_6 = Gini_I(UCM) =  1 - 2 \sum_{i=1}^{|I|}    \frac{|I| +1 - i}{|I|+1} \times \frac{|C_i|}{|C|}
		\label{eq:gini_i}
	\end{equation}

	\begin{equation}
		ev_7 = Gini_U(UCM) =  1 - 2 \sum_{u=1}^{|U|}    \frac{|U| +1 -u}{|U|+1} \times \frac{|C_u|}{|C|}
		\label{eq:gini_u}
	\end{equation}

\end{definition}
The Gini coefficient captures the frequency distribution of the check-ins for users or items.
It is scaled between $[0,1]$, where a score of $0$ would correspond to a uniform popularity distribution, and $1$ to the extreme case of all check-ins being concentrated on one user/item.

\subsubsection{EVs Based on the Item and User Properties}

The following two EVs, Popularity Bias and Long Tail Items, have been motivated by \cite{DBLP:journals/ipm/DeldjooBN21} to be included in their explanatory framework.
Arguably, the \ac{POI} recommendation domain is even more severely impacted by popularity bias \citep{DBLP:journals/jitt/MassimoR21,Sanchez2022a}, thus, it is imperative to include them in our framework.
Given the distribution of the metrics with outliers, we augment the aggregation methods used in \citep{DBLP:journals/ipm/DeldjooBN21} (mean, standard deviation, skewness, and kurtosis) with the median value, as the median is more robust against outliers compared to the mean value. Note that a higher kurtosis is related to heavier tails, and hence, more outliers, while the skewness is related to the symmetry of the distribution. If the skewness is positive, it means that the right-hand tail is longer than the left-hand tail. If the skewness is negative, then the right-hand tail is shorter than the left-hand tail.

\begin{definition}{(Popularity Bias)}
	\label{def:pop_bias}
	We follow the commonly accepted definition of popularity bias proposed by \cite{DBLP:conf/flairs/AbdollahpouriBM19}, which should not be confused with the popularity bias produced by the recommendation algorithm, as this applies to the bias that exists in the original data:

	\begin{equation}
		\label{eq:pop_bias}
		ev_{8:12} =   f\Bigg(\Bigg\{\frac{\sum_{i\in C_u}\phi(i)}{|C_u|}\Bigg\}_u\Bigg)
	\end{equation}
\end{definition}
where $\phi(i)$ is the popularity scoring function for an item $i$.
The notation $\{\cdot\}_u$ aims to indicate that we iterate over the users and compute the value inside the brackets, which is then processed by the outer function $f$.
An item's popularity score is thus defined as the number of users who visited $i$ over the entire number of users,
and $|C_u|$ is the number of  \checkins of user $u$.
The term $f$ is an aggregation operator over users to capture inter-user differences in the popularity profiles of users.
They include average popularity bias ($ev_8$, APB), median popularity bias ($ev_9$, MedPB), standard deviation of popularity bias scores ($ev_{10}$, StPB), skewness popularity bias ($ev_{11}$, SkPB), and kurtosis popularity bias ($ev_{12}$, KuPB).

\begin{definition}{(Long Tail Items)}
	\label{def:long_tail_items}
Analyzing the popularity of items, they can be separated into a short-head and a long-tail.

	\begin{equation}
		ev_{13:17} =   f\Bigg(\Bigg\{\frac{|i,i \in (C_u \cap \Gamma)|}{|C_u|}\Bigg\}_u\Bigg)
	\end{equation}
\end{definition}
where $C_u$ are, again, the \checkins of user $u$, $\Gamma$, on the other hand, represents long-tail items, which is determined by splitting the items into short-head and long-tail items.
We define the split between short-head and long-tail in terms of the number of different users that have visited the item.
In the literature, typical cutoffs for separating the short-head from the long-tail are at 20\%--80\%, cf.\, \citep{Yin2012, Abdollahpouri2017, DBLP:journals/ipm/DeldjooBN21}, which we also use in our experiments.

As before, $f$ is an aggregation operator over users to capture inter-user differences in long-tail profiles of users.
They include average long tail items ($ev_{13}$, ALT), median long tail items ($ev_{14}$, MedLT), standard deviation of long-tail items scores ($ev_{14}$, StLT), skewness long tail items ($ev_{15}$, SkLT), and kurtosis long tail items ($ev_{17}$, KuLT).

\subsubsection{EVs Based on the User Activity and Mobility}
\label{subsec:evs_mobility}

In this section, we introduce a family of data characteristics that are specific to the \ac{POI} recommendation domain.
The Radius of Gyration captures the size of a user's activity area.
The distance from the city center is defined similarly, but the information it captures is more about how central the check-ins of a user are.
Finally, the user's activity duration is interesting to include as well, as one can assume that the longer a user is within a city, the more familiar she becomes with the city, and it might result increasingly difficult for recommendation algorithms to propose interesting items.

\begin{definition}{(Radius of Gyration)}
	\label{def:radius Gyration}
This is a common metric to capture the geographic extent of user mobility.

	\begin{equation}
		ev_{18:22} = f\Bigg(\Bigg\{\sqrt{ \frac{\sum_{i \in C_u} \text{distance}((\text{lat}_i, \text{lon}_i), (c_u^x, c_u^y)) }{|C_u|}} \Bigg\}_u \Bigg)
	\end{equation}
	\begin{equation}
			c_u = (c_u^x, c_u^y)= \left( \frac{1}{|C_u|} \sum_{i \in C_u} \text{lat}_i, \frac{1}{|C_u|} \sum_{i \in C_u} \text{lon}_i \right)
	\end{equation}
\end{definition}
where $c_u$ is the centroid of all the user's visited venues
\citep{Gonzalez2008} and $\text{lat}_i$ and $\text{lon}_i$ represent the latitude and the longitude of item $i$ respectively.

Again, $f$ is an aggregation operator over users to capture inter-user differences in the Radius of Gyration of users.
They include average Radius of Gyration ($ev_{18}$, ARG), median Radius of Gyration ($ev_{19}$, MedRG),  standard deviation of Radius of Gyration scores ($ev_{20}$, StRG), skewness of Radius of Gyration ($ev_{21}$, SkRG), and kurtosis of the Radius of Gyration ($ev_{22}$, KuRG).

\begin{definition}{(Distance to City Center)}
	\label{def:dist_ts}
This \ac{EV} is very similar to the Radius of Gyration, however, the center is not set to the centroid of the venues visited by the user, but to the center of the city.
It is useful to differentiate between users who perform activities near the center of a city --  typically of historic significance -- and users who are more active in the outskirts.
	\begin{equation}
		ev_{23:27} = f\Bigg( \Bigg\{ \sqrt{ \frac{\sum_{i \in C_u} \text{distance}((\text{lat}_i, \text{lon}_i), (cc^x, cc^y))}{|C_u|}} \Bigg\}_u \Bigg)
	\end{equation}
\end{definition}
where $cc=(cc^x,cc^y)$ is the geographic location of the city center.
Again, we use different aggregation functions:
the average distance to the city center
($ev_{23}$, ADCC), median distance to the city center ($ev_{24}$, MedDCC), standard deviation of the distances to the city center ($ev_{25}$, StDCC), skewness of distance to the city center ($ev_{26}$, SkDCC), and kurtosis of the distance to the city center ($ev_{27}$, KuDCC).

\begin{definition}{(Duration Active)}
	\label{def:duration_activity}
This EV is useful to provide insights into the effects of how long the duration of user activity was. In this context, users who have been active for a shorter duration may correspond to tourists, whereas those who have been performing \checkins for a longer period can be considered local residents of the city~\citep{DBLP:journals/ipm/SanchezB21}.

	\begin{equation}
		ev_{28:32} = f\Bigg( \Bigg\{ t(i)_{l} - t(i)_0 | i \in C_u \Bigg\}_u \Bigg) 
	\end{equation}
\end{definition}
where $t(i)_0, t(i)_{l}$ is the time of the first and last check-in of the items $i$ the user $u$ has visited, respectively.
Note that in this case, we took the duration from all check-ins of the user, including repeated check-ins.
Again, we use different aggregation functions:
the average duration active  ($ev_{28}$, ADA), median duration active ($ev_{29}$, MedDA), standard deviation of duration active ($ev_{30}$, StDA), skewness of duration active ($ev_{31}$, SkDA), and kurtosis of the duration active ($ev_{32}$, KuDA).

\subsection{Dependent Variables}
\label{subsec:depented_vars}
For the dependent variables, we decided to analyze three different dimensions of the recommendations:

\begin{description}
	\item[\textbf{Ranking accuracy:}] We will focus on measuring how many recommended items actually match the ground truth of the user. For this purpose, we will use the \NDCG metric~\citep{DBLP:journals/tois/JarvelinK02} that is defined in Equations~\ref{f:ndcg} and~\ref{f:dcg}.
	\begin{equation}
		\label{f:ndcg}
		\mbox{nDCG@k} = \frac{1}{U}\sum_{u \in U} \frac{\mbox{DCG}(u)@k}{\mbox{IDCG}(u)@k}
	\end{equation}
	\begin{equation}
		\label{f:dcg}
		\mbox{DCG}(u)@k = \sum_{i_n \in RL_u@k}\frac{2^{rel_n}-1}{\log_2(n + 1)}
	\end{equation}
	where $RL_u$ is the recommendation list for user $u$, $rel_n$ denotes the real relevance of item $i_n$, and $k$ denotes the first $k$ items of $RL_u$. In explicit rating data sets, this relevance value is normally bounded in the $[0,5]$ interval, with $0$ representing a non-relevant value; in our experiments, as we do not have explicit ratings but \checkins, the relevance of the items appearing in the test set will always be $1$.
IDCG represents the ideal DCG, and it is computed in the same way as DCG but using the ground truth as the ranking.
	Higher values in \NDCG mean that more relevant recommendations are being provided to the users; that is, more recommended venues are actually visited by the user according to the test set.

	\item[\textbf{Novelty:}]
	The novelty of a recommendation can be assessed by measuring the proportion of popular venues being recommended. If a high percentage of the recommended venues are already well-known or frequently consumed/visited, it indicates that the recommendations lack novelty. To measure novelty, popularity is often used as a proxy, especially in offline evaluation where direct feedback from users is not possible.
	This is because it is generally assumed that whatever is popular within a community is likely to be known by most users and, thus, not novel.
	To measure this dimension, we use the Expected Popularity Complement (EPC)\footnote{Please, note that the original definition of the metric provided by  \cite{DBLP:conf/recsys/VargasC11} also incorporates a discount model (as the one used in the \NDCG metric) and a relevance model, in order to measure both the relevance and the novelty of the recommendations together. However, in our work, as we are evaluating ranking accuracy with \NDCG, we will use the pure definition of EPC.} metric~\citep{DBLP:conf/recsys/VargasC11}, defined in the following equation:
	\begin{equation}
		\label{f:EPC}
		EPC@k = \frac{1}{U} \sum_{u \in U} Z(u) \sum_{i_n \in RL_u@k} (1-p(\mbox{seen}\mid i_n))
	\end{equation}
	\noindent where $RL_u$ is again the recommendation list for user $u$, $Z(u)$ is a normalizing constant (generally
	$Z(u) = 1 / \sum_{i \in RL_u@k}{1}$), and $p(\mbox{seen}\mid i_n)$ represents the probability of item $i_n$ to be consumed. This probability is estimated as $|U_i|/|U|$, that is, the number of users who have visited POI $i$ in the training set, divided by the total number of users in the training set.
	Higher values in EPC imply that more novel recommendations are provided to the users.

	\item[\textbf{Item exposure:}]
	In traditional recommendation domains, many algorithms tend to emphasize only a few items from the entire catalog \citep{DBLP:conf/recsys/LiuZ20, DBLP:conf/flairs/AbdollahpouriBM19}, leading to a popularity bias which we discussed in \autoref{ss:Offline_evaluation}. This bias results in models favoring the most popular items, regardless of their relevance.
	In our study, to account for this effect, we measure item exposure by means of the so-called \textit{expected exposure loss}, which slightly differs from plain popularity \citep{DBLP:conf/sigir/ShihHJLLC16,Ekstrand2022}. While popularity bias means recommending the most popular items without considering the distribution in the ground truth, poor performance with respect to expected exposure loss occurs when items are over- or under-represented in recommendations compared to the test set.%
	Thus, we compare the number of times an item is recommended against the number of actual interactions in the test set \citep{Ekstrand2022}: %

	\begin{equation}
\label{eq:i_e}
		IE@k = \sum_{i \in I} \left| \frac{U_{test}(i)}{U_{test}} - \frac{Rec@k(i)}{U_{test}} \right|
	\end{equation}
	where $U_{test}$ denotes the number of users in the test set, $U_{test}(i)$ refers to the number of users that visited item $i$ in the test set, %
	and $Rec@k(i)$ is the number of times item $i$ has been recommended considering all recommendations (i.e., rankings) until position $k$, i.e., at cutoff $@k$.
	As we are comparing the recommended exposure and the exposure of the items in the test set, the lower the values obtained in this metric, the better the performance of the recommenders in terms of item exposure.

\end{description}

\section{Constructing Subsamples with Different Data Characteristics}
\label{sec:subsampling}

To apply the explanatory framework described in the previous section, it is necessary to obtain several subsamples from a larger recommendation data set.
We now discuss our approach to constructing various data sets with different characteristics that will serve as inputs for the explanatory framework.
This aspect poses a challenge since the recommendation data sets, specifically the user-item interaction matrices, exhibit interdependencies that are complex to disentangle.
The approach proposed by \cite{DBLP:conf/sigir/DeldjooNSM20} constructs subsamples by selecting a random number of users and items while enforcing certain constraints, such as predefined data set densities.

As opposed to prior studies \citep{Adomavicius2012, DBLP:journals/ipm/DeldjooBN21}, we propose to use subsamples created by exploiting different data characteristics that are grounded in the domain instead of randomly sampling users with constraints.
Generating the subsamples in such a way has two advantages:
first, the subsamples retain meaningful semantics, which provides additional analytic insights into the domain;
second, it allows providers of recommender systems to understand in which real-world situations different recommendation models are advantageous or unfavorable.
In the following section, we describe our proposed domain-driven subsampling procedure, which is based on subsetting a recommendation data set by factors that might be relevant to recommending \aclp{POI} within a city.

\subsection{Data Characteristics for Creating Domain-driven Subsamples}
\label{subsec:subsampling_variables}

Our design goals when developing the methods to construct subsamples are that
\begin{inparaenum}[\itshape i)]
	\item the subsamples must have a common basis to enable fair and meaningful comparisons,
	\item while at the same time, they show variability in terms of their resulting data characteristics.
	\item Further, we need to generate a sufficiently large number of subsamples to obtain robust results from the regression model, but
	\item each subsample should be a tractable recommendation problem, i.e., there is sufficient signal for the individual recommendation models to produce sensible recommendations.
\end{inparaenum}

As mentioned before, we take a different approach to construct subsamples compared to the approaches in the literature \citep{DBLP:journals/ipm/DeldjooBN21,DBLP:conf/sigir/DeldjooNSM20}.
We formulate hypotheses regarding relevant factors that might have an influence on the outcome of POI recommendations.
The core idea is to introduce a number of data characteristics relevant to the \ac{POI} recommendation domain and use them as filters to include an interaction in the UCM or not.
Thus, each data characteristic represents the explicit hypothesis that changing its value has an influence on the recommendation outcome.

The set of all subsamples is the cross-product of the data characteristics values applied to the original data set.
This means that the generation of the subsamples is not only controlled by the outcome of the random processes that define the number of items and users in the interaction matrix but by meaningful subsetting of groups of users, items, or interactions.
In the following subsections, we propose different data characteristics for subsampling,
as the \ac{POI} recommendation domain offers possibility for more complex hypotheses, since -- unlike most classic recommendation data sets -- there is the temporal (when a venue is visited) and the geographical (where the venue is and where the user is from) aspects to analyze.
We leverage these aspects to formulate hypotheses along with common strategies employed in the evaluation of \ac{POI} recommendation data sets to shape the recommendation outcome.

\subsubsection{Enforce a Minimum $k$-core}
To mitigate the extreme sparsity of typical POI recommendation data sets, it is common to remove interactions from the UCM until all users and venues have at least $k$ interactions.
This is done to achieve a certain -- higher -- level of density in the UCM and, thus, fewer `cold' items/users which usually results in higher accuracy metrics for interaction-based algorithms.
Typical values for $k$ are $2$ (cf.\, \cite{DBLP:conf/aaai/GaoTHL15}), %
$5$ (cf.\, \cite{DBLP:conf/sigir/YuanCMSM13, DBLP:conf/cikm/YuanCS14, DBLP:conf/sigir/YaoSQWSH15}),
or $10$ (cf.\, to \cite{DBLP:conf/la-web/NunesM14, DBLP:conf/ijcai/FengLZCCY15, DBLP:conf/kdd/LiGHZ16}).

\subsubsection{Drop Top $n$\% Popular Venues}
As previously discussed, popularity bias plays a large role in \ac{POI} recommendation with a substantial interplay between the popularity of items and recommendation accuracy \citep{DBLP:conf/flairs/AbdollahpouriBM19,DBLP:journals/jitt/MassimoR21,DBLP:journals/datamine/SanchezBB23}.
As check-in-based data sets usually do not come with rating information, we limit the scope of popularity to the number of people that have visited a venue \citep{Jannach2015}.

The method to analyze the impact of popularity bias is to generate different subsamples by removing the most popular venues in the data set.
We propose to drop the most $n$\% popular items from the data set to obtain different distributions of the item popularity \citep{Abdollahpouri2017}.
The concrete values depend on the data set at hand, but as a general guideline, we propose values for $n$ between $0$\% and $5$\% for the specific point-of-interest recommendation domain after considering the popularity bias discussed in Section~\ref{ss:Considerations}.

\subsubsection{Filter by Season}
Seasonality is also a relevant factor that undoubtedly has an influence on the visited venues both by locals and tourists \citep{Liu2011}.
The exact split between seasons can be tricky to make as a high granularity (e.g., weeks or months) can result in very small subsamples.
Further, seasons are not the same in different regions, potentially requiring different segmentations for destinations in different climate zones \citep{Trattner2018}.
In the context of an explainability study, we recommend using broad season categories, such as a two-season (warmer and colder months) or a four-season model.

\subsubsection{Filter by  User Residence}
\label{subsubsection:user_residence}
In the context of \ac{POI} recommendation, different groups of users exhibit different behavior~\citep{Sanchez2022a}.
Due to the differences in behavior between locals and visitors, we argue that it is very promising to use such a division as a subsampling variable.
If the information about the user's home is available in the recommendation data set, the typical groups to analyze would be the locals of the city, domestic visitors, or international travelers.

\begin{figure*}[t]

\begin{subfigure}[b]{0.31\textwidth}
  \includegraphics[width=\textwidth]{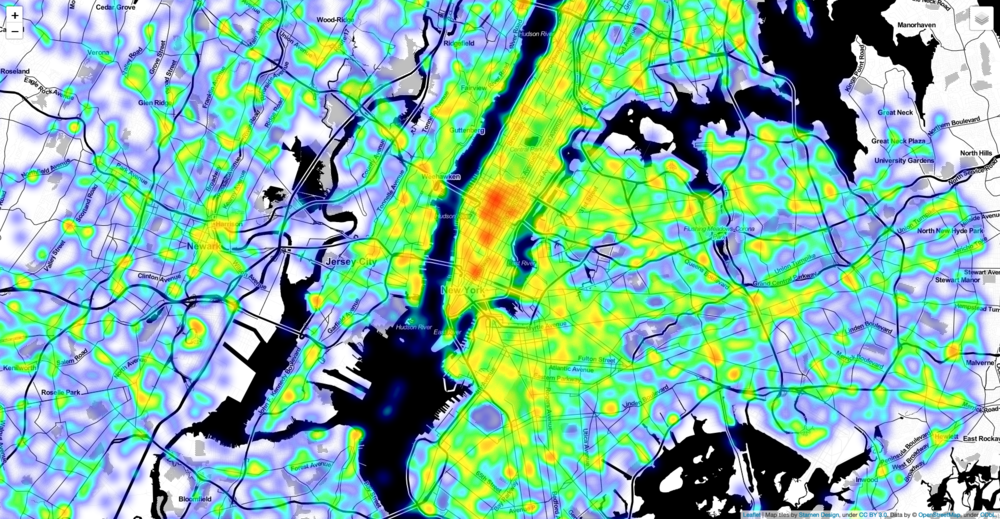}
  \subcaption{\scriptsize{$k$-core $ = 2$, origin $ = $ NYC}}
  \label{fig:/Users/dietzl/git/subsampling/paper/figures/heatmap_subsample_k2-homeNYC.html}
\end{subfigure}~\hspace{5pt}
\begin{subfigure}[b]{0.31\textwidth}
  \includegraphics[width=\textwidth]{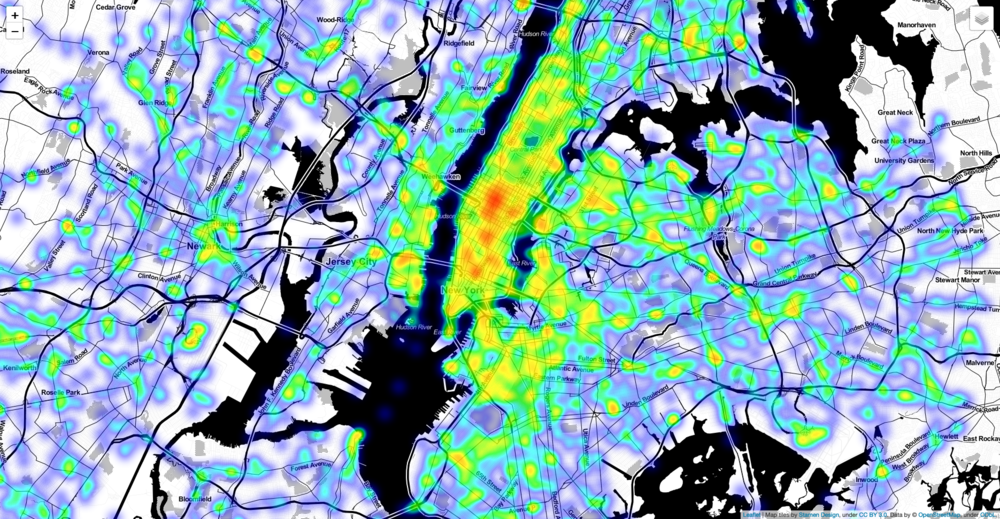}
  \subcaption{\scriptsize{$k$-core $ = 5$, origin $ = $ NYC}}
  \label{fig:/Users/dietzl/git/subsampling/paper/figures/heatmap_subsample_k5-homeNYC.html}
\end{subfigure}~\hspace{5pt}
\begin{subfigure}[b]{0.31\textwidth}
  \includegraphics[width=\textwidth]{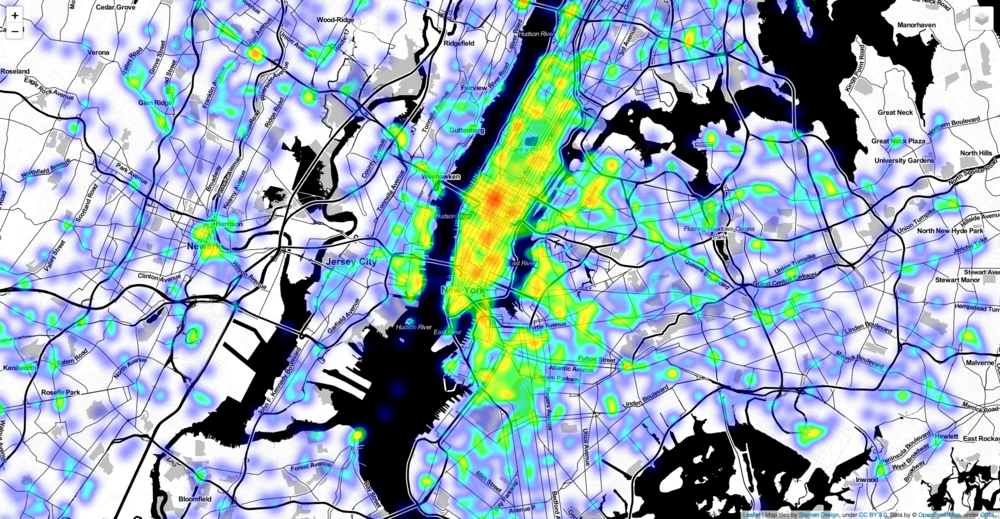}
  \subcaption{\scriptsize{$k$-core $ = 10$, origin $ = $ NYC}}
  \label{fig:/Users/dietzl/git/subsampling/paper/figures/heatmap_subsample_k10-homeNYC.html}
\end{subfigure}~\hspace{5pt}

\begin{subfigure}[b]{0.31\textwidth}
  \includegraphics[width=\textwidth]{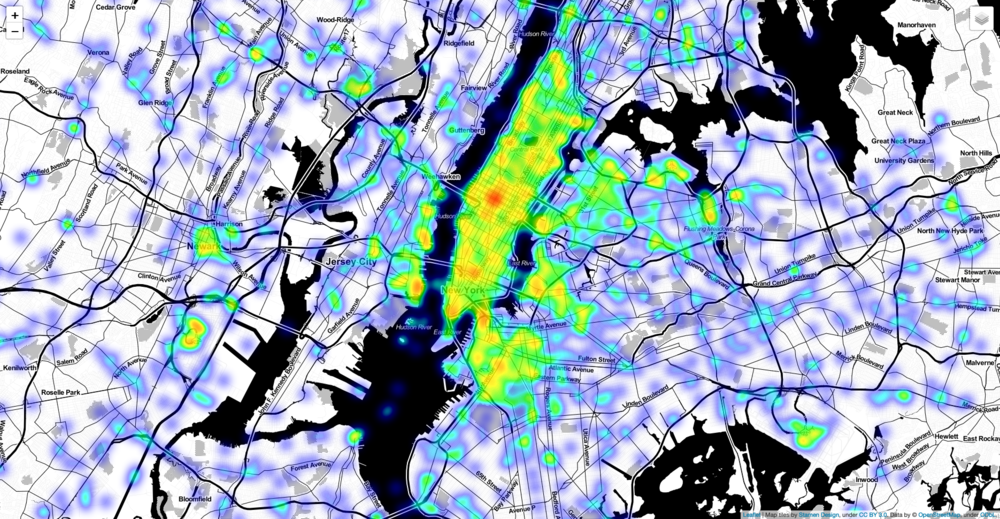}
  \subcaption{\scriptsize{$k$-core $ = 2$, origin $ = $ US}}
  \label{fig:/Users/dietzl/git/subsampling/paper/figures/heatmap_subsample_k2-homeUS.html}
\end{subfigure}~\hspace{5pt}
\begin{subfigure}[b]{0.31\textwidth}
  \includegraphics[width=\textwidth]{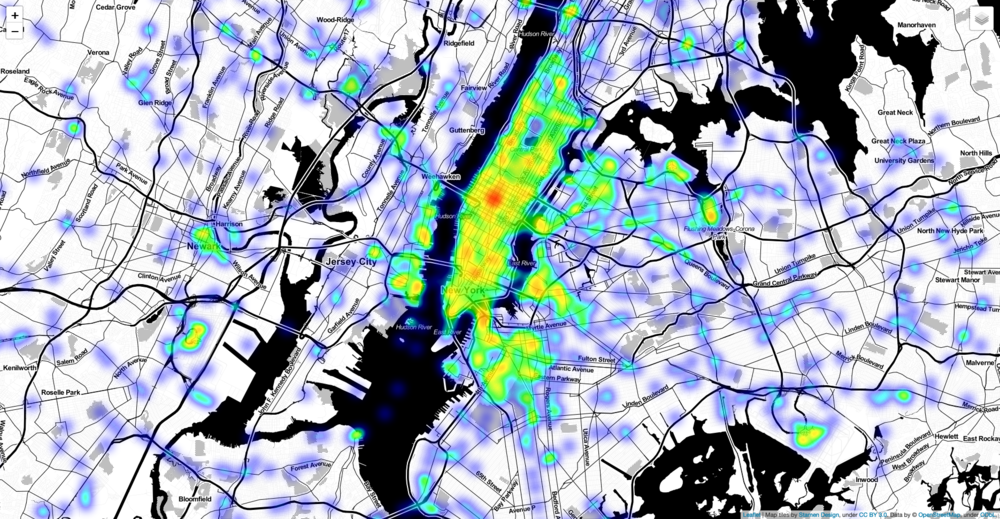}
  \subcaption{\scriptsize{$k$-core $ = 5$, origin $ = $ US}}
  \label{fig:/Users/dietzl/git/subsampling/paper/figures/heatmap_subsample_k5-homeUS.html}
\end{subfigure}~\hspace{5pt}
\begin{subfigure}[b]{0.31\textwidth}
  \includegraphics[width=\textwidth]{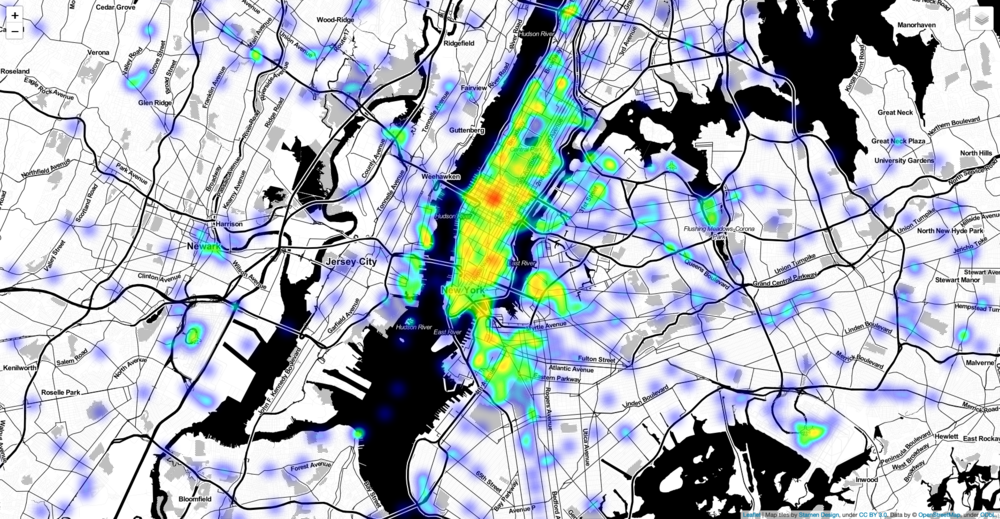}
  \subcaption{\scriptsize{$k$-core $ = 10$, origin $ = $ US}}
  \label{fig:/Users/dietzl/git/subsampling/paper/figures/heatmap_subsample_k10-homeUS.html}
\end{subfigure}~\hspace{5pt}

\begin{subfigure}[b]{0.31\textwidth}
  \includegraphics[width=\textwidth]{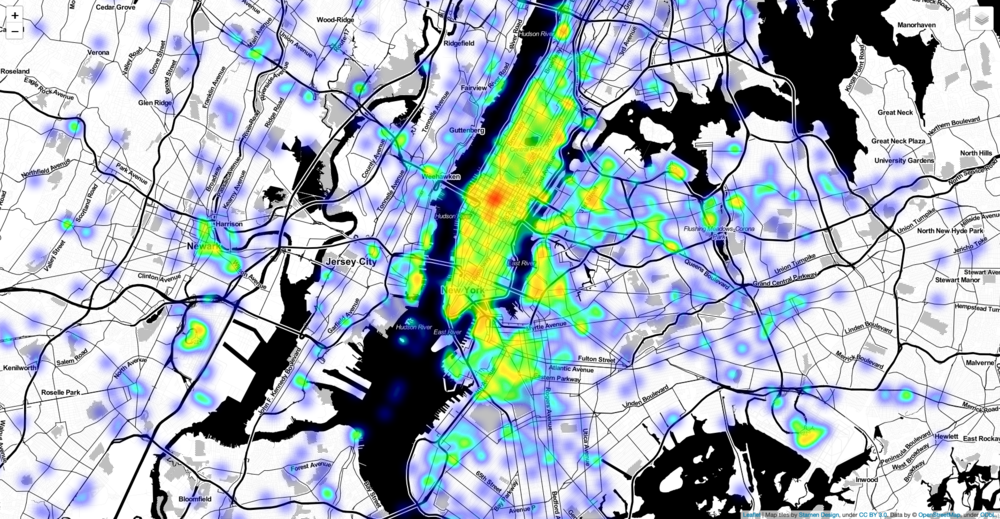}
  \subcaption{\scriptsize{$k$-core $ = 2$, origin $ = $ other}}
  \label{fig:/Users/dietzl/git/subsampling/paper/figures/heatmap_subsample_k2-homeother.html}
\end{subfigure}~\hspace{5pt}
\begin{subfigure}[b]{0.31\textwidth}
  \includegraphics[width=\textwidth]{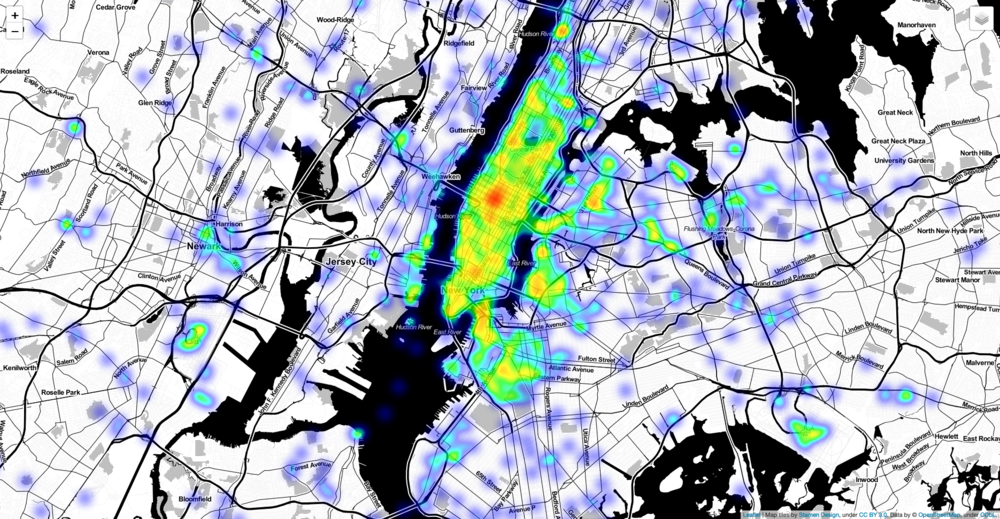}
  \subcaption{\scriptsize{$k$-core $ = 5$, origin $ = $ other}}
  \label{fig:/Users/dietzl/git/subsampling/paper/figures/heatmap_subsample_k5-homeother.html}
\end{subfigure}~\hspace{5pt}
\begin{subfigure}[b]{0.31\textwidth}
  \includegraphics[width=\textwidth]{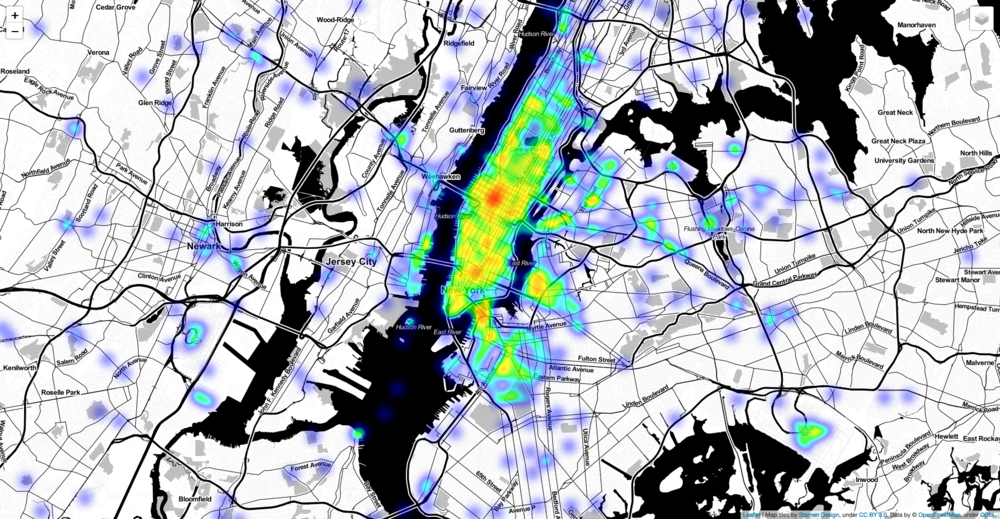}
  \subcaption{\scriptsize{$k$-core $ = 10$, origin $ = $ other}}
  \label{fig:/Users/dietzl/git/subsampling/paper/figures/heatmap_subsample_k10-homeother.html}
\end{subfigure}~\hspace{5pt}\caption{Heat map of the visited venues in New York with different parameters for the $k$-core and the origin of the users (better viewed in color).}
\label{fig:heatmaps_nyc}
\end{figure*}

\subsection{Discussion}

In this section, we have formulated various hypotheses of what influences \ac{POI} recommendations.
These hypotheses are manifested in different data characteristics to enable a rigorous computational analysis.
In the choice of data characteristics, we discussed the ones that we deem to be most relevant based on the literature on the analysis \ac{POI} recommendation algorithms.

When setting up the experiments, it is still necessary to analyze the statistics of the resulting subsamples to understand which value ranges of the data characteristics to test.
Here, it is important to retain tractable recommendation problems, i.e., not result in too sparse or small subsamples.
Also, depending on the data set, it is not always possible to test all data characteristics discussed in \autoref{subsec:subsampling_variables}.
For example, removing resident users in a city that is not very active as a tourist destination could be counterproductive in making interesting recommendations that may attract more tourists.

\subsection{Visualizing the Subsampling Variations}

To exemplify the effect of subsampling variables,  \autoref{fig:heatmaps_nyc} visualizes the interplay of two data characteristics: %
the $k$-core and the origin of the users.
In this heat map showing the density of check-ins in New York City, USA, the difference between the behavior of the locals and travelers becomes apparent:
travelers tend to visit venues in Manhattan (with the exception of the airports), while the locals naturally have check-ins all over the map.
A higher value for $k$-core leads to a higher density in the UCM but eliminates many venues, which can be observed in the map visualization.

\section{Experimental Setup}
\label{sec:exp_setup}

\begin{figure*}[t]
	\includegraphics[width=\textwidth]{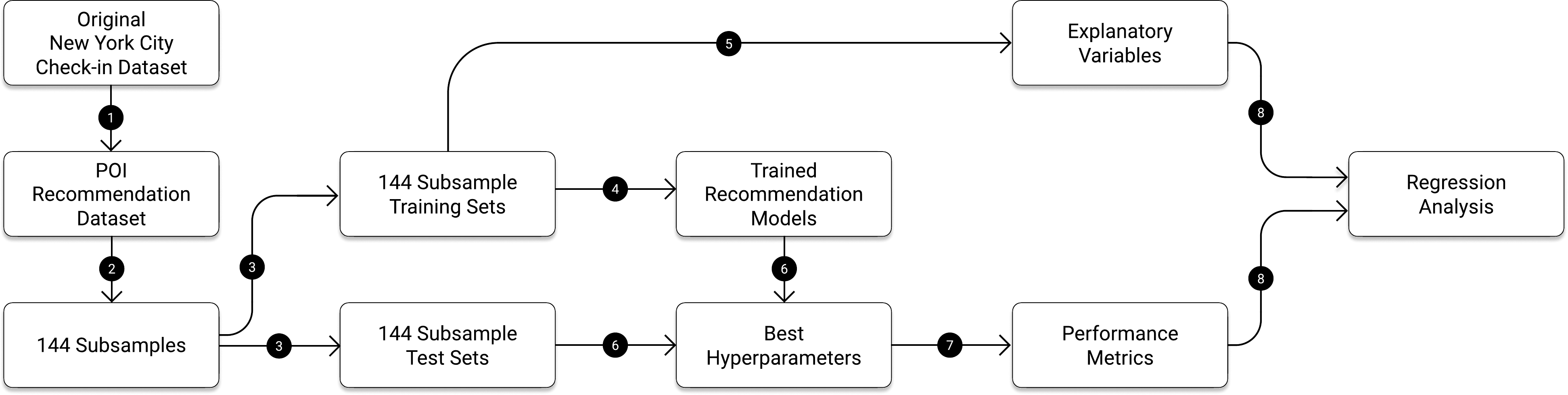}
    \caption{Diagram representing the methodology followed in the paper. Each number corresponds to a step in the process: Initially, we clean the original check-in data set~(1) to obtain a recommendation data set, which we subsequently subdivide into the 144~subsamples~(2).
    Each subsample is further split into training and test sets (3), where a recommendation model is trained on each subsample individually~(4) and the explanatory variables are computed based on the training sets~(5).
    We determine the best hyperparameters of each recommender in each subsample using the test set~(6),  and record the metrics of the best performing recommendation configuration~(7). Finally, we perform the regression analysis towards the performance metrics of each recommendation algorithm using the explanatory variables~(8).}
\label{fig:methodology}
\end{figure*}

In this section, we describe the used data set and the process of selecting the subsampling variables to obtain the subsampled recommendation data sets.
We provide details about the data preprocessing, how we conducted the recommendation experiments, and give an overview of the outcomes.
Finally, we outline the variable selection process for the regression model, which is the core of the explanatory framework. The full process explained in this section is shown in Figure~\ref{fig:methodology}.

\subsection{Selecting a Suitable Data Set}
\label{subsec:data_set}

To evaluate our proposed approach, %
it is essential to have a sufficiently large data set for \ac{POI} recommendation that enables us to perform meaningful subsampling.
Revisiting the literature \citep{Bao2015,Sanchez2022}, we decided to use the  Foursquare data set published by \citeauthor{Yang2015}, which has about 33 million check-ins in 415 different cities of the world \citep{Yang2015} and has been frequently used to benchmark \ac{POI} recommendation performance\footnote{Data set is available from \url{https://sites.google.com/site/yangdingqi/home/foursquare-dataset}}. %
Although the data set contains check-ins from many cities, we conclude that it is infeasible to include multiple cities in the scope of the analysis.
The number of check-ins per city is influenced by the number of people and the popularity of Foursquare in the city, which results in a few cities having a large number of check-ins but many not having sufficient interactions to further subsample them.
Furthermore, the behavior of users is influenced by the topological realities of cities, such as centralized vs. decentralized cities.
Therefore, each city needs to be analyzed separately to account for the different geographic influences.

In this study, we focus on New York City (NYC), NY, USA, as it is one of the most active cities on Foursquare and, hence, has been a common subject of analysis in the POI recommendation domain, e.g., in \citep{DBLP:journals/ijgi/AlbannaSMM16, DBLP:conf/bigdataconf/MaroulisBK16, DBLP:journals/fgcs/JiaoXZWH19}.
Concretely, the scope of our analysis is the New York City Metropolitan Area, which comprises the five boroughs of New York City and Newark, NJ. The complete data set from New York City Metropolitan Area consists of 17,467 users, 71,310 venues, and 608,131 \checkins.
The geographical scope of the analysis is visualized in~\autoref{fig:heatmaps_nyc}. %
To adapt the data set for the \ac{POI} recommendation domain, we eliminated venues of the ``Residences'' category, as we do not consider them as interesting \acp{POI} to recommend.
Finally, we also removed duplicated \checkins, i.e., \checkins at the same venue and identical timestamps.%

\subsection{Generation of Subsampled Recommendation Data Sets}

In \autoref{sec:subsampling}, we discussed subsampling data characteristics to be used in an explanatory framework for \ac{POI} recommendation.
In the context of this study we instantiated them as follows: We imposed a minimum $k$-core of the recommendation data sets, excluded varying levels of the most popular venues, and subdivided by season of the year and the origin of the user.

\subsubsection{Subsampling Data Characteristics}

\paragraph{\ac{UCM} Density}

Traditionally, density (i.e., the inverse of sparsity) has been a key metric to quantify the difficulty of a recommendation problem. The sparsity is normally referred to as the situation where most of the user-item interactions are not observed in the training data \citep{DBLP:journals/snam/IdrissiZ20}.

However, density is a dependent variable, which is typically adjusted by enforcing a $k$-core, i.e., requiring at least $k$ interactions for each user and venue and discarding users and venues that do not fulfill this threshold.

Following the practice in literature, we create subsamples using the following values for $k$:
\begin{itemize}
	\item \textbf{2:} Enforce a $k=2$-core.
	\item \textbf{5:} Enforce a $k=5$-core.
	\item \textbf{10:} Enforce a $k=10$-core.
\end{itemize}

\paragraph{Item Popularity}

Due to the large popularity bias in \ac{POI} recommendation, we argue that it is important to analyze the effect of disregarding the most popular venues.
We used the following values to analyze this effect:
\begin{itemize}
	\item \textbf{0.5:} Drop the most popular 0.5\% venues from the current data set.
	\item \textbf{1:} Drop the most popular 1\% venues from the current data set.
	\item \textbf{2:} Drop the most popular 2\% venues from the current data set.
	\item \textbf{5:} Drop the most popular 5\% venues from the current data set.
\end{itemize}

\paragraph{Season}

The effect of seasonality on the recommendation outcome has not been analyzed in depth so far, providing us with the opportunity to analyze this within the explanatory study.
The oceanic climate of  New York City comes with relatively similar precipitation throughout the year, thus, we used the temperature aspect of the climate diagram to subdivide the year into the following groups:
\begin{itemize}
	\item \textbf{all:} All \checkins irrespective of the season.
	\item\textbf{summer:} \checkins during the warmer months in New York City from May to October.
	\item \textbf{winter:} \checkins during the colder months in New York City from November to April.
\end{itemize}

By using only two groups, we hope to achieve a clear separation and keep the size of the resulting subsamples larger.

\paragraph{User Origin}

The Foursquare data set contains \checkins from users from all around the world.
Although we are only running the recommendation experiments with check-ins in New York City, we can use the complete data set to determine the home city of the people in the recommendation data set.
To achieve this, we use the open-source tripmining library\footnote{\url{https://github.com/LinusDietz/tripmining}} to obtain the residence of the different users \citep{Dietz2020}. %
This library converts the check-in stream of users into periods of being at home and on travel.
It uses the plurality strategy, i.e., selecting the city with the most check-ins as home city, to determine which is the user's home city, which has been shown to be accurate in a ground-truth study~\citep{Kariryaa2018}.

We use this user home label as a subsampling data characteristic with the following values:
\begin{itemize}
	\item \textbf{all:} considering all users.
	\item \textbf{US:} only domestic visitors from the United States, but not citizens of New York City.
	\item \textbf{NYC:} citizens of New York City. %
	\item  \textbf{other:} travelers from outside of the US.
\end{itemize}

The intuition behind using the users' home as a subsampling data characteristic is that the behavior of locals is different than the one of visitors.
This also has a significant influence on the recommendation outcome, as shown by \cite{Sanchez2022a}.

\subsubsection{Summary}
\label{subsec:summary_subsampling}

Using these aspects, we generate 144 recommendation subsamples in the form of \aclp{UCM}.
The result of the cross-product of applying the aforementioned subsampling data characteristics as filters on the original data set is 144:

\begin{center}
 \{origin = [all, NYC, US, international]\} $\times$ \{season= [all, summer,winter]\} $\times$ \newline $\times$ \{k-core = [2, 5, 10]\} $\times$ \{drop top venues = [0.5, 1, 2, 5]\}
\end{center}

\subsubsection{Training and Test Set Generation}

For each of the subsamples, we perform a temporal split per user, where 80\% of the oldest interactions of each user are sent to the training set and the rest to the test set.
Foursquare users might have performed \checkins at the same venue more than once, but the algorithms we use are meant to recommend new items, which means that we discard all duplicate check-ins of a user in the same venue both in the training and the test set.
We decided to proceed like this in the test set because the goal of recommender systems is to recommend new venues to users to explore, not venues that the user already knows, which is common practice in the \ac{POI} recommendation domain.

\autoref{tab:subsampling_outcome} (in the Appendix) tabulates statistics regarding the subsamples with the average
values of different explanatory variables.
For space reasons, we only tabulate the 12 subsampling data characteristics independently to get an impression of their individual impact.
The experiments for the explanatory analysis used the cross-product of the subsampling data characteristics, resulting in 144 subsamples.

\subsection{Algorithms for POI Recommendation}
\label{ss:baselines}

In this section, we explain in detail the
algorithms that we have used in our experiments.
Due to the nature of the application of the different models, we will divide them into two main groups: classical recommendation algorithms and those specifically designed for %
point-of-interest recommendation.

\subsubsection{Classical Recommendation Algorithms}
	\begin{description}
		\item\textbf{Random:} performs recommendations of venues randomly.
		\item \textbf{\Popularity:} recommends to the target user the venues ordered by decreasing popularity. The popularity is measured by the number of different users that have visited that venue.
		\item \textbf{\UB:} user-based neighborhood. Non-normalized $k$-nn algorithm that recommends to the target user venues that other similar users visited before \citep{DBLP:reference/sp/NikolakopoulosNDK22, DBLP:conf/recsys/Aiolli13}. We used the cosine similarity and Jaccard Index as similarity metrics. %
		\item \textbf{\IB: }item-based neighborhood. Non-normalized $k$-nn that recommends to the target user venues similar to the ones that she visited previously \citep{DBLP:reference/sp/NikolakopoulosNDK22, DBLP:conf/recsys/Aiolli13}. We use the item variations of the same similarity metrics used for \UB. %
		\item \textbf{\HKV:} matrix factorization (MF) algorithm that uses Alternate Least Squares for optimization proposed by \cite{DBLP:conf/icdm/HuKV08}.
		\item \textbf{\BPRMF:} matrix factorization (MF) algorithm that uses the pairwise Bayesian Personalized Ranking loss proposed by \cite{DBLP:conf/uai/RendleFGS09} as optimization algorithm. For our experiments. we used the version from MyMedialite's\footnote{MyMedialite library: \url{http://www.mymedialite.net/}} library.
	\end{description}
\subsubsection{Point-of-Interest Recommendation Algorithms}

	\begin{description}
		\item \textbf{\IRENMF:} weighted matrix factorization method proposed by \cite{DBLP:conf/cikm/LiuWSM14}.
		This algorithm incorporates geographical information by assuming that users tend to visit neighboring venues (instance-level influence) and also by considering that the users \checkins are shared in the same geographical region (region-level influence).
		\item \textbf{\GeoBPRMF:} geographical Bayesian Personalized Ranking matrix factorization. Algorithm proposed by \cite{DBLP:conf/ictai/YuanJGCYA16} that assumes that the target user will prefer to visit new venues that are close to the ones she visited previously.
		\item \textbf{\RankGeoFM:} a ranking-based matrix factorization model proposed by \cite{DBLP:conf/sigir/LiCLPK15}. They also incorporate the geographical influence in the recommendations by exploiting the neighboring venues (by geographical distance) with respect to the candidate POIs to recommend.
		\item \textbf{\PopGeoNN:} hybrid algorithm combining popularity (\Popularity), a user-based neighborhood method (\UB), and a simple geographical component that recommends to the target user the venues closer to the average geographical position of all the venues visited by the user in the training set.
				This recommender has been used in previous works such as \citep{Sanchez2022a}.
				The final score is an aggregation of every item score provided by each recommender after normalizing its values by the maximum score of each method.
	\end{description}

To achieve optimal parameters for each recommendation subsample, we systematically chose the optimal hyperparameters for each recommendation model by \NDCGAux@5.
We do this as it is a standard procedure in the area, despite the models could be optimized independently for each evaluation dimension, however, this would not be practical, as accuracy is typically understood as a first order-objective of any recommender system.
The tested hyperparameter ranges are listed in the Appendix, \autoref{tab:Parameters}.

\subsection{Recommendation Results on Subsamples}
\label{subsec:recommendations}

\begin{figure*}[p]
	\includegraphics[width=1\textwidth]{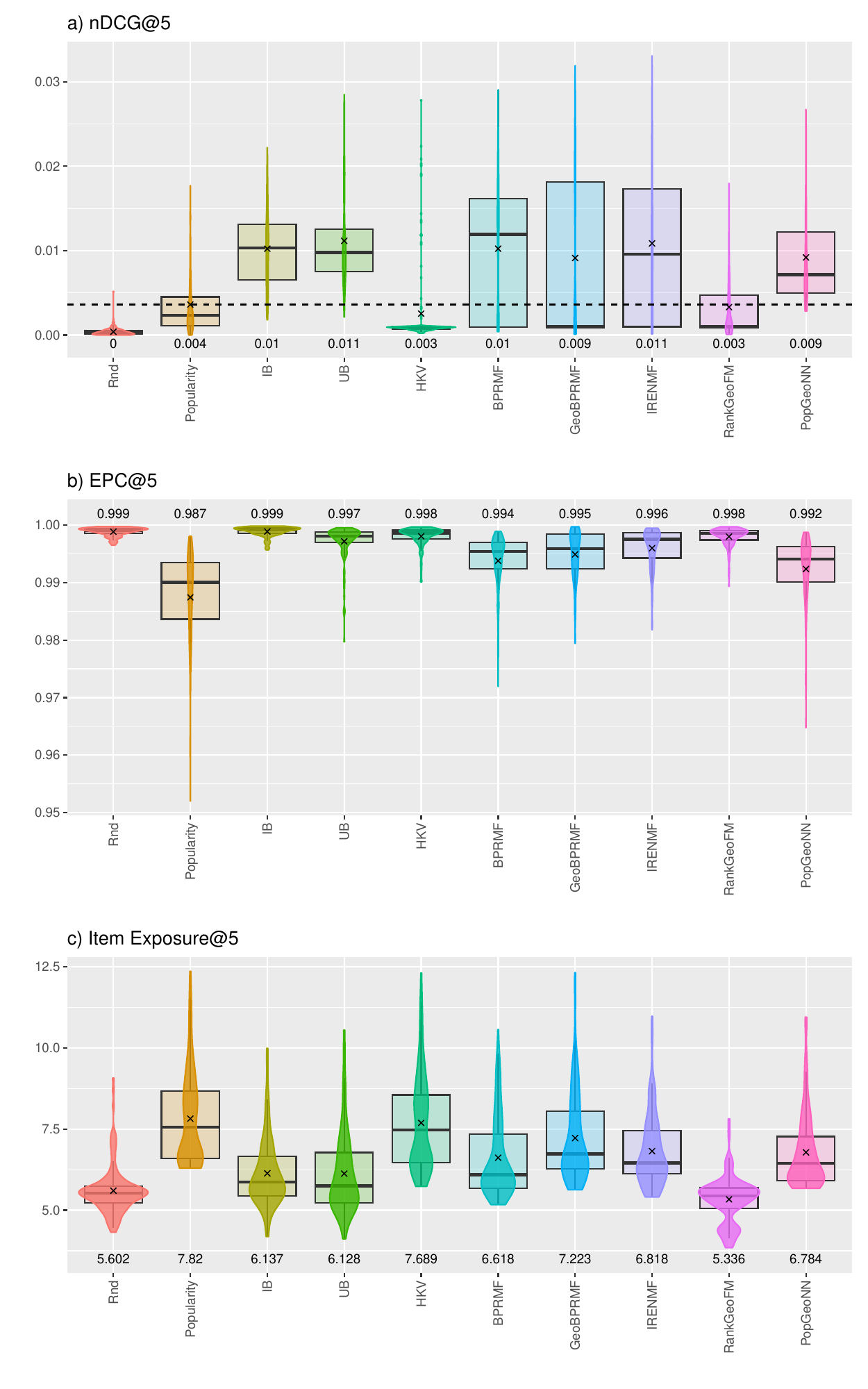}
	\caption{The recommendation outcomes using the following metrics: a) \NDCGAux@5, b) EPC@5, and c) Item Exposure@5.
		The boxplot indicates the 25\%, the median, and the 75\% quantiles.
		Overlayed is a violin plot emphasizing the density of values and the overall range of the outcomes and the mean value with an x.
		The dashed line in the nDCG plot of \autoref{fig:recommendation_outcome}a) indicates the mean value of the Popularity algorithm.
	}
	\label{fig:recommendation_outcome}
\end{figure*}

Using the recommendation models presented in \autoref{ss:baselines} with the optimal parameters, we achieve the following recommendation outcomes regarding \NDCGAux ~(\autoref{fig:recommendation_outcome}a), EPC~(\autoref{fig:recommendation_outcome}b), and Item Exposure~(\autoref{fig:recommendation_outcome}c)  in the 144 subsamples. We also report in Table~\ref{t:ResultsMeanAllRecommenders} the average (denoted as avg) and the standard deviation (denoted as std) by each recommender in nDCG, EPC, and Item Exposure, all of them measured at a ranking cutoff of 5.

\pgfplotstableread{
Family	ShortRec	aggrdiv@20	RealAD@20	epc@5-rec	giniITr@10	Recall@10-1-rec	gini-rel@20	epc@20-rec	Recall@20-1-rec	Recall@5-1-rec	RelUsers@20	epcMixMaxNorm@20-rec	RealAD@5	giniITr-rel@10	efd@5-rec	RealAD@10	epcMixMaxNorm@10-rec	MRR@20-1-rec	ARHR@20-1-rec	gini-rel@10	MAP@10-1-rec	MRR@5-1-rec	Precision@20-1-rec	gini@10	Success@10-1-rec	aggrdiv@5	AverageItemExposureSimpleDifference@5	NDCG@20-1-rec	MAP@5-1-rec	giniITr-rel@20	RelUsers@5	AverageItemExposureSimpleDifference@20	RealAD-rel@5	gini-rel@5	gini@20	giniITr-rel@5	ARHR@5-1-rec	NDCG@5-1-rec	Success@5-1-rec	efd@10-rec	usercov-rel	RealAD-rel@10	aggrdiv@10	AverageItemExposureSimpleDifference@10	Precision@10-1-rec	epc@10-rec	usercov	giniITr@20	MRR@10-1-rec	ARHR@10-1-rec	epcMixMaxNorm@5-rec	efd@20-rec	gini@5	Success@20-1-rec	RelUsers@10	giniITr@5	MAP@20-1-rec	RealAD-rel@20	NDCG@10-1-rec	Precision@5-1-rec	k	dtv	aggrdiv@20-std	RealAD@20-std	epc@5-rec-std	giniITr@10-std	Recall@10-1-rec-std	gini-rel@20-std	epc@20-rec-std	Recall@20-1-rec-std	Recall@5-1-rec-std	RelUsers@20-std	epcMixMaxNorm@20-rec-std	RealAD@5-std	giniITr-rel@10-std	efd@5-rec-std	RealAD@10-std	epcMixMaxNorm@10-rec-std	MRR@20-1-rec-std	ARHR@20-1-rec-std	gini-rel@10-std	MAP@10-1-rec-std	MRR@5-1-rec-std	Precision@20-1-rec-std	gini@10-std	Success@10-1-rec-std	aggrdiv@5-std	AverageItemExposureSimpleDifference@5-std	NDCG@20-1-rec-std	MAP@5-1-rec-std	giniITr-rel@20-std	RelUsers@5-std	AverageItemExposureSimpleDifference@20-std	RealAD-rel@5-std	gini-rel@5-std	gini@20-std	giniITr-rel@5-std	ARHR@5-1-rec-std	NDCG@5-1-rec-std	Success@5-1-rec-std	efd@10-rec-std	usercov-rel-std	RealAD-rel@10-std	aggrdiv@10-std	AverageItemExposureSimpleDifference@10-std	Precision@10-1-rec-std	epc@10-rec-std	usercov-std	giniITr@20-std	MRR@10-1-rec-std	ARHR@10-1-rec-std	epcMixMaxNorm@5-rec-std	efd@20-rec-std	gini@5-std	Success@20-1-rec-std	RelUsers@10-std	giniITr@5-std	MAP@20-1-rec-std	RealAD-rel@20-std	NDCG@10-1-rec-std	Precision@5-1-rec-std	k-std	dtv-std
Classic	{\Rnd}	12529.9375	12529.9375	0.998838128329313	0.645267297757935	0.00103361086466242	0.690799083782581	0.99883724249371	0.00209599827423502	0.000480908772445887	13.5208333333333	0.945512990623301	8660.80555555556	0.645267297757935	13.5979609377158	11254.5138888889	0.945502183386357	0.000915172404185658	0.000254575232776358	0.599084963596936	0.000293640912898072	0.000587957398469034	0.000251567243477869	0.599084963596936	0.00252735808589834	8660.80555555556	5.60177574830172	0.000907849085513301	0.00023431285994197	0.744617563704996	3.51388888888889	18.5723644912926	8660.80555555556	0.480084768022881	0.690799083782581	0.516420510192143	0.000257499590570368	0.000390943081610337	0.00129971642685442	13.597477565942	3158.94444444444	11254.5138888889	11254.5138888889	9.48448548188373	0.000252884799030573	0.998837610063271	3158.94444444444	0.744617563704996	0.000747676931948991	0.00025535450865229	0.945566083617693	13.5972069189759	0.480084768022881	0.00502036118316587	6.90277777777778	0.516420510192143	0.000363326323107632	12529.9375	0.000585586789743504	0.000259943285370856	5.66666666666667	2.125	10240.8350205998	10240.8350205998	0.000756163424361777	0.0532885777734727	0.00109717402994456	0.0643567609984103	0.000758585169117875	0.00186654711826781	0.000750224675551263	11.657686770503	0.0293785946153198	6480.36763965932	0.0532885777734727	1.40826668879333	8785.61174421483	0.0293538590797297	0.000815419808117414	0.000227038792234221	0.0720759675504136	0.000403413299768465	0.00070222603073056	0.000174171596911341	0.0720759675504136	0.00196699751683793	6480.36763965932	0.736864081351786	0.000798654790385402	0.000386618182866551	0.0399376322912103	3.55768863268674	1.28505736571906	6480.36763965932	0.0792357887934537	0.0643567609984103	0.0658009507091234	0.000307544247035272	0.000534240540848522	0.00142582444291717	1.40797607872613	2346.12845729954	8785.61174421483	8785.61174421483	0.852811034990782	0.00019674420093651	0.000757382804696054	2346.12845729954	0.0399376322912103	0.000740211098327943	0.00025279647629537	0.0292739873290658	1.4079618037151	0.0792357887934537	0.00345650636543303	6.37666138444017	0.0658009507091234	0.000438338997177602	10240.8350205998	0.000606223898000099	0.000285164888583421	3.31134941753886	1.75162262436343
Classic	{\Popularity}	25.1944444444452	25.1944444444444	0.987424544940013	0.00144955683602149	0.00982014948393368	0.00273415293672339	0.98983838934161	0.0180418844115087	0.00505693355272596	89.5416666666667	0.280920119703792	7.93055555555556	0.00144955683602149	9.74140157216673	13.9236111111111	0.200381076633879	0.00665531334956334	0.00190938656882055	0.00130107427543345	0.00298665097986482	0.00437700847678776	0.00184060417430286	0.00130107427543345	0.0187869818294838	7.93055555555574	7.82030355982778	0.00774483513849185	0.00241340460296649	0.00304606166640605	21.2777777777778	22.7620157519873	7.93055555555556	0.000583025194138506	0.00273415293672339	0.00064960694880561	0.00194821484459355	0.00360705041467	0.00958659203064564	9.87304844745969	3158.94444444444	13.9236111111111	13.9236111111115	12.799700703818	0.00194893553818318	0.988589009139079	3158.94444444444	0.00304606166640605	0.00558051361234228	0.00194877958951997	0.125501116291107	10.0299391301741	0.000583025194138506	0.0347911119748995	44.7152777777778	0.00064960694880561	0.00355091557798841	25.1944444444444	0.0053209713329257	0.00196762227363553	5.66666666666667	2.125	2.15249985673795	2.15249985673766	0.00876635501580505	0.00118781925762199	0.00898000669296275	0.00213660339458949	0.00663871301028502	0.0147974046530509	0.00514725871897124	87.2312155425738	0.112004821876434	1.10720954471725	0.00118781925762199	1.17455010694288	1.58700959546546	0.0944217232795881	0.00584642937895609	0.00170603152979646	0.00102013654663039	0.00313466899207334	0.00448582185428898	0.00135055245220948	0.00102013654663039	0.0151525634358204	1.10720954471732	1.42960742621225	0.00660370790220702	0.00265562496416144	0.00248808550151486	20.958648746161	1.42835440331347	1.10720954471725	0.000459953039889308	0.00213660339458949	0.000535452407308043	0.00203656889381669	0.00368686523985484	0.00888007232420947	1.13651014994946	2346.12845729954	1.58700959546546	1.58700959546562	1.42991693537613	0.00162064435937884	0.00773360959083526	2346.12845729954	0.00248808550151486	0.00525078060500832	0.00186897795599809	0.075194302909815	1.08720631735943	0.000459953039889308	0.0248317871314893	43.457826706132	0.000535452407308043	0.00353765744906082	2.15249985673766	0.00499899516400904	0.00191221586356036	3.31134941753886	1.75162262436343
Classic	{\UB}	8779.98463628294	8349.27083333333	0.99716476271092	0.251570449557466	0.0232805067033249	0.303963311868356	0.997347243329883	0.0348709020388312	0.0144693929916234	232.159722222222	0.852459573909603	3906.85416666667	0.251570449557466	12.7805606814093	5961.59027777778	0.850765596673083	0.017580901894642	0.00548518661494387	0.231874591344283	0.00901091904636532	0.0136649106606824	0.00400534618077476	0.231874591344283	0.0439946751231501	3954.65921630205	6.127561775739	0.0179616461159182	0.00797056846606467	0.329407077641452	82.6041666666667	17.8033725941584	3906.85416666667	0.162621265832078	0.303963311868356	0.176443088514447	0.00638277745816017	0.0111543637198152	0.0267280904102111	12.7483283874325	3122.65972222222	5961.59027777778	6117.68787701461	9.89538549467816	0.00496147575050743	0.997246449206171	3122.65972222222	0.329407077641452	0.0159205443636294	0.00595130395389552	0.849716696472952	12.718982598515	0.162621265832078	0.0684232449390766	142.25	0.176443088514447	0.00989497850336706	8349.27083333333	0.0143178344243993	0.00582600044179402	5.66666666666667	2.125	5597.10061142863	5853.39558045083	0.00306719364140887	0.0786095379829665	0.012051922563664	0.0778730831535723	0.00241413705098793	0.0175377789334877	0.00775092581942061	210.472185202649	0.0445249014325953	2518.86198071363	0.0786095379829665	1.49583320087732	3933.34153997776	0.0447961519494008	0.00801712852663742	0.00254071948007735	0.0699204580458744	0.00505154199202587	0.0063909575706472	0.00206875059674038	0.0699204580458744	0.0202930509526408	2490.48239488261	1.23005069576039	0.00862761416335936	0.00454164613105818	0.0869337766067348	68.3667832024774	2.28555266780773	2518.86198071363	0.0534334690435109	0.0778730831535723	0.0598385885133234	0.00303936859837186	0.00563895823777711	0.011929475393751	1.43296798914544	2369.87325547964	3933.34153997776	3839.14906158986	1.29903080432623	0.00240116741277525	0.00275906174246611	2369.87325547964	0.0869337766067348	0.00733024081517555	0.0027882415668592	0.047358921776256	1.35433012031897	0.0534334690435109	0.0330130069167323	122.667125898603	0.0598385885133234	0.0054226439926634	5853.39558045083	0.00707953776446033	0.00273742876810941	3.31134941753886	1.75162262436343
Classic	{\IB}	11439.8991500213	10940.1805555556	0.998876339766251	0.401677772522665	0.0178661003833658	0.450661924872837	0.998785126409237	0.0258211913388077	0.0121940455432468	194.805555555556	0.945604425446891	6145.40277777778	0.401677772522665	13.8262412189842	8748.34722222222	0.951748805033729	0.0162218870388702	0.00512972271438514	0.373239770731358	0.00791810627954458	0.013541259093796	0.00320080048890406	0.373239770731358	0.0362446752344935	6202.22810007992	6.13730205590815	0.0146384996654012	0.00735185933681742	0.485125409590765	85.7708333333333	17.8522928251979	6145.40277777778	0.279004179275273	0.450661924872837	0.299693388440281	0.00639915738043875	0.0102250670080941	0.0245791173830397	13.771619130656	3122.72916666667	8748.34722222222	8940.6609924707	9.92346393119661	0.00424512288493981	0.998846565553097	3122.72916666667	0.485125409590765	0.0150740563146768	0.00572443514128115	0.955235770923584	13.6788108759409	0.279004179275273	0.0531527437227552	130.229166666667	0.299693388440281	0.00851284863984351	10940.1805555556	0.0121601416271644	0.00548184594552776	5.66666666666667	2.125	8463.95101983331	8770.03681938255	0.000887949150198794	0.0516278515954949	0.00844801045540281	0.0629567458143916	0.000917447079008579	0.0116253954670869	0.00623306536735912	197.021644396062	0.0249135586084563	4537.41902132667	0.0516278515954949	1.53200396174732	6648.10390584633	0.0252919478854485	0.00833845783269505	0.00257057809399275	0.0598152781145063	0.0036964936280625	0.00683308460825322	0.00188469582891445	0.0598152781145063	0.0199375202258137	4495.70816109795	1.02552611830438	0.00640337949518816	0.00350491320648162	0.0521685701544727	83.0494602610418	2.2581549126968	4537.41902132667	0.0583353253645057	0.0629567458143916	0.0514534942208084	0.00320817173724942	0.00469224165566292	0.0129167981278852	1.5495267704703	2369.97971493709	6648.10390584633	6517.39680013096	1.15324095599052	0.00238516786937256	0.000892440570621737	2369.97971493709	0.0521685701544727	0.00768937917125631	0.00287307874202367	0.0264585025133878	1.56834617218791	0.0583353253645057	0.0301864227278435	129.753489719566	0.0514534942208084	0.00391317152753772	8770.03681938255	0.00540517014047226	0.00294591130066138	3.31134941753886	1.75162262436343
Classic	{\HKV}	624.125	624.125	0.998035806633804	0.0225105603317796	0.00548669063771777	0.0289794360380583	0.998091840360124	0.00861877576130752	0.00336933594350248	31.7430555555556	0.90114998967581	304.583333333333	0.0225105603317796	13.2183712629816	435.3125	0.898625030582865	0.00405422542241172	0.00118125006783696	0.020126783299419	0.00203074076708938	0.00315896768584099	0.000872388956341314	0.020126783299419	0.00995960022136763	304.583333333333	7.68933513234468	0.0041999643767045	0.00178970901863174	0.0324813245668116	12.1597222222222	22.5353617374252	304.583333333333	0.0141270665967218	0.0289794360380583	0.015761710015099	0.00141098323839002	0.00253450858249274	0.00608930128351891	13.2421064041192	3158.94444444444	435.3125	435.3125	12.621600339878	0.00107039268115196	0.998065652843016	3158.94444444444	0.0324813245668116	0.00366690632840937	0.00129507498950164	0.896060387272912	13.2643060738603	0.0141270665967218	0.0156983376563162	19.7291666666667	0.015761710015099	0.00225931483023227	624.125	0.00327100205387302	0.00126504876763472	5.66666666666667	2.125	586.869886760272	586.869886760272	0.00175043793096087	0.03675984278734	0.0107717868552791	0.0462170583656693	0.00171454561997786	0.015567171267398	0.00710460316407949	26.7539282855841	0.037726920663565	301.442742942548	0.03675984278734	1.48409701356068	416.813410464871	0.0430615271456483	0.00673038973631793	0.00202073130002543	0.0320378405186284	0.0042319394990135	0.00554965757174838	0.0014067387336576	0.0320378405186284	0.0164806988595287	301.442742942548	1.48220795476787	0.00773400400955553	0.00374999974386069	0.0530367544414977	11.2639428404937	1.5861596333669	301.442742942548	0.0219352228952339	0.0462170583656693	0.0251453825067081	0.00250649966171364	0.00505054307390291	0.0109227429810938	1.51068786067146	2346.12845729954	416.813410464871	416.813410464871	1.5267760813229	0.00186461171819138	0.00174306391072326	2346.12845729954	0.0530367544414977	0.00627814618146615	0.00226635636068637	0.0462352462719066	1.53784518664747	0.0219352228952339	0.0233719363808425	17.39150936239	0.0251453825067081	0.00461059736488934	586.869886760272	0.00635422937930115	0.00232144148080168	3.31134941753886	1.75162262436343
Classic	{\BPRMF}	3154.88194444444	3154.88194444444	0.993773724318485	0.0789981689558283	0.0194613432216978	0.0961306679369421	0.994689813782266	0.0305923699279787	0.0120958377257953	260.916666666667	0.632736839222807	1574.61111111111	0.0789981689558283	10.9779695616496	2251.22916666667	0.598742403472608	0.0184958791595673	0.00555151334598497	0.0718994621838752	0.00755163392216898	0.0145782731333419	0.00412940481231206	0.0718994621838752	0.0450014985084599	1574.61111111111	6.61750817702892	0.0162858383512829	0.00686329498101898	0.105946034244476	101.645833333333	20.7537507325897	1574.61111111111	0.0540193658507125	0.0961306679369421	0.0591788673065594	0.00662082588978324	0.010227235719114	0.0281281740939433	11.0876327207767	3158.94444444444	2251.22916666667	2251.22916666667	11.2117268201295	0.00504681582608123	0.994201079205558	3158.94444444444	0.105946034244476	0.0167880417184	0.00607468670742092	0.568685061425734	11.2223545843265	0.0540193658507125	0.0700515983242714	164.875	0.0591788673065594	0.00836078734825734	3154.88194444444	0.0127612740382051	0.00600022043822854	5.66666666666667	2.125	2389.25932724319	2389.25932724319	0.00478228678228357	0.0637557722344831	0.014313422519388	0.0765037713166916	0.00393029000080038	0.0215559603485594	0.0097071009278694	320.643420012329	0.159092263632938	1290.20287942845	0.0637557722344831	1.08136043181246	1782.72689184059	0.169209004390198	0.0154421876665466	0.00469270109553989	0.0560924787067715	0.00591213781690548	0.0123918423108295	0.0034951691647616	0.0560924787067715	0.0373107397281799	1290.20287942845	1.2231435355546	0.0118449506325612	0.00543115539443281	0.0876203141658143	127.656689335764	1.21729356386578	1290.20287942845	0.0415003545185336	0.0765037713166916	0.0467322700070971	0.00565223744412637	0.0079411531401514	0.0237955127641803	1.04121043997181	2346.12845729954	1782.72689184059	1782.72689184059	1.19737300598	0.00423812265222675	0.00438667429683427	2346.12845729954	0.0876203141658143	0.0141287342179801	0.00514537212688448	0.178059634999708	0.9929303324772	0.0415003545185336	0.057165085229895	205.261077995464	0.0467322700070971	0.00642928606627598	2389.25932724319	0.00952792372687615	0.00510954984532247	3.31134941753886	1.75162262436343
POI	{\IRENMF}	4870.19444444444	4870.19444444444	0.995975121943979	0.105200022362283	0.0202115524492997	0.122541930891223	0.996449159434376	0.0315299744715941	0.0127053634346094	287.201388888889	0.775238538349963	2248.24305555556	0.105200022362283	12.1461704771829	3469.40277777778	0.764443756111111	0.0197560293318005	0.00583230161164174	0.096744249467208	0.00790752465184033	0.0157086573058995	0.00416913372625466	0.096744249467208	0.0477523124486977	2248.24305555556	6.81831691765449	0.0169175227398792	0.00727479757732645	0.13335537965745	116.652777777778	20.8691768380329	2248.24305555556	0.0695332588194805	0.122541930891223	0.0756614490645779	0.0070937226555879	0.0108570104480135	0.0300536812535548	12.225238256305	3158.94444444444	3469.40277777778	3469.40277777778	11.3775981208055	0.00521865424835857	0.996224475493282	3158.94444444444	0.13335537965745	0.0180266365830987	0.00644013074142449	0.750775974673946	12.2751939394584	0.0695332588194805	0.073062812656978	186.930555555556	0.0756614490645779	0.00869664564479761	4870.19444444444	0.0133854554636699	0.00635377996452363	5.66666666666667	2.125	3126.55055412852	3126.55055412852	0.00360197212089753	0.0763972752129473	0.015541531318038	0.0840202747357813	0.0028888635674217	0.0228013874374605	0.0105460263294743	321.232146626127	0.137150853404662	1388.27439926109	0.0763972752129473	1.06002672983476	2206.5075632876	0.144797225619325	0.0167761648143379	0.00505855663766914	0.0680203380878503	0.00646178430749929	0.013714839448402	0.00355281810656283	0.0680203380878503	0.0396658324119077	1388.27439926109	1.08936868181145	0.01268784499328	0.00601893030635478	0.0947389828094982	134.547847804688	0.970300294401177	1388.27439926109	0.0484276842717284	0.0840202747357813	0.0544613575669745	0.00623353919079	0.00875349856308991	0.0256791259478036	1.0270355005902	2346.12845729954	2206.5075632876	2206.5075632876	1.00833468606266	0.00446765328490282	0.00324693694452537	2346.12845729954	0.0947389828094982	0.0155139422607223	0.00561252846836025	0.152572137320923	0.986236817622123	0.0484276842717284	0.0584334254935718	212.762931625354	0.0544613575669745	0.00699022208997436	3126.55055412852	0.0103720686554745	0.00549616715314601	3.31134941753886	1.75162262436343
POI	{\GeoBPRMF}	1490.54861111111	1490.54861111111	0.994905051284304	0.0359336495145099	0.0171081283266081	0.0504990792689549	0.995937727527769	0.0267623181405662	0.0108022562289932	229.826388888889	0.715906009840966	577.326388888889	0.0359336495145099	11.6753043725836	935.277777777778	0.678884328331504	0.0165285648074245	0.00490420502990152	0.0327894971071946	0.00666651108227816	0.0132424440458699	0.00348493405028414	0.0327894971071946	0.0394364451488225	577.326388888889	7.22306640477979	0.0142954373649304	0.00609282862527597	0.0553099978706202	93.1597222222222	21.615334496302	577.326388888889	0.021181104087951	0.0504990792689549	0.0231989605087716	0.0059926313793888	0.00912632466366087	0.0251608452040858	11.804107886208	3158.94444444444	935.277777777778	935.277777777778	11.9588573574731	0.00434862395933623	0.995410701507907	3158.94444444444	0.0553099978706202	0.0151097719632466	0.00541996928267803	0.643973538696949	11.9556235551156	0.021181104087951	0.060415475274855	148.611111111111	0.0231989605087716	0.00735717410620212	1490.54861111111	0.0112759324333734	0.00533447101271625	5.66666666666667	2.125	1526.42702614698	1526.42702614698	0.00416539635255589	0.0488958246645277	0.0164623661748403	0.0633007803969939	0.00303333469565276	0.0240975969415708	0.0112280537336598	336.358061999684	0.187108237961324	607.636828297094	0.0488958246645277	1.87357431798961	964.86526067165	0.207187860707381	0.0189867444733844	0.00571484556063019	0.0432819225488867	0.00694475703030654	0.0156981471853646	0.0038849492479008	0.0432819225488867	0.0438953533612859	607.636828297094	1.34765937939865	0.0140163800418233	0.00652707220866825	0.0714615311615452	143.968608815942	1.55845290893696	607.636828297094	0.0289420339974878	0.0633007803969939	0.0326738821312127	0.00713661564392637	0.00978643211039022	0.0289173786294812	1.78665138939521	2346.12845729954	964.86526067165	964.86526067165	1.40634463757809	0.00495749827131565	0.00359666906918255	2346.12845729954	0.0714615311615452	0.0176387895804812	0.00638190030002737	0.227379340471819	1.69385279722016	0.0289420339974878	0.0640709864516583	224.885615403787	0.0326738821312127	0.00748810576098502	1526.42702614698	0.0115244194083912	0.00619294418335109	3.31134941753886	1.75162262436343
POI	{\RankGeoFM}	11061.7638888889	11061.7638888889	0.997983627566095	0.464643654562761	0.00520735900587647	0.502817218147897	0.998100830432476	0.00838784514444292	0.00321546119203865	127.729166666667	0.887750459745115	6832.09722222222	0.464643654562761	12.9286491391443	9228.25694444445	0.882622347899961	0.00716022594113136	0.00218021599401125	0.428301777008866	0.00206479873776948	0.00562657480656367	0.00161630006752408	0.428301777008866	0.0173437676148454	6832.09722222222	5.33581594063229	0.00495032073840241	0.00202110125064794	0.54601245829761	52.3125	18.5804129016553	6832.09722222222	0.342384979939277	0.502817218147897	0.370907993321335	0.00260879346541504	0.00330156047086417	0.010775421585635	12.9635770612241	3158.94444444444	9228.25694444445	9228.25694444445	9.34595593106934	0.00198366720191213	0.998036111730488	3158.94444444444	0.54601245829761	0.00648446806995396	0.00238870222939015	0.878300542557768	13.0050763549937	0.342384979939277	0.0272636126016574	82.9444444444444	0.370907993321335	0.00226832149236998	11061.7638888889	0.0038945745733874	0.00237368950618452	5.66666666666667	2.125	8916.49521599539	8916.49521599539	0.00166931147270859	0.128979842931777	0.00474755772388926	0.125864834870334	0.0015460106160906	0.00725640503383943	0.00291126685313926	204.073797797641	0.0887169481198224	5176.66811643209	0.128979842931777	1.07364463976738	7246.00626315529	0.0924383702385346	0.00974819080825969	0.00300498186660277	0.114900075272523	0.0020534112876111	0.00775593329054191	0.00222689377672607	0.114900075272523	0.0235908728280401	5176.66811643209	0.708187111312754	0.00520698226200205	0.00220201704026675	0.144016659297227	82.0185437628529	1.38137818159798	5176.66811643209	0.0969450226665486	0.125864834870334	0.106467681155431	0.00358886291143752	0.00386723056125764	0.0146275176532186	1.07543073507318	2346.12845729954	7246.00626315529	7246.00626315529	1.00852496942613	0.0027424761635394	0.00161742022429062	2346.12845729954	0.144016659297227	0.00891902751742917	0.00329925849149375	0.0951696109317256	1.08113228466231	0.0969450226665486	0.0358248687940712	132.85576477765	0.106467681155431	0.00217860999583867	8916.49521599539	0.00427108571345169	0.00321713549426655	3.31134941753886	1.75162262436343
POI	{\PopGeoNN}	5987.72222222222	5987.72222222222	0.99234838108521	0.0975983306896514	0.0174403357068595	0.0911200412793698	0.992128194235872	0.0263925809766478	0.0116204511003454	133.638888888889	0.442016031926917	3512.29861111111	0.0975983306896514	11.3741388698206	4695.875	0.449799568743351	0.0141420668059162	0.00407072399450793	0.0892368582990253	0.00731620504170132	0.0116020455104347	0.00262401227860607	0.0892368582990253	0.0318877046224946	3512.29861111111	6.78445123746336	0.0137644601974357	0.00671200977784687	0.100110069975909	58.2430555555556	20.9008699194028	3512.29861111111	0.0825827438343669	0.0911200412793698	0.0896261083338544	0.005190509079572	0.00919230553265832	0.0211329353149158	11.1185587755965	3158.94444444444	4695.875	4695.875	11.3764091205144	0.00336270460313203	0.992040250858931	3158.94444444444	0.100110069975909	0.0130016966171426	0.00456382640878041	0.49082716064754	10.9529226234367	0.0825827438343669	0.0486800600574191	86.5833333333333	0.0896261083338544	0.00793345602873795	5987.72222222222	0.0111364456248071	0.00439452807741295	5.66666666666667	2.125	3989.99217876872	3989.99217876872	0.00581828827307651	0.0408495051524879	0.0118208986826403	0.0428691036821241	0.00519224649141299	0.0176059190194221	0.00786807714535109	113.577096277184	0.136384712511976	2423.86695346711	0.0408495051524879	1.32629670901638	3115.75265379817	0.114967381919541	0.00827174298349378	0.00241904705260975	0.0347314797739707	0.00500661304111532	0.0069610953252993	0.00157923162401673	0.0347314797739707	0.0188452944413771	2423.86695346711	1.12318670618119	0.00865682148513023	0.00451237505219902	0.0501144767670003	49.0272147241232	1.10193145277462	2423.86695346711	0.0264334493215975	0.0428691036821241	0.0301821840686556	0.00312802248484959	0.00576653609590812	0.0127379975467924	1.17458805900591	2346.12845729954	3115.75265379817	3115.75265379817	1.08584909344579	0.00206579948163177	0.00555855350529535	2346.12845729954	0.0501144767670003	0.00769944928744708	0.00273492247683379	0.0810291172281441	1.06302435251591	0.0264334493215975	0.0279964176031181	71.7823536372714	0.0301821840686556	0.00540157871844063	3989.99217876872	0.00707521056450662	0.00269585883125413	3.31134941753886	1.75162262436343
}\ResultsMeanAllRecommenders

\begin{table*}[t!]
\caption{Mean value of the performance results of the recommenders in all subsamples. All values are reported at a cutoff of 5. We represent in bold the best value obtained when computing the mean in each metric.} 
\centering
\label{t:ResultsMeanAllRecommenders}
\begingroup \small %
\begin {tabular}{cccccccc}%
\toprule \multirow {2}{*}{\textbf {Family}} & \multirow {2}{*}{\textbf {Recommender}} & \multicolumn {2}{c}{\textbf {nDCG}} & \multicolumn {2}{c}{\textbf {EPC}} & \multicolumn {2}{c}{\textbf {Item Exposure}}\\\cmidrule {3-8} & & \textbf {mean} & \textbf {std} & \textbf {mean} & \textbf {std} & \textbf {mean} & \textbf {std}\\\midrule %
\multirow {6}{*}{Classic}&\Rnd &\pgfutilensuremath {0.000}&\pgfutilensuremath {0.001}&\pgfutilensuremath {\textbf{0.999}}&\pgfutilensuremath {0.001}&\pgfutilensuremath {5.602}&\pgfutilensuremath {0.737}\\%
&\Popularity &\pgfutilensuremath {0.004}&\pgfutilensuremath {0.004}&\pgfutilensuremath {0.987}&\pgfutilensuremath {0.009}&\pgfutilensuremath {7.820}&\pgfutilensuremath {1.430}\\%
&\UB &\pgfutilensuremath {\textbf{0.011}}&\pgfutilensuremath {0.006}&\pgfutilensuremath {0.997}&\pgfutilensuremath {0.003}&\pgfutilensuremath {6.128}&\pgfutilensuremath {1.230}\\%
&\IB &\pgfutilensuremath {0.010}&\pgfutilensuremath {0.005}&\pgfutilensuremath {\textbf{0.999}}&\pgfutilensuremath {0.001}&\pgfutilensuremath {6.137}&\pgfutilensuremath {1.026}\\%
&\HKV &\pgfutilensuremath {0.003}&\pgfutilensuremath {0.005}&\pgfutilensuremath {0.998}&\pgfutilensuremath {0.002}&\pgfutilensuremath {7.689}&\pgfutilensuremath {1.482}\\%
&\BPRMF &\pgfutilensuremath {0.010}&\pgfutilensuremath {0.008}&\pgfutilensuremath {0.994}&\pgfutilensuremath {0.005}&\pgfutilensuremath {6.618}&\pgfutilensuremath {1.223}\\%
\midrule \multirow {4}{*}{POI}&\IRENMF &\pgfutilensuremath {\textbf{0.011}}&\pgfutilensuremath {0.009}&\pgfutilensuremath {0.996}&\pgfutilensuremath {0.004}&\pgfutilensuremath {6.818}&\pgfutilensuremath {1.089}\\%
&\GeoBPRMF &\pgfutilensuremath {0.009}&\pgfutilensuremath {0.010}&\pgfutilensuremath {0.995}&\pgfutilensuremath {0.004}&\pgfutilensuremath {7.223}&\pgfutilensuremath {1.348}\\%
&\RankGeoFM &\pgfutilensuremath {0.003}&\pgfutilensuremath {0.004}&\pgfutilensuremath {0.998}&\pgfutilensuremath {0.002}&\pgfutilensuremath {\textbf{5.336}}&\pgfutilensuremath {0.708}\\%
&\PopGeoNN &\pgfutilensuremath {0.009}&\pgfutilensuremath {0.006}&\pgfutilensuremath {0.992}&\pgfutilensuremath {0.006}&\pgfutilensuremath {6.784}&\pgfutilensuremath {1.123}\\\bottomrule %
\end {tabular}%
\endgroup %

\end{table*}

Table~\ref{t:ResultsMeanAllRecommenders} reveals that the performance of the recommenders, in terms of nDCG, is relatively low. This is a common phenomenon in the POI recommendation domain, given the vast number of candidate POIs and the scarcity of user visits to train the models. However, some recommenders, like \HKV or \RankGeoFM, consistently show a lower performance overall, with limited standard deviation. With respect to novelty, we generally obtain high results, even in the \Popularity recommender. This can be attributed to the data sparsity, where although certain popular POIs have a significant number of visits, they have been explored by only a small percentage of users relative to the total number of potential users. Notably, the \Popularity algorithm exhibits the highest deviation, indicating that each subsample may feature different popular POIs. This also shows that the popularity bias is different in each subsample, being more exaggerated in those subsamples where we remove a smaller percentage of popular items (dtv=0.5). This popularity bias impacts the performance of other recommenders with a bias towards popularity, such as \BPRMF or \PopGeoNN, %
 while others like \IB or \RankGeoFM are less affected. In terms of Item Exposure, we can observe notable variations among the algorithms. While the \Popularity recommender achieves higher exposure results by focusing solely on recommending popular venues, other models such as \RankGeoFM, \UB, or \IB can offer recommendations that encompass a more diverse range of POIs. The behavior of the \IB recommender is particularly interesting: as derived from the results, it ranks
fourth %
in terms of Item Exposure, highlighting its ability to provide fair recommendations for items appearing in the test sets; simultaneously, its performance in novelty stands out while it also obtains a competitive level of relevance, falling slightly behind the \UB and \IRENMF recommenders. However, both \UB and \IRENMF do not obtain results as competitive as \IB in both Item Exposure and Novelty.

Visualizing the distribution of recommendation results in \autoref{fig:recommendation_outcome}, we see that the recommendation accuracy varies substantially between the 144 subsamples, which is an expected and desired outcome (\autoref{fig:recommendation_outcome}a).
The random recommender is the one with the least variance and performance, however, the \HKV matrix factorization model is also consistently low in performance, indicating that it can not deal well with many of the
smaller subsamples.
Furthermore, the \Popularity, \IB, \UB, and \RankGeoFM models seem to be more robust regarding their outcomes compared to the purely matrix factorization-based approaches that do not consider the geographical component, as their interquartile ranges are smaller.

There are no surprises regarding the novelty of the recommendations, which we measure using the EPC metric, cf. \autoref{subsec:depented_vars}.
The models that involve some aspect of popularity (Popularity, PopGeoNN) and the BPR models produce relatively fewer novel recommendations, which is expected due to how the recommendations are computed~(\autoref{fig:recommendation_outcome}b).
On the contrary, as mentioned before, the behavior of \IB is interesting in terms of novelty since it is the second-best model (after \Rnd) in all subsamples, obtaining relatively competitive accuracy results.
This makes the \IB a model that should be considered to try to achieve a balance between accuracy and novelty.

Finally, the Item Exposure (\autoref{fig:recommendation_outcome}c) shows a low variation between most models, with the quantiles all being between $5$ and $7.5$.
Again, the recommenders that generated fewer novel recommendations or exhibited a higher popularity bias, such as \Popularity, \BPRMF, \GeoBPRMF, or \PopGeoNN, also achieved higher scores in terms of item exposure. This indicates a notable disparity in the distribution of recommended POIs compared to the POIs that the user has visited during the test set.
This is consistent with previous work~\citep{DBLP:journals/datamine/SanchezBB23}, where different biases are analyzed in the POI recommendation domain, and the effect on the performance of a set of recommenders in different cities around the world is compared.

The plots in \autoref{fig:recommendation_outcome} also tell much about the outcome of the subsampling process.
With respect to the \NDCG and the EPC metrics, the mass of the density plots is relatively compact, with some outliers to the top or bottom, respectively.
In the Item Exposure plot, the shapes of the density plots are strung out, generally matching the interquartile ranges better.
An interesting observation is that in the recommendation outcomes of the worse-performing models in terms of accuracy, i.e., Random, \Popularity, \HKV, and \RankGeoFM two groups become visible, i.e., the density plot looks tapered just below the mean value, meaning that the subsamples could be divided into 2 groups according to their performance.
Even though a deep analysis of this aspect is out of the scope of this paper, it might be interesting to understand in the future which samples belong to each group and the impact the number (and frequency) of these groups may have on the explanatory power of the methodology followed in this work.

\subsection{Excluding Low-performing Recommendation Models from the Explanatory Study}
\label{subsec:low-performing-rec}

The pure Popularity-based recommendation algorithm is a very simple, parameter-free model, but nevertheless a useful baseline in POI recommendation due to the inherent popularity bias of the domain \citep{DBLP:journals/ir/BelloginCC17}.
Even though computing the recommendations solely on the popularity of the items contradicts the principle of personalization, visitors tend to visit the popular highlights of a destination.
Thus, we argue that any model in POI recommendation should at least outperform the Popularity model in terms of accuracy.

When it comes to the explanatory analysis, we remove the Random, \HKV, and \RankGeoFM models from the pool of algorithms for the explanatory study since the mean recommendation accuracy over all subsamples is lower than the simple Popularity baseline.
The reason for this is that the purpose of the explanatory study is that we want to learn what the success factors of recommendation models are in terms of their data characteristics measured using EVs.
By including models that are not ``successful'' (by outperforming the popularity baseline in terms of \NDCG), we would analyze the factors contributing to poor recommendations, which is a meaningless endeavor.

This sets the final pool of 7 recommendation models to \Popularity, \IB, \UB, \BPRMF, \GeoBPRMF, \IRENMF, and \PopGeoNN.

\subsection{Selecting Relevant Explanatory Variables}
\label{subsec:vif_analysis}

In \autoref{subsec:evs}, we defined 32 potential explanatory variables that can be used to explain the dependent variables (\autoref{subsec:depented_vars}) using the regression model.
Naturally, not all independent variables possess equal levels of informative signal, i.e., they can be noisy.
The regression model is useful in identifying noisy or unrelated independent variables as these will not be statistically significant coefficients with respect to the target variable.
However, when using multiple variables in a regression model, multicollinearity can arise, i.e., two or more explanatory variables being correlated with each other.
While this does not impede the outcome of the regression model, multicollinearity in the explanatory variables decreases the predictive contribution of the individual coefficients.
To obtain meaningful results in the significance analysis of the coefficients, we mitigate collinearity by eliminating such redundant variables.

To do this in a reproducible way, we propose a procedure to remove highly correlated variables until the collinearity is mitigated to an acceptable level.
The multicollinearity of a regression model is measured by the \ac{VIF}, but the scientific literature is divided on what maximum \ac{VIF} value is acceptable \citep{Robinson2009Interaction, Obrien2007,Stine1995Graphical}.
Reflecting on this, we systematically analyzed the outcome of applying \autoref{alg:corr} with VIF thresholds between 5 and 25, ultimately choosing 12, which retains 8 EVs and explains on average $R^2 = 0.79$ of the variance in our regression models towards the nDCG@5.
The choice of the VIF threshold is a trade-off between the number of variables, the resulting variance the regression model can explain, and the level of multicollinearity, which any analyst or researcher must consider carefully on a case-by-case basis.
Our proposed procedure is useful with an increasing amount of variables, as it removes human judgement from the process of choosing which variables to eliminate.
This is an improvement on the previous work \citep{Adomavicius2012,DBLP:conf/sigir/DeldjooNSM20,DBLP:journals/ipm/DeldjooBN21}, where this step was not precisely specified.

\begin{algorithm*}
	\caption{Elimination of EVs Based on Correlation Analysis}
	\label{alg:corr}
	$\text{features} \gets \{EVs\}$\;
	\While{max(VIF(\text{features})) $ >$ TR\_VIF}{
		correlation\_matrix $\gets$ PCC(features)\;
			$c_1, c_2 \gets$ argmax( | correlation\_matrix  | )\;
			\eIf{max(| PCC($c_1$, \{ features \textbackslash ~{$c_2$} \})|) > max(| PCC($c_2$, \{ features \textbackslash ~{$c_1$} \})|)}
			{features $\gets$ features \textbackslash ~$c_1$ \;}{features $\gets$ features \textbackslash ~$c_2$ \;}
	}
	\Return features\;
\end{algorithm*}

The goal is to determine a set of input variables for the regression analysis that have low collinearity, which is measured by the \acf{VIF}.
To determine variables that cause unwanted collinearity, a correlation analysis is required to discard correlated variables.
\autoref{alg:corr} describes our proposed procedure: %
while there is still an EV with a \ac{VIF} over the threshold (TR\_VIF), we compute all pairwise Pearson Correlation Coefficients
(PCC) of the features, obtaining the correlation matrix, and determining the two different features with the highest positive or negative correlation as our candidates for elimination.
From these two candidates $(c_1, c_2)$, we eliminate the candidate that has the highest correlation to any other feature in the remaining features. %
This elimination of EVs is repeated until the \ac{VIF} values of all remaining EVs satisfy the threshold.

\begin{table}[ht]
\centering
\caption{Final EVs after controlling for multicollinearity.} 
\label{tab:vif_elimination}
\begin{tabular}{lrr}
  \toprule
EV & VIF before& VIF after \\ 
  \midrule
SpaceSize & 25.90 & - \\ 
  Shape & 71.79 & 1.51 \\ 
  Density & 30.95 & 5.10 \\ 
  $Cp_u$ & 69.71 & - \\ 
  $Cp_i$ & 304.45 & - \\ 
  $Gini_I$ & 103.34 & - \\ 
  $Gini_U$ & 130.78 & 6.13 \\ 
  APB & 9138.49 & - \\ 
  MedPB & 6802.78 & - \\ 
  StPB & 316.46 & 3.27 \\ 
  SkPB & 31.19 & - \\ 
  KuPB & 15.47 & 1.36 \\ 
  ALT & 519.42 & - \\ 
  StLT & 661.18 & - \\ 
  SkLT & 567.80 & - \\ 
  KuLT & 449.43 & - \\ 
  ADCC & 5711.40 & - \\ 
  MedDCC & 1229.74 & - \\ 
  StDCC & 1731.52 & - \\ 
  SkDCC & 354.85 & - \\ 
  KuDCC & 225.33 & - \\ 
  ARG & 2749.51 & - \\ 
  MedRG & 561.53 & - \\ 
  StRG & 129.15 & 11.32 \\ 
  SkRG & 1002.16 & - \\ 
  KuRG & 204.44 & - \\ 
  ADA & 561.36 & - \\ 
  MedDA & 66.90 & 2.77 \\ 
  StDA & 52.75 & - \\ 
  SkDA & 429.65 & - \\ 
  KuDA & 146.96 & 4.95 \\ 
   \bottomrule
\end{tabular}
\end{table}

\autoref{tab:vif_elimination} shows the outcome of applying \autoref{alg:corr} to the data from the experiments comprising the recommendation outcome of the well-performing recommenders as established previously in  \autoref{subsec:low-performing-rec} on the 144 subsamples.
The target metric of the linear model to compute the \ac{VIF} was  nDCG@5.
We retain 8 EVs, namely Shape, Density, Gini$_U$, StPB, KuPB, StRG, MedDA, and KuDA.
This outcome is interesting as it puts emphasis on the EVs that capture the structure of the \acl{UCM}, such as Shape, and Density. %
This is not surprising, as these are very common metrics to quantify the difficulty of a recommendation problem.
Furthermore, some aspect of most families of EVs was included, with the exception of the distance to city center and long-tail items. %
While we will come to the explanatory power of the EVs in the following section, this result alone underlines that the newly introduced EVs regarding mobility and user activity broaden the perspective of  \ac{POI} recommendation problems.
We plot the pairwise correlations of the EVs in the Appendix, \autoref{fig:correlation_vif_all}, where we find that the maximum (in absolute terms) pairwise PCC is $-0.69$ between Gini$_U$ and Density, after removing highly correlated variables.

\section{Results}
\label{sec:results}

Recall from \autoref{sec:framework} that in its core, the explanatory framework is a linear regression (cf. \autoref{eq:regression}) with the
data characteristics of the 144 subsamples quantified in the explanatory variables as input variables and  \NDCG, EPC, and Item Exposure as dependent variables. %
What we are interested in are the coefficients of the model ($\theta_{ev}$ in \autoref{eq:regression}), as these coefficients quantify the impact of the individual EV on the outcome variable.
We run the explanatory framework for each recommendation model that produced competitive results (cf. \autoref{subsec:low-performing-rec}) independently using the EVs that did not suffer from multicollinearity (cf. \autoref{subsec:vif_analysis}).

In the analysis of the results, we assessed three key aspects: Accuracy, Novelty, and Item Exposure.
Our findings revealed notably high $R^2$ values for each of these aspects.
This suggests that, even after discarding numerous explanatory variables, our regression models still have substantial explanatory power.
In this section, we make a detailed discussion of our results, using a cutoff value of 5 for each metric (i.e., we evaluate the top 5 recommendations), as in the POI recommendation domain, it is common to report small cutoffs~\citep{DBLP:conf/cikm/LiuWSM14, DBLP:conf/sigir/LiCLPK15, DBLP:conf/ictai/YuanJGCYA16}.
For additional results at cutoff values of 10 and 20, please refer to  \autoref{app:result_tables}.

Our structure for presenting the experimental results follows this format:
firstly, we tabulate the regression coefficients ($\theta_{ev}$) and then visualize these coefficients through coefficient plots.
The result tables, i.e., Tables \ref{tab:regression_ndcg5}, \ref{tab:regression_epc5}, and \ref{tab:regression_ie5}, provide an overview of the goodness-of-fit with the $R^2$ and adjusted $R^2$ values.
Subsequently, we present the coefficients $\theta_{ev}$ for each model.
These coefficients are annotated with stars, which indicate the significance level of the relationship between the explanatory variable and the outcome variable.
The significance levels and the corresponding p-values used are ***  to indicate that $p< 0.001$, ** for $p<0.01$, and * for $p<0.05$.

Furthermore, we visualize the values of the tables in coefficient plots (Figures \ref{fig:coef_plot_ndcg5}, \ref{fig:coef_plot_epc5}, and \ref{fig:coef_ie5}).
The dots show the coefficient $\theta_{ev}$; the whiskers span the 95\% confidence interval.

\subsection{Accuracy}

First, we analyze the recommendation accuracy in terms of \NDCG in \autoref{tab:regression_ndcg5}.
The $R^2$ coefficients of determination of the regression models are all between $0.68$ and $0.88$, indicating that the 8 EVs could explain 68\%--88\% of the \NDCGAux@5 variation.
This result is consistent with the two previous studies of \cite{Adomavicius2012} and \cite{DBLP:journals/ipm/DeldjooBN21}.
The least predictable algorithm was the \IB recommendation model, while the \GeoBPRMF algorithm had the highest  $R^2$.

\begin{table*}[t!]
	\caption{Coefficients of the Regression Model for \NDCGAux@5.}
	\label{tab:regression_ndcg5}
	\begingroup \fontTables %
	\begin{tabular}{l}%
  \toprule%
name \\ %
  \midrule%
$R^2$ \\ %
  $R^2$ (adj) \\ %
  Shape\ \\ %
  Density\ \\ %
  $Gini_U$\ \\ %
  StPB\ \\ %
  KuPB\ \\ %
  StRG\ \\ %
  MedDA\ \\ %
  KuDA\ \\ %
\addlinespace[-0.04cm]
   \bottomrule%
\end{tabular}%
	\endgroup %
	\begingroup \fontTables %
	\begin{tabular}{l}%
  \toprule%
  Popularity \\ %
 \midrule%
0.768 \\ %
  0.754 \\ %
  -0.003*** \\ %
  0.006*** \\ %
  -0.001 \\ %
  0.005*** \\ %
  -0.002. \\ %
  -0.006** \\ %
  -0.004*** \\ %
  0.001 \\ %
   \bottomrule%
\end{tabular}%
	\endgroup %
	\begingroup \fontTables %
	\begin{tabular}{l}%
  \toprule%
  IB \\ %
 \midrule%
0.698 \\ %
  0.68 \\ %
  0.002 \\ %
  0.009*** \\ %
  0.015*** \\ %
  0.003. \\ %
  0.002 \\ %
  -0.009** \\ %
  0.01*** \\ %
  0.011*** \\ %
   \bottomrule%
\end{tabular}%
	\endgroup %
	\begingroup \fontTables %
	\begin{tabular}{l}%
  \toprule%
  UB \\ %
 \midrule%
0.72 \\ %
  0.704 \\ %
  -0.002. \\ %
  0.009*** \\ %
  0.014*** \\ %
  0.005* \\ %
  0.002 \\ %
  -0.014*** \\ %
  0.003* \\ %
  0.011*** \\ %
   \bottomrule%
\end{tabular}%
	\endgroup %
	\begingroup \fontTables %
	\begin{tabular}{l}%
  \toprule%
  BPRMF \\ %
 \midrule%
0.845 \\ %
  0.836 \\ %
  0.001 \\ %
  0.018*** \\ %
  0.015*** \\ %
  0.006*** \\ %
  0.005*** \\ %
  -0.009** \\ %
  0.011*** \\ %
  0.004. \\ %
   \bottomrule%
\end{tabular}%
	\endgroup%
	\begingroup \fontTables %
	\begin{tabular}{l}%
  \toprule%
  GeoBPRMF \\ %
 \midrule%
0.886 \\ %
  0.879 \\ %
  0 \\ %
  0.019*** \\ %
  0.012*** \\ %
  0.011*** \\ %
  0.006*** \\ %
  -0.009*** \\ %
  0.01*** \\ %
  0.002 \\ %
   \bottomrule%
\end{tabular}%
	\endgroup %
	\begingroup \fontTables %
	\begin{tabular}{l}%
  \toprule%
  IRENMF \\ %
 \midrule%
0.798 \\ %
  0.786 \\ %
  0.003* \\ %
  0.009*** \\ %
  0.006 \\ %
  0.007** \\ %
  -0.004* \\ %
  -0.018*** \\ %
  0.01*** \\ %
  0.009** \\ %
   \bottomrule%
\end{tabular}%
	\endgroup
	\begingroup \fontTables %
	\begin{tabular}{l}%
  \toprule%
  PopGeoNN \\ %
 \midrule%
0.799 \\ %
  0.787 \\ %
  -0.003* \\ %
  0.011*** \\ %
  0.003 \\ %
  0.009*** \\ %
  -0.002 \\ %
  -0.01** \\ %
  -0.001 \\ %
  0.012*** \\ %
   \bottomrule%
\end{tabular}%
	\endgroup \\ %
	***:  $p< 0.001$, **: $p<0.01$, *: $p<0.05$
\end{table*}

\begin{figure*}[htb]
	\includegraphics[width=\textwidth]{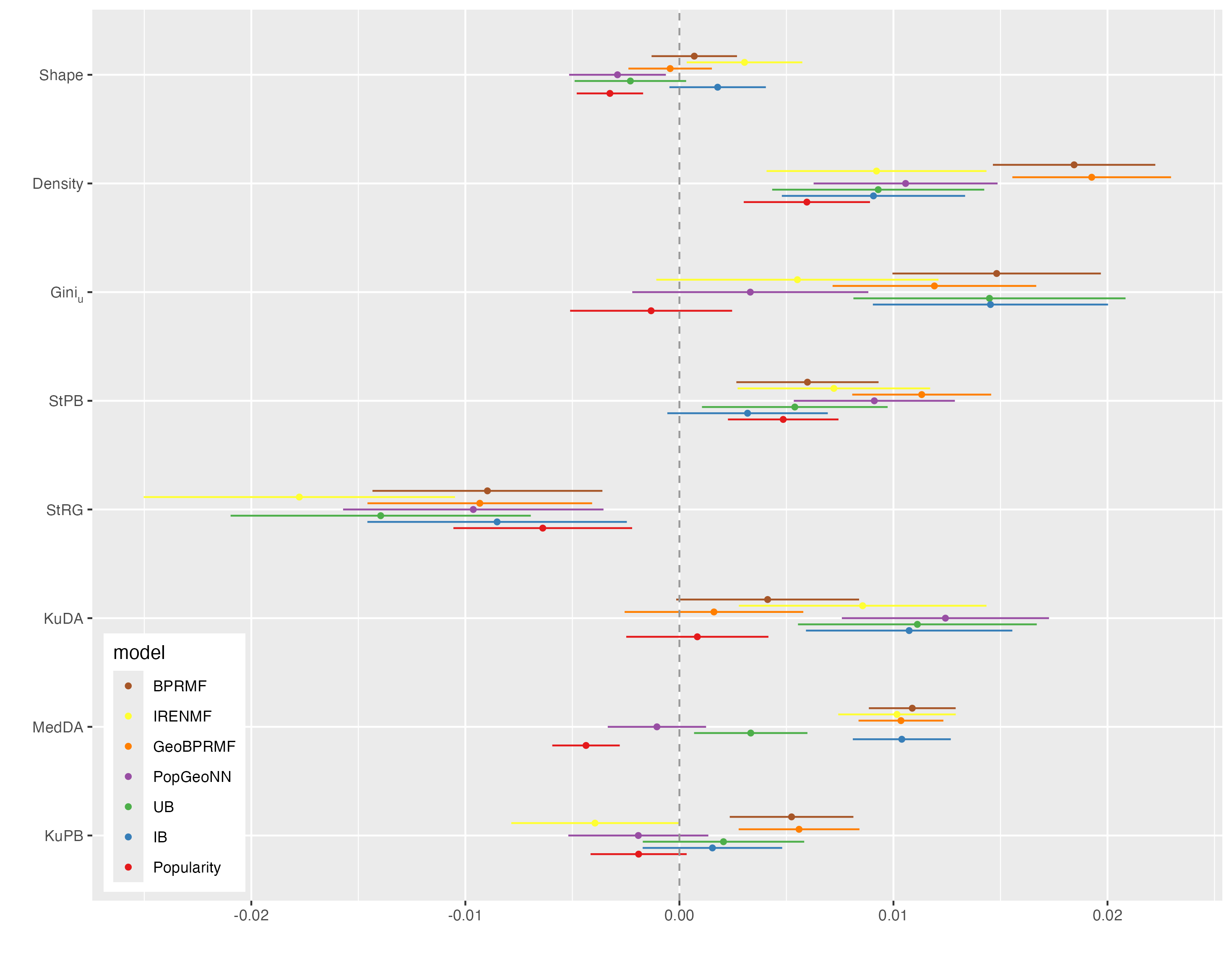}
	\caption{Coefficient plot for \NDCGAux@5. StRG has a negative influence on \NDCGAux@5, Shape, KuPB, and MedDA are mostly neutral, whereas the other EVs have a positive impact.}
	\label{fig:coef_plot_ndcg5}
\end{figure*}

When it comes to the general patterns in the coefficients (cf. \autoref{fig:coef_plot_ndcg5}), 
we observe that an increase in the standard deviation of the Radius of Gyration has clearly the most negative influence on the recommendation accuracy of all variables.
On the contrary, a higher Density, Gini$_U$, standard deviation of the Popularity Bias, and kurtosis of the Duration Active generally tend to improve the accuracy.
The EVs Shape, median of Duration Active, and the kurtosis of the Popularity Bias have mixed influences, with Shape overall having the smallest (but still statistically significant in three cases) influence.

Turning our attention to the significance levels of the coefficients regarding the outcome variable, we observe that Density and StRG are always highly significant with $p<0.01$  for all recommendation models. %
These consistently low p-values across the board of all recommendation models underline their importance for the success of POI recommendation algorithms.
All EVs were a significant predictor towards the \NDCGAux@5 in some of the recommendation models, although the Shape was only significant towards the accuracy of the Popularity and the \PopGeoNN models.

\subsection{Novelty}

The second aspect of our analysis is novelty, which we measure using the EPC metric.
Again, we tabulate the coefficients in \autoref{tab:regression_epc5}.
Comparing the $R^2$ values to the ones in \autoref{tab:regression_ndcg5} (accuracy), we see a slightly better regression fit with values ranging from $0.73$ (\GeoBPRMF model) to $0.95$ (Item-based model).
This indicates that despite the fact that the explanatory variables were selected based on their collinearity with respect to the \NDCGAux@5 (cf. \autoref{subsec:vif_analysis}), the regression model for the EPC metric is similarly accurate and even slightly more expressive compared to the one for nDCG@5.

\begin{table*}[htb]
    \caption{Coefficients ($\theta_{ev}$) of the Regression Model for EPC@5.}
	\label{tab:regression_epc5}
	\begingroup \fontTables %
	\begin{tabular}{l}%
  \toprule%
name \\ %
  \midrule%
$R^2$ \\ %
  $R^2$ (adj) \\ %
  Shape\ \\ %
  Density\ \\ %
  $Gini_U$\ \\ %
  StPB\ \\ %
  KuPB\ \\ %
  StRG\ \\ %
  MedDA\ \\ %
  KuDA\ \\ %
\addlinespace[-0.04cm]
   \bottomrule%
\end{tabular}%
	\endgroup %
	\begingroup \fontTables %
	\begin{tabular}{l}%
  \toprule%
  Popularity \\ %
 \midrule%
0.88 \\ %
  0.873 \\ %
  0.01*** \\ %
  -0.024*** \\ %
  0.009** \\ %
  -0.021*** \\ %
  -0.004* \\ %
  0.005 \\ %
  0.004** \\ %
  -0.002 \\ %
   \bottomrule%
\end{tabular}%
	\endgroup %
	\begingroup \fontTables %
	\begin{tabular}{l}%
  \toprule%
  IB \\ %
 \midrule%
0.953 \\ %
  0.951 \\ %
  0 \\ %
  -0.004*** \\ %
  0. \\ %
  -0.001*** \\ %
  0 \\ %
  0 \\ %
  0 \\ %
  -0.001*** \\ %
   \bottomrule%
\end{tabular}%
	\endgroup %
	\begingroup \fontTables %
	\begin{tabular}{l}%
  \toprule%
  UB \\ %
 \midrule%
0.803 \\ %
  0.791 \\ %
  0.003*** \\ %
  -0.008*** \\ %
  0.001 \\ %
  -0.004*** \\ %
  0.002. \\ %
  0.003. \\ %
  0.003*** \\ %
  0.001 \\ %
   \bottomrule%
\end{tabular}%
	\endgroup %
	\begingroup \fontTables %
	\begin{tabular}{l}%
  \toprule%
  BPRMF \\ %
 \midrule%
0.769 \\ %
  0.755 \\ %
  0.004*** \\ %
  -0.013*** \\ %
  0.009*** \\ %
  -0.01*** \\ %
  0.003. \\ %
  -0.001 \\ %
  0.004*** \\ %
  0 \\ %
   \bottomrule%
\end{tabular}%
	\endgroup%
	\begingroup \fontTables %
	\begin{tabular}{l}%
  \toprule%
  GeoBPRMF \\ %
 \midrule%
0.749 \\ %
  0.734 \\ %
  0.006*** \\ %
  -0.017*** \\ %
  0.007* \\ %
  -0.013*** \\ %
  0.002 \\ %
  -0.001 \\ %
  0.004** \\ %
  -0.005. \\ %
   \bottomrule%
\end{tabular}%
	\endgroup %
	\begingroup \fontTables %
	\begin{tabular}{l}%
  \toprule%
  IRENMF \\ %
 \midrule%
0.861 \\ %
  0.852 \\ %
  0.004*** \\ %
  -0.01*** \\ %
  0.006*** \\ %
  -0.011*** \\ %
  0.001 \\ %
  -0.001 \\ %
  0.001* \\ %
  -0.003* \\ %
   \bottomrule%
\end{tabular}%
	\begingroup \fontTables %
	\begin{tabular}{l}%
  \toprule%
  PopGeoNN \\ %
 \midrule%
0.869 \\ %
  0.861 \\ %
  0.006*** \\ %
  -0.017*** \\ %
  0.006** \\ %
  -0.014*** \\ %
  0 \\ %
  0.002 \\ %
  0.003*** \\ %
  -0.002 \\ %
   \bottomrule%
\end{tabular}%
	\endgroup %
\endgroup \\ %
***:  $p< 0.001$, **: $p<0.01$, *: $p<0.05$
\end{table*}

\begin{figure*}[htb]
	\includegraphics[width=\textwidth]{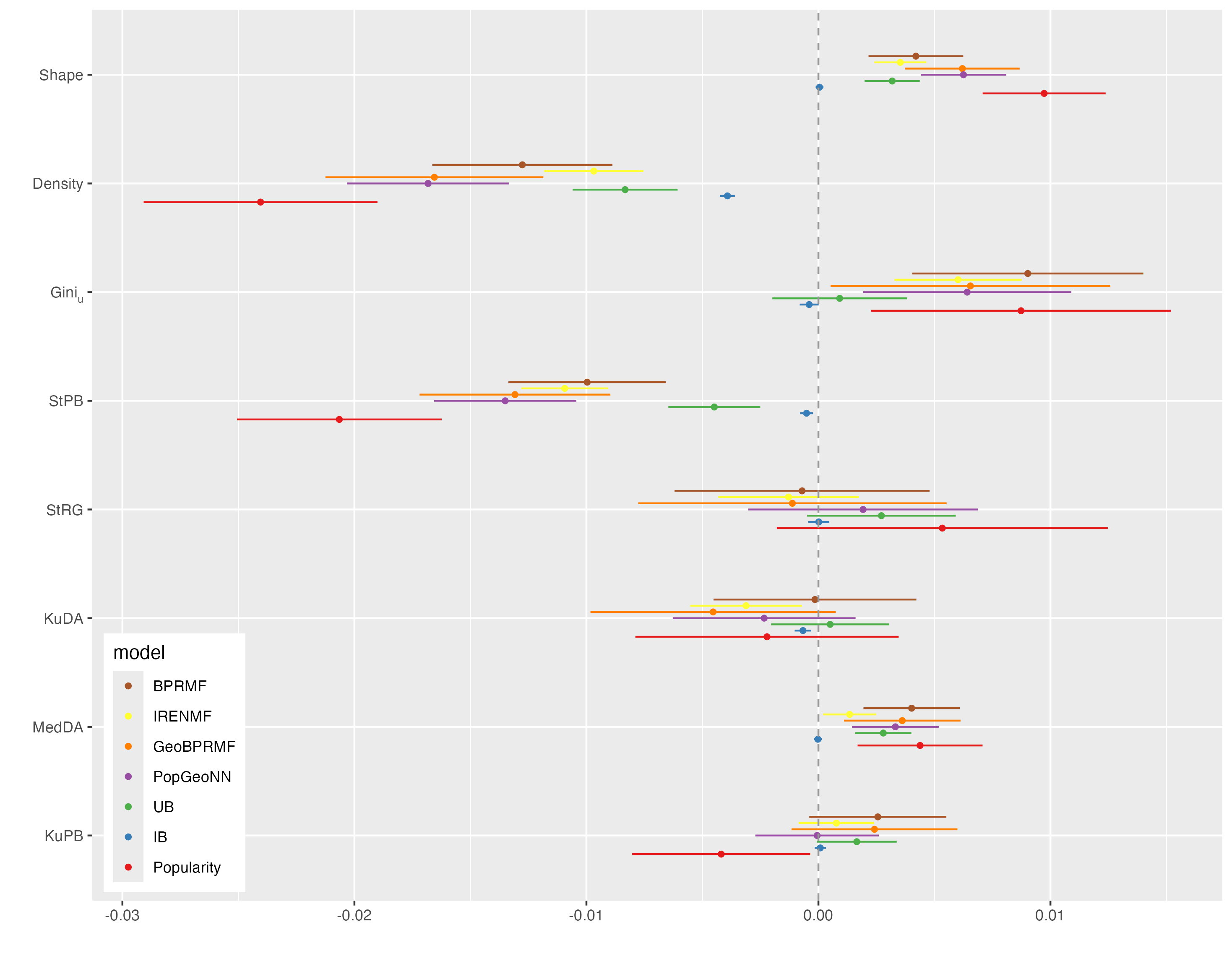}
	\caption{Coefficient plot for novelty measured using EPC@5. We observe an inverse relationship compared to \NDCGAux @5 with Density, StPB, and KuDA
having a negative impact and Shape, Gini$_U$, and MedDA having a positive impact on the EPC metric.}
			\label{fig:coef_plot_epc5}
\end{figure*}

Analyzing the patterns within the coefficients presented in \autoref{fig:coef_plot_epc5}, we can discern a noteworthy observation: our experiments reveal an inverse relationship between novelty and accuracy, which is a well-known trade-off in recommender systems.
For all considered EVs, except for the statistically non-significant StRG and KuPB (except in the \Popularity model), and across various recommender models a distinct change in the sign of coefficients is evident.
Specifically, many coefficients switch from a positive association to a negative one and vice versa.
The coefficients for the Item-based model converge near zero, indicating that this model produces recommendations with a stable Novelty regardless of the data characteristics.
Among the EVs, the Shape, Gini$_U$, and MedDA display positive coefficients for EPC@5.
Conversely, the other EVs exhibit a negative impact (Density, StPB) or a neutral influence (StRG, KuDA) on this particular outcome variable.
The observation of a neutral influence of the standard deviation of the Radius of Gyration is bolstered by the finding that its coefficients do not achieve statistical significance across any of the recommendation models.
In contrast, we note that certain
EVs
exhibit a high level of significance ($p < 0.001$) across all recommendation models.
Specifically, these influential EVs are Density, Shape, StPB, and MedDA.

\subsection{Item Exposure}

Lastly, we shift our focus to the assessment of item exposure, as measured by the metric defined in \autoref{subsec:depented_vars}.
Notably, the $R^2$ values for this analysis are remarkably high, exceeding $0.86$ in all cases, except for \BPRMF, where the EVs can still account for $0.65$ of the variance, as detailed in \autoref{tab:regression_ie5}.
We would like to emphasize that higher values in the item exposure metric indicate a larger disparity between the number of times items are recommended and the number of times they should ideally be recommended (as defined in \autoref{eq:i_e}).

\begin{table*}[htb]
	\caption{Coefficients ($\theta_{ev}$) of the Regression Model for Average Item Exposure@5.}
	\label{tab:regression_ie5}
	\begingroup \fontTables %
	\begin{tabular}{l}%
  \toprule%
name \\ %
  \midrule%
$R^2$ \\ %
  $R^2$ (adj) \\ %
  Shape\ \\ %
  Density\ \\ %
  $Gini_U$\ \\ %
  StPB\ \\ %
  KuPB\ \\ %
  StRG\ \\ %
  MedDA\ \\ %
  KuDA\ \\ %
\addlinespace[-0.04cm]
   \bottomrule%
\end{tabular}%
	\endgroup %
	\begingroup \fontTables %
	\begin{tabular}{l}%
  \toprule%
  Popularity \\ %
 \midrule%
0.932 \\ %
  0.928 \\ %
  -1.609*** \\ %
  0.547. \\ %
  1.813*** \\ %
  1.573*** \\ %
  -0.514* \\ %
  0.688 \\ %
  3.388*** \\ %
  0.496 \\ %
   \bottomrule%
\end{tabular}%
	\endgroup %
	\begingroup \fontTables %
	\begin{tabular}{l}%
  \toprule%
  IB \\ %
 \midrule%
0.907 \\ %
  0.902 \\ %
  -2.631*** \\ %
  -0.069 \\ %
  1.302*** \\ %
  0.985*** \\ %
  -0.264 \\ %
  0.193 \\ %
  1.719*** \\ %
  0.019 \\ %
   \bottomrule%
\end{tabular}%
	\endgroup %
	\begingroup \fontTables %
	\begin{tabular}{l}%
  \toprule%
  UB \\ %
 \midrule%
0.902 \\ %
  0.896 \\ %
  -2.421*** \\ %
  0.273 \\ %
  1.052* \\ %
  1.222*** \\ %
  -0.326 \\ %
  0.477 \\ %
  2.595*** \\ %
  0.422 \\ %
   \bottomrule%
\end{tabular}%
	\endgroup %
	\begingroup \fontTables %
	\begin{tabular}{l}%
  \toprule%
  PopGeoNN \\ %
 \midrule%
0.873 \\ %
  0.866 \\ %
  -1.924*** \\ %
  0.36 \\ %
  0.839. \\ %
  1.232*** \\ %
  -0.889*** \\ %
  0.934. \\ %
  2.553*** \\ %
  1.348*** \\ %
   \bottomrule%
\end{tabular}%
	\endgroup %
	\begingroup \fontTables %
	\begin{tabular}{l}%
  \toprule%
  GeoBPRMF \\ %
 \midrule%
0.874 \\ %
  0.866 \\ %
  -2.165*** \\ %
  0.106 \\ %
  0.351 \\ %
  1.743*** \\ %
  -0.525* \\ %
  1.371** \\ %
  1.988*** \\ %
  1.583*** \\ %
   \bottomrule%
\end{tabular}%
	\endgroup %
	\begingroup \fontTables %
	\begin{tabular}{l}%
  \toprule%
  BPRMF \\ %
 \midrule%
0.666 \\ %
  0.646 \\ %
  -1.515*** \\ %
  0.561 \\ %
  -1.486* \\ %
  2.26*** \\ %
  -1.418*** \\ %
  2.249** \\ %
  1.503*** \\ %
  2.517*** \\ %
   \bottomrule%
\end{tabular}%
	\endgroup
	\begingroup \fontTables %
	\begin{tabular}{l}%
  \toprule%
  IRENMF \\ %
 \midrule%
0.917 \\ %
  0.912 \\ %
  -2.347*** \\ %
  -0.361 \\ %
  2.126*** \\ %
  0.814** \\ %
  -0.015 \\ %
  0.849* \\ %
  2.23*** \\ %
  0.563. \\ %
   \bottomrule%
\end{tabular}%
	\endgroup \\ %
	***:  $p< 0.001$, **: $p<0.01$, *: $p<0.05$
\end{table*}

The coefficients that exhibit high levels of significance across all recommendation models include Shape, StPB, and MedDA.
In contrast, Density fails to achieve significance in any of the models.
Lastly, the significance of the coefficient for Gini$_U$ in the \IB and \UB models is noteworthy as this EV quantifies the inequality in the frequency distribution of item check-ins.

\begin{figure*}[htb]
	\includegraphics[width=\textwidth]{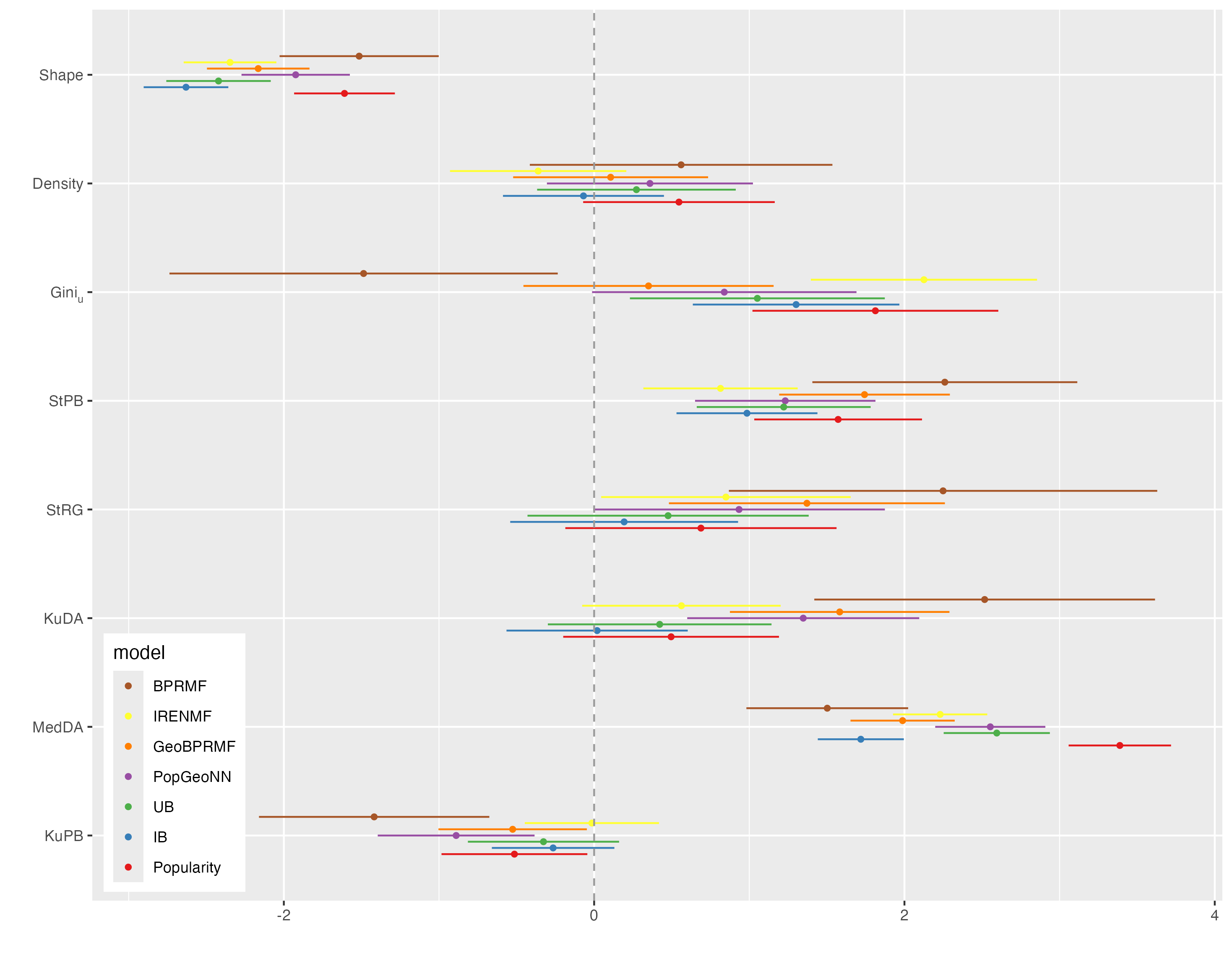}
	\caption{Coefficient plot for Average Item Exposure@5.
	 Shape and KuPB exert a negative influence on average item exposure, meaning that the difference to the expected item exposure is smaller, which is better. StPB and MedDA have a clearly positive influence on the item exposure metrics, whereas the remaining EVs are mostly slightly positive or neutral (Density).}
	\label{fig:coef_ie5}
\end{figure*}

Upon analyzing the grouping of coefficients and models in \autoref{fig:coef_ie5}, again certain patterns emerge:
Shape and KuPB consistently exert a negative influence on Item Exposure within all recommendation models.
As their values increase, the item exposure metric decreases, which means that the item exposure is closer to the expected item exposure from the test set.
This observation is in line with the intuition that a relatively higher number of check-ins within the \acl{UCM} results in a more dispersed distribution, which in turn helps mitigate the influence of popularity bias.
Conversely, the median of the Duration Active (MedDA) consistently exhibits a positive impact on the item exposure metric across all the recommendation models.
One plausible explanation for this trend is that when users spend longer periods in a city, they are more likely to explore and visit a large number of popular POIs, further accentuating the effect of popularity bias.
 The other EVs typically have an overall positive influence on the Item Exposure metric, with many (but not all) being significant predictors.

\section{Discussion}

The obtained results shed light on the strengths and weaknesses of POI recommendation models in terms of the data characteristics of the recommendation problem.
Using the domain-driven subsampling approach to create specific subsamples from a large POI recommendation data set, we noticed that several established POI recommendation models are unsuited for smaller problems (cf. \autoref{fig:recommendation_outcome}a), which warranted their exclusion from the explanatory analysis.

In terms of the quality of the linear model, it is striking that it was possible to explain 68--88\% variation of the \NDCG with the remaining 8 out of 32 EVs after the elimination of the collinear EVs. Besides, we also obtained high results in terms of explaining item exposure (65--86\%) and novelty (73--95\%) variations.
This result confirms that the collinearity analysis is necessary and helps to focus on the interesting variables without losing explainability.
In terms of accuracy, we find a clearly positive influence of Density on the \NDCGAux@5, which confirms the findings from other domains that a higher density creates an easier recommendation problem \citep{DBLP:journals/ipm/DeldjooBN21}. Furthermore, we provide evidence that a higher standard deviation of the Radius of Gyration leads to diminished accuracy.
This shows that geographic information is determinant in predicting the results in terms of ranking accuracy in POI recommendation. In this scenario, an increase in the standard deviation of the Radius of Gyration suggests a greater diversity in user movement patterns, making it more difficult (for algorithms) to identify consistent global movement trends among users.

In terms of novelty of recommendations, we see a similar trend as for the accuracy; however, our results showed once more that these two concepts are inverse due to the popularity bias and related trade-offs discussed in the literature \citep{Kaminskas2016,DBLP:conf/aaai/ZhaoZLXLZSZ19}.
Unsurprisingly, we could show that a higher variance in the popularity bias in the interaction data helps to promote novel recommendations.
Thus, these two targets still need to be balanced in any POI recommender system in accordance with business needs.
Emphasizing these dimensions is also important from the user's point of view, as it would allow the system to surprise them with recommendations that are different from what they may be familiar with beforehand.

Regarding Item Exposure, we observe negative coefficients of the Shape (which is the ratio between the number of users and items) on this metric, signifying that relatively more users compared to items tend to yield lower values of the item exposure metric.
In this context, lower values are preferable, as it means that they are closer to the exposed values in the unbiased test set.
However, we should also consider that an increased Duration Active and
standard deviation
of the Popularity Bias leads to a higher item exposure than desired, as this variable always obtains positive coefficients in the regression model.
Platforms need to monitor such effects closely, as the exposure of small local travel enterprises in the recommendations can determine whether they have a means of existence in the market.

\subsection{Practical and Theoretical Implications}

In this paper, we go beyond evaluating recommendation models by employing a framework that reveals the associations between various data characteristics and different dimensions of performance.
While this framework is not without its imperfections, as discussed below, its application carries significant implications for the training and evaluation of POI recommender systems.
For instance, our observations highlight the impact of explanatory variables such as
Density, Gini$_U$, and StPB which have a positive influence on nDCG@5, as opposed to EVs like StRG, which displays a negative influence.
Based on our findings, developers, analysts, and researchers of POI recommender systems should be aware of the importance of these data characteristics.
In practice this means that countermeasures against adverse data characteristics can be undertaken: The Density can be addressed by defining a minimum number of interactions before a user is served by the main recommendation model (for \emph{``warm''} users) instead of computing cold-start recommendations to a user with too little interactions. By directly asking \emph{``cold''} users for venues they have visited, relevant information can be collected.
At the same time, the geographic scope of candidate items can be adjusted to deal with the adversary effects of the Radius of Gyration.

Furthermore, it is worth noting that the effect of these explanatory variables is dependent on the specific recommendation algorithm employed;  however, we see systematic common effects on related recommendation models. %
Nevertheless, researchers should be mindful of this when comparing baseline methods against proposed models.
Unintentionally, they might report comparisons that are not fair because some methods are either positively or negatively impacted by the data characteristics.
Notably, this observation extends to beyond-accuracy metrics, as we have observed similar trends in novelty and item exposure.

From a more theoretical perspective, this work introduces the prospect of learning these data dependencies directly from the data themselves and incorporating them into recommendation models and user profiles. Specifically, our work gives guidance to the choice of models for recommendations in automated ways, e.g., through Auto-RecSys \citep{Anand2020Auto,Vente2023Introducing}, analogous to Auto-ML \citep{Karmaker2021AutoML}. %
As our approach is entirely data-driven, the only ad-hoc decisions we made were related to the selection of the original explanatory variables, which could vary depending on the specific target domain.
However, once the explanatory and dependent variables are established, it becomes conceivable to integrate the methodology presented here into a reinforcement learning framework.
In such a setup, data could be fed to the recommenders in accordance with the predicted effects they are anticipated to have, thus optimizing recommendation outcomes, as addressed in some works based on specific user characteristics to improve overall performance \citep{DBLP:journals/umuai/SaidB18,DBLP:conf/recsys/PenhaS20}.

Even in the absence of a reinforcement learning framework, the current methodology can provide valuable insights into designing and enhancing existing recommendation algorithms, at least in the POI recommendation domain.
For instance, we observed that \IB is less sensitive to Density than \GeoBPRMF across the three metrics analyzed.
Given that their accuracy metrics are similar, one way to leverage this insight is to explore how to make GeoBPRMF more resilient in scenarios where data sets exhibit varying levels of density.
To achieve this, researchers and practitioners could consider incorporating strategies from \IB or similar robust methods into an improved version of the \GeoBPRMF algorithm.
This integration might involve adapting the underlying model or introducing additional mechanisms that allow \GeoBPRMF to handle data sets with different density levels more effectively.
By doing so, \GeoBPRMF could become more versatile and capable of delivering consistent performance across a wider range of data set characteristics, enhancing its practical applicability in diverse POI recommendation scenarios.

Most importantly, whether it is hotels, restaurants, or other \acp{POI}, this work gives guidance to recommendation platforms of the e-tourism sector to characterize their recommendation data and understand the benefits and drawbacks of their recommendation model from the perspective of different users groups and businesses.
It can also inform methods to self-audit biases in the recommendations of platforms, e.g., with regards to the expected and achieved item exposure of items in certain recommendation models \citep{Srba2023}.

\subsection{Limitations \& Future Work}

While being a widespread drawback of offline analyses of POI recommendation algorithms using \acl{LBSN} data, we still acknowledge that there is a gap between actual user behavior and what is recorded in \acp{LBSN}.
Although this is universal for all recommendation models in all studies, it is worth mentioning that such studies analyze a proxy concept of check-ins that have been actively submitted instead of the actual ground truth of the temporal visitation of all POIs, venues, and other places.
Thus, the generalization of the analyses in this study is -- just as in all other studies --  constrained by the limited availability of high-quality POI recommendation data sets.
This limitation can only be overcome by collecting data sets that reflect user interactions from actual \ac{POI} recommender systems, which would be a crucial next step in advancing the field.

Although we utilized the widely employed global Foursquare data set containing 33~million check-ins \citep{Yang2015}, its sparsity posed a significant challenge since the actual number of check-ins per destination dwindled, limiting our ability to create meaningful subsamples for most of the cities within the data set.
To prevent biases stemming from geographic influences of the topological features of different cities, we focused our study on a single destination, the New York City metropolitan area, due to its prominence and the number of check-ins in the data set.
This decision to focus on one city was essential because many less popular destinations on Foursquare lacked the volume of check-in data to support the explanatory analysis with subsamples of sufficient size, which would lead to a deterioration in recommendation quality and lower statistical significance.
In fact, while New York City provided ample interaction data for addressing the recommendation problem, it is a city less susceptible to seasonal variations in travel behavior, as evidenced by the minor differences in data characteristics between the summer and winter subsamples shown in \autoref{tab:subsampling_outcome}, hence the obtained results cannot be considered \textit{universal}, since this analysis in only based on one data set from one city.
As future work, analyzing the generalizability of the explanatory framework \emph{between} cities and to further data sources would be an obvious extension to our work. If one would repeat the experiments in many cities, how robust would the set of \emph{significant}  explanatory variables (i.e., the coefficients in the respective linear regression models for the individual algorithms) be? %
Such an analysis would be interesting, however, it would involve computing explanatory variables whose values are comparable, i.e., normalized between different cities. 
Generalizing the method towards incorporating different data sources would be a further step to understand the impact of different data collection methods on the data characteristics and the performance of POI recommender systems. Herein, the challenge lies in establishing comparable check-ins data sets, both in terms of geographic and temporal coverage. 

In this context, another distinct line of future work would be to compute a \emph{universal} regression model, with the aim to \emph{predict} the performance of recommendation algorithms in different cities. The relevant research questions are two-fold: a) under which circumstances is it permissible to mix data characteristics of samples from different cities, as they have different sizes, which might result in incomparable values of the same explanatory variables?, and b) are other influences, such aspects that \emph{cannot} be directly quantified as explanatory variables, like cultural or climatic aspects, small enough so that they do not have a practical influence on the predictions?

In our study, we incorporated EVs capturing both spatial and temporal aspects, additional variables capturing more of the users' context were omitted for the subsampling for practical reasons.
With the currently used data sets, a finer temporal subdivision of check-ins by different times of the day or weekdays and weekends would be possible in theory but are practically prevented by the size of the resulting subsamples.
While further contextual information is not available in the data set, the performance of a \ac{POI} recommendation is known to be influenced by many contextual factors, such as the weather on a given day \citep{Trattner2018}.
User context has been the focus of various approaches in  literature~\citep{Woerndl2011a,Cai2017,Yang2017c,DBLP:conf/aaai/ZhaoZLXLZSZ19},
this gives opportunity for future works to analyze the importance of different contextual factors, including more information about the users and their needs, which has been analyzed in literature on traveler types \citep{Gibson2002,Neidhardt2014,Dietz2020}, which offers various roles that one could use as subsampling data characteristics.
Specifically, since it has been shown that locals and tourists showcase different behavior in the same city \citep{Sanchez2022a}, it would be worthwhile identifying which explanatory variables are more relevant for various user groups.

\section{Conclusions}
\label{s:conclusion_future}

In both the classical recommendation domain and specialized fields, such as \acl{POI} recommendation, a multitude of algorithms have been proposed.
However, the effectiveness of these algorithms often varies significantly depending on the data set under evaluation.
While addressing this challenge has been explored previously in classical recommendation scenarios, there remains a research gap within the POI recommendation domain.
This gap is particularly significant, given the relevance of POI recommendations in the tourism sector, affecting a multitude of businesses and consumers.
Unlike the traditional recommendation domain, like purchasing a book or watching a movie, recommending \acp{POI} involves algorithms that utilize various signals, such as popularity and geographic locations of venues.

In this paper, we expanded upon the framework introduced by~\cite{DBLP:journals/ipm/DeldjooBN21}, incorporating additional explanatory variables specific to the \ac{POI} recommendation domain.
Our objective was to investigate which data characteristics affect the performance of both classical and state-of-the-art \ac{POI} recommendation algorithms.
We assessed these algorithms across three dimensions: accuracy, novelty, and item exposure.
The results we obtained shed light on the robustness of recommendation models concerning the data characteristics inherent in recommendation data sets, offering valuable insights for the field.
Among the various explanatory variables we analyzed, it became evident that certain factors pertaining to the data structure (Shape, Density), as well as those associated with the distribution of check-ins (Gini$_U$, StPB, and KuPB), along with geographical and temporal variables (StRG, MedDA, and KuDA) are critical for explaining the performance of POI recommender systems.
In terms of ranking accuracy, Density, Gini$_U$, StPB, and KuDA generally tend to be conducive to higher accuracy, whereas an increase in the standard deviation of the Radius of Gyration is detrimental.
The significance of newly introduced spatio-temporal explanatory variables  (Radius of Gyration and Duration Active) in the coefficient analysis of the regression models underlines our conclusion that the well-known data characteristics from the analysis of classical recommendation domains are insufficient to explain algorithmic performance in the \ac{POI} recommendation domain.

\section*{Data Availability}

The Foursquare data set used in this study is available from Dingqi Yang's web page \url{https://sites.google.com/site/yangdingqi/home/foursquare-dataset}.
Further intermediate results are available in the supplementary material shared with this paper.

\section*{Conflict of Interest}

The authors have no competing interests to declare that are relevant to the content of this article.

\bibliography{complete_bibliography_condensed.bib}

\clearpage
\appendix

\normalsize
\section*{\Large Appendix}

\section{Statistics of Subsamples}

Table~\ref{tab:subsampling_outcome} presents the values obtained by independent subsamples in the explanatory variables defined in Section~\ref{subsec:evs}.

\begin{table}[htb]
	\caption{Summary statistics of the impact of the subsampling on the explanatory variables. Note that for space reasons the table only contains the independent subsamples and not the combination of the subsampling data characteristics. $Cp_u$ and $Cp_i$ correspond to \checkins per user and \checkins per item (definition~\ref{def:rpu}), Gini$_U$ and Gini$_I$ refers to Gini for users and for items (definition~\ref{def:gini}), and APB, ALT, ARG, ADCC, and ADA terms represent the averages of the popularity bias, long tail items, distance to the city center and duration active (definitions~\ref{def:pop_bias}, \ref{def:long_tail_items}, \ref{def:dist_ts}, and \ref{def:duration_activity}, respectively). %
	}
	\label{tab:subsampling_outcome}
	\scriptsize
	\begin{tabular}{lrrrrrrrrrrrr}
  \toprule
subsample & SpaceSize & Shape & Density & $Cp_u$ & $Cp_i$ & $Gini_U$ & $Gini_I$ & APB & ALT & ARG & ADCC & ADA \\ 
  \midrule
k = 2 & 5.57E+08 & 0.36 & 4.45E-04 & 17.44 & 6.32 & 0.62 & 0.66 & 254.24 & 0.02 & 5.12 & 6.39 & 119.12 \\ 
  k = 5 & 1.97E+08 & 0.47 & 1.04E-03 & 21.28 & 10.07 & 0.51 & 0.62 & 176.72 & 0.02 & 5.74 & 5.93 & 158.40 \\ 
  k = 10 & 7.04E+07 & 0.55 & 2.24E-03 & 25.27 & 13.96 & 0.40 & 0.59 & 120.08 & 0.02 & 5.96 & 5.96 & 200.77 \\ 
  dtv = 0.5\% & 5.29E+08 & 0.35 & 3.85E-04 & 15.02 & 5.22 & 0.65 & 0.59 & 22.50 & 0.02 & 3.93 & 6.03 & 110.57 \\ 
  dtv = 1\% & 5.15E+08 & 0.34 & 3.68E-04 & 14.27 & 4.88 & 0.65 & 0.57 & 17.15 & 0.03 & 3.89 & 6.18 & 109.86 \\ 
  dtv = 2\% & 4.98E+08 & 0.34 & 3.43E-04 & 13.18 & 4.45 & 0.65 & 0.54 & 13.00 & 0.03 & 3.83 & 6.35 & 108.96 \\ 
  dtv = 5\% & 4.56E+08 & 0.33 & 3.02E-04 & 11.25 & 3.70 & 0.64 & 0.48 & 8.41 & 0.03 & 3.77 & 6.77 & 108.86 \\ 
  o = NYC & 1.99E+08 & 0.14 & 9.66E-04 & 36.41 & 5.09 & 0.44 & 0.62 & 71.39 & 0.04 & 6.38 & 7.41 & 215.42 \\ 
  o = US & 5.09E+07 & 0.46 & 6.01E-04 & 6.31 & 2.91 & 0.46 & 0.56 & 104.55 & 0.06 & 4.68 & 6.55 & 81.60 \\ 
  o = other & 3.27E+07 & 0.51 & 7.81E-04 & 6.23 & 3.20 & 0.48 & 0.61 & 88.59 & 0.05 & 4.03 & 4.89 & 39.66 \\ 
  s = summer & 3.67E+08 & 0.33 & 3.80E-04 & 12.62 & 4.22 & 0.59 & 0.61 & 154.02 & 0.03 & 4.75 & 6.96 & 108.58 \\ 
  s = winter & 3.29E+08 & 0.32 & 4.02E-04 & 12.84 & 4.15 & 0.57 & 0.61 & 140.04 & 0.04 & 4.70 & 6.71 & 105.17 \\ 
   \bottomrule
\end{tabular}
\\
	Key for the subsample column: k: value for the k-core, dtv: drop top venues (in terms of popularity), o: origin, s: season (cf. \autoref{subsec:summary_subsampling})
	The definition of the EV acronyms in the header row can be found in \autoref{subsec:evs}. %
\end{table}

\clearpage
\section{Hyperparameters}

\autoref{tab:Parameters} shows the search space of the hyperparameters for each recommendation model. Refer to \autoref{ss:baselines} for context.

\begin{table*}[h]
	\centering
	\caption{Hyperparameters tested in the recommenders. The best configurations are selected by maximizing \NDCGAux@5.}
	\label{tab:Parameters}
	\fontTables
	\begin{tabular}{ll}
		\toprule
		\textbf{Rec} & \textbf{Hyperparameters} \\
		\midrule
		Random & None \\
		\midrule
		\Popularity & None \\
		\midrule
		\UB             & Sim = \{Vector Cosine, Set Jaccard\}, $k=\{20, 40, 60, 80, 100, 120\}$\\ %
			\midrule
			\IB               & Sim = \{Vector Cosine, Set Jaccard\}, $k=\{20, 40, 60, 80, 100, 120\}$\\ %
				\midrule
				\HKV                     & Iter = $20$, Factors = $\{10, 50, 100\}$, $\lambda = \{0.1, 1\}$, $\alpha = \{0.1, 1\}$\\
				\midrule
				\BPRMF                     & \pbox{0.8\columnwidth}{Factors = $\{10, 50, 100\}$, BiasReg = $\{0, 0.5, 1\}$, LearnRate = $0.05$, Iter = $50$,  RegU = RegI = $\{0.0025, 0.001, 0.005, 0.01, 0.1\}$, RegJ = RegU$/10$}\\
				\midrule
				\GeoBPRMF                     & \pbox{0.8\columnwidth}{Factors = $\{10, 50, 100\}$, BiasReg = $\{0, 0.5, 1\}$, LearnRate = $0.05$, Iter = $50$,  RegU = RegI = $\{0.0025, 0.001, 0.005, 0.01, 0.1\}$, maxDist=$\{1,4\}$}\\
				\midrule
				\IRENMF & Factors = $\{50, 100\}$, geo-$\alpha = \{0.4, 0.6\}$, $\lambda_{3} = \{0.1, 1\}$, clusters = \{5, 50\} \\
				\midrule
				\RankGeoFM & \pbox{0.8\columnwidth}{Factors = $\{50, 100\}$, $\alpha = \{0.1, 0.2\}$, c = 1, $\epsilon = 0.3$, neighs = $\{10, 50, 100, 200\}$ iter = 120, decay = 1, boldDriver = True, learnRate = 0.001} \\
				\midrule
				\PopGeoNN & Sim = \{Vector Cosine, Set Jaccard\}, $k = \{20, 40, 60, 80, 100, 120\}$\\
				\bottomrule
			\end{tabular}
		\end{table*}

\clearpage
\section{Supplementary Result Tables}
\label{app:result_tables}

The following tables correspond to the results presented in \autoref{sec:results} at cutoffs of 10 and 20, respectively.
The results generally follow the same patterns as presented for a cutoff of 5, however, we can observe a tendency that with a higher cutoff, the absolute impact of coefficients becomes smaller in the item exposure metric.


\begin{table*}[htb]
	\caption{Coefficients of the Regression Model for \NDCGAux@10}
	\begingroup \fontTables %
	\begin{tabular}{l}%
  \toprule%
name \\ %
  \midrule%
$R^2$ \\ %
  $R^2$ (adj) \\ %
  Shape\ \\ %
  Density\ \\ %
  $Gini_U$\ \\ %
  StPB\ \\ %
  KuPB\ \\ %
  StRG\ \\ %
  MedDA\ \\ %
  KuDA\ \\ %
\addlinespace[-0.04cm]
   \bottomrule%
\end{tabular}%
	\endgroup %
	\begingroup \fontTables %
	\begin{tabular}{l}%
  \toprule%
  Popularity \\ %
 \midrule%
0.801 \\ %
  0.789 \\ %
  -0.004*** \\ %
  0.008*** \\ %
  -0.002 \\ %
  0.007*** \\ %
  -0.001 \\ %
  -0.009** \\ %
  -0.006*** \\ %
  0.002 \\ %
   \bottomrule%
\end{tabular}%
	\endgroup %
	\begingroup \fontTables %
	\begin{tabular}{l}%
  \toprule%
  IB \\ %
 \midrule%
0.717 \\ %
  0.701 \\ %
  0.003** \\ %
  0.012*** \\ %
  0.018*** \\ %
  0.003 \\ %
  0.003. \\ %
  -0.011** \\ %
  0.012*** \\ %
  0.009** \\ %
   \bottomrule%
\end{tabular}%
	\endgroup %
	\begingroup \fontTables %
	\begin{tabular}{l}%
  \toprule%
  UB \\ %
 \midrule%
0.724 \\ %
  0.708 \\ %
  -0.003. \\ %
  0.012*** \\ %
  0.018*** \\ %
  0.008** \\ %
  0.002 \\ %
  -0.019*** \\ %
  0.004* \\ %
  0.011** \\ %
   \bottomrule%
\end{tabular}%
	\endgroup %
	\begingroup \fontTables %
	\begin{tabular}{l}%
  \toprule%
  PopGeoNN \\ %
 \midrule%
0.827 \\ %
  0.817 \\ %
  -0.004** \\ %
  0.012*** \\ %
  0.003 \\ %
  0.011*** \\ %
  -0.001 \\ %
  -0.013*** \\ %
  -0.003* \\ %
  0.013*** \\ %
   \bottomrule%
\end{tabular}%
	\endgroup %
	\begingroup \fontTables %
	\begin{tabular}{l}%
  \toprule%
  GeoBPRMF \\ %
 \midrule%
0.861 \\ %
  0.852 \\ %
  0 \\ %
  0.022*** \\ %
  0.014*** \\ %
  0.013*** \\ %
  0.007*** \\ %
  -0.013*** \\ %
  0.012*** \\ %
  -0.001 \\ %
   \bottomrule%
\end{tabular}%
	\endgroup %
	\begingroup \fontTables %
	\begin{tabular}{l}%
  \toprule%
  BPRMF \\ %
 \midrule%
0.825 \\ %
  0.815 \\ %
  0.002 \\ %
  0.023*** \\ %
  0.019*** \\ %
  0.007** \\ %
  0.007*** \\ %
  -0.012*** \\ %
  0.012*** \\ %
  0.002 \\ %
   \bottomrule%
\end{tabular}%
	\endgroup%
	\begingroup \fontTables %
	\begin{tabular}{l}%
  \toprule%
  IRENMF \\ %
 \midrule%
0.817 \\ %
  0.806 \\ %
  0.005*** \\ %
  0.011*** \\ %
  0.005 \\ %
  0.008** \\ %
  -0.004. \\ %
  -0.021*** \\ %
  0.011*** \\ %
  0.008* \\ %
   \bottomrule%
\end{tabular}%
	\endgroup
\end{table*}

\begin{table*}[htb]
	\caption{Coefficients of the Regression Model for EPC@10}
	\begingroup \fontTables %
	\begin{tabular}{l}%
  \toprule%
name \\ %
  \midrule%
$R^2$ \\ %
  $R^2$ (adj) \\ %
  Shape\ \\ %
  Density\ \\ %
  $Gini_U$\ \\ %
  StPB\ \\ %
  KuPB\ \\ %
  StRG\ \\ %
  MedDA\ \\ %
  KuDA\ \\ %
\addlinespace[-0.04cm]
   \bottomrule%
\end{tabular}%
	\endgroup %
	\begingroup \fontTables %
	\begin{tabular}{l}%
  \toprule%
  Popularity \\ %
 \midrule%
0.892 \\ %
  0.886 \\ %
  0.008*** \\ %
  -0.021*** \\ %
  0.009** \\ %
  -0.019*** \\ %
  -0.002 \\ %
  0.004 \\ %
  0.003** \\ %
  -0.002 \\ %
   \bottomrule%
\end{tabular}%
	\endgroup %
	\begingroup \fontTables %
	\begin{tabular}{l}%
  \toprule%
  IB \\ %
 \midrule%
0.955 \\ %
  0.953 \\ %
  0 \\ %
  -0.004*** \\ %
  0. \\ %
  0** \\ %
  0 \\ %
  0 \\ %
  0 \\ %
  -0.001** \\ %
   \bottomrule%
\end{tabular}%
	\endgroup %
	\begingroup \fontTables %
	\begin{tabular}{l}%
  \toprule%
  UB \\ %
 \midrule%
0.82 \\ %
  0.809 \\ %
  0.003*** \\ %
  -0.008*** \\ %
  0.001 \\ %
  -0.004*** \\ %
  0.002* \\ %
  0.002. \\ %
  0.002*** \\ %
  0.001 \\ %
   \bottomrule%
\end{tabular}%
	\endgroup %
	\begingroup \fontTables %
	\begin{tabular}{l}%
  \toprule%
  PopGeoNN \\ %
 \midrule%
0.896 \\ %
  0.89 \\ %
  0.005*** \\ %
  -0.017*** \\ %
  0.007*** \\ %
  -0.014*** \\ %
  0 \\ %
  0.001 \\ %
  0.002* \\ %
  -0.002 \\ %
   \bottomrule%
\end{tabular}%
	\endgroup %
	\begingroup \fontTables %
	\begin{tabular}{l}%
  \toprule%
  GeoBPRMF \\ %
 \midrule%
0.739 \\ %
  0.723 \\ %
  0.005*** \\ %
  -0.014*** \\ %
  0.006* \\ %
  -0.012*** \\ %
  0.002 \\ %
  -0.002 \\ %
  0.003** \\ %
  -0.004. \\ %
   \bottomrule%
\end{tabular}%
	\endgroup %
	\begingroup \fontTables %
	\begin{tabular}{l}%
  \toprule%
  BPRMF \\ %
 \midrule%
0.764 \\ %
  0.75 \\ %
  0.004*** \\ %
  -0.011*** \\ %
  0.009*** \\ %
  -0.01*** \\ %
  0.002. \\ %
  -0.001 \\ %
  0.003*** \\ %
  0 \\ %
   \bottomrule%
\end{tabular}%
	\endgroup
	\begingroup \fontTables %
	\begin{tabular}{l}%
  \toprule%
  IRENMF \\ %
 \midrule%
0.873 \\ %
  0.865 \\ %
  0.003*** \\ %
  -0.009*** \\ %
  0.006*** \\ %
  -0.01*** \\ %
  0.001 \\ %
  -0.001 \\ %
  0.001* \\ %
  -0.003** \\ %
   \bottomrule%
\end{tabular}%
	\endgroup
\end{table*}

\begin{table*}[htb]
	\caption{Coefficients of the Regression Model for ItemExposure@10}
	\begingroup \fontTables %
	\begin{tabular}{l}%
  \toprule%
name \\ %
  \midrule%
$R^2$ \\ %
  $R^2$ (adj) \\ %
  Shape\ \\ %
  Density\ \\ %
  $Gini_U$\ \\ %
  StPB\ \\ %
  KuPB\ \\ %
  StRG\ \\ %
  MedDA\ \\ %
  KuDA\ \\ %
\addlinespace[-0.04cm]
   \bottomrule%
\end{tabular}%
	\endgroup %
	\begingroup \fontTables %
	\begin{tabular}{l}%
  \toprule%
  Popularity \\ %
 \midrule%
0.933 \\ %
  0.929 \\ %
  -1.593*** \\ %
  0.518. \\ %
  1.843*** \\ %
  1.521*** \\ %
  -0.515* \\ %
  0.697 \\ %
  3.403*** \\ %
  0.483 \\ %
   \bottomrule%
\end{tabular}%
	\endgroup %
	\begingroup \fontTables %
	\begin{tabular}{l}%
  \toprule%
  IB \\ %
 \midrule%
0.851 \\ %
  0.842 \\ %
  -3.224*** \\ %
  -0.277 \\ %
  2.059*** \\ %
  0.894** \\ %
  -0.015 \\ %
  -0.968. \\ %
  1.048*** \\ %
  -1.908*** \\ %
   \bottomrule%
\end{tabular}%
	\endgroup %
	\begingroup \fontTables %
	\begin{tabular}{l}%
  \toprule%
  UB \\ %
 \midrule%
0.835 \\ %
  0.825 \\ %
  -3.208*** \\ %
  -0.026 \\ %
  1.125. \\ %
  1.45*** \\ %
  -0.258 \\ %
  -0.574 \\ %
  1.844*** \\ %
  -1.037* \\ %
   \bottomrule%
\end{tabular}%
	\endgroup %
	\begingroup \fontTables %
	\begin{tabular}{l}%
  \toprule%
  PopGeoNN \\ %
 \midrule%
0.818 \\ %
  0.807 \\ %
  -1.712*** \\ %
  0.537 \\ %
  1.083* \\ %
  0.758* \\ %
  -1.241*** \\ %
  1.009. \\ %
  2.569*** \\ %
  1.598*** \\ %
   \bottomrule%
\end{tabular}%
	\endgroup %
	\begingroup \fontTables %
	\begin{tabular}{l}%
  \toprule%
  GeoBPRMF \\ %
 \midrule%
0.811 \\ %
  0.799 \\ %
  -2.339*** \\ %
  -0.179 \\ %
  -0.188 \\ %
  1.8*** \\ %
  -0.526* \\ %
  1.53** \\ %
  1.308*** \\ %
  1.926*** \\ %
   \bottomrule%
\end{tabular}%
	\endgroup %
	\begingroup \fontTables %
	\begin{tabular}{l}%
  \toprule%
  BPRMF \\ %
 \midrule%
0.476 \\ %
  0.445 \\ %
  -1.509*** \\ %
  0.394 \\ %
  -2.266** \\ %
  2.282*** \\ %
  -1.613*** \\ %
  2.4** \\ %
  0.77** \\ %
  2.902*** \\ %
   \bottomrule%
\end{tabular}%
	\endgroup
	\begingroup \fontTables %
	\begin{tabular}{l}%
  \toprule%
  IRENMF \\ %
 \midrule%
0.897 \\ %
  0.891 \\ %
  -2.662*** \\ %
  -0.626* \\ %
  2.123*** \\ %
  0.687** \\ %
  -0.074 \\ %
  0.817. \\ %
  1.638*** \\ %
  0.839* \\ %
   \bottomrule%
\end{tabular}%
	\endgroup
\end{table*}


\begin{table*}[htb]
	\caption{Coefficients of the Regression Model for \NDCGAux@20}
	\begingroup \fontTables %
	\begin{tabular}{l}%
  \toprule%
name \\ %
  \midrule%
$R^2$ \\ %
  $R^2$ (adj) \\ %
  Shape\ \\ %
  Density\ \\ %
  $Gini_U$\ \\ %
  StPB\ \\ %
  KuPB\ \\ %
  StRG\ \\ %
  MedDA\ \\ %
  KuDA\ \\ %
\addlinespace[-0.04cm]
   \bottomrule%
\end{tabular}%
	\endgroup %
	\begingroup \fontTables %
	\begin{tabular}{l}%
  \toprule%
  Popularity \\ %
 \midrule%
0.834 \\ %
  0.824 \\ %
  -0.005*** \\ %
  0.011*** \\ %
  -0.005. \\ %
  0.011*** \\ %
  -0.002 \\ %
  -0.01** \\ %
  -0.009*** \\ %
  0.003 \\ %
   \bottomrule%
\end{tabular}%
	\endgroup %
	\begingroup \fontTables %
	\begin{tabular}{l}%
  \toprule%
  IB \\ %
 \midrule%
0.746 \\ %
  0.731 \\ %
  0.005*** \\ %
  0.014*** \\ %
  0.021*** \\ %
  0.003 \\ %
  0.004. \\ %
  -0.015*** \\ %
  0.014*** \\ %
  0.007* \\ %
   \bottomrule%
\end{tabular}%
	\endgroup %
	\begingroup \fontTables %
	\begin{tabular}{l}%
  \toprule%
  UB \\ %
 \midrule%
0.747 \\ %
  0.732 \\ %
  -0.003 \\ %
  0.017*** \\ %
  0.024*** \\ %
  0.01** \\ %
  0.004 \\ %
  -0.025*** \\ %
  0.006** \\ %
  0.009* \\ %
   \bottomrule%
\end{tabular}%
	\endgroup %
	\begingroup \fontTables %
	\begin{tabular}{l}%
  \toprule%
  PopGeoNN \\ %
 \midrule%
0.831 \\ %
  0.821 \\ %
  -0.005*** \\ %
  0.014*** \\ %
  0.001 \\ %
  0.015*** \\ %
  -0.001 \\ %
  -0.016*** \\ %
  -0.006*** \\ %
  0.013*** \\ %
   \bottomrule%
\end{tabular}%
	\endgroup %
	\begingroup \fontTables %
	\begin{tabular}{l}%
  \toprule%
  GeoBPRMF \\ %
 \midrule%
0.858 \\ %
  0.85 \\ %
  0.001 \\ %
  0.026*** \\ %
  0.018*** \\ %
  0.017*** \\ %
  0.009*** \\ %
  -0.017*** \\ %
  0.015*** \\ %
  -0.005 \\ %
   \bottomrule%
\end{tabular}%
	\endgroup %
	\begingroup \fontTables %
	\begin{tabular}{l}%
  \toprule%
  BPRMF \\ %
 \midrule%
0.844 \\ %
  0.835 \\ %
  0.003. \\ %
  0.029*** \\ %
  0.021*** \\ %
  0.012*** \\ %
  0.008*** \\ %
  -0.016*** \\ %
  0.015*** \\ %
  0 \\ %
   \bottomrule%
\end{tabular}%
	\endgroup%
	\begingroup \fontTables %
	\begin{tabular}{l}%
  \toprule%
  IRENMF \\ %
 \midrule%
0.821 \\ %
  0.811 \\ %
  0.008*** \\ %
  0.014*** \\ %
  0.005 \\ %
  0.011*** \\ %
  -0.003 \\ %
  -0.025*** \\ %
  0.014*** \\ %
  0.007. \\ %
   \bottomrule%
\end{tabular}%
	\endgroup
\end{table*}

\begin{table*}[h]
	\caption{Coefficients of the Regression Model for EPC@20}
	\begingroup \fontTables %
	\begin{tabular}{l}%
  \toprule%
name \\ %
  \midrule%
$R^2$ \\ %
  $R^2$ (adj) \\ %
  Shape\ \\ %
  Density\ \\ %
  $Gini_U$\ \\ %
  StPB\ \\ %
  KuPB\ \\ %
  StRG\ \\ %
  MedDA\ \\ %
  KuDA\ \\ %
\addlinespace[-0.04cm]
   \bottomrule%
\end{tabular}%
	\endgroup %
	\begingroup \fontTables %
	\begin{tabular}{l}%
  \toprule%
  Popularity \\ %
 \midrule%
0.903 \\ %
  0.897 \\ %
  0.007*** \\ %
  -0.018*** \\ %
  0.009*** \\ %
  -0.017*** \\ %
  0 \\ %
  0.003 \\ %
  0.002* \\ %
  -0.002 \\ %
   \bottomrule%
\end{tabular}%
	\endgroup %
	\begingroup \fontTables %
	\begin{tabular}{l}%
  \toprule%
  IB \\ %
 \midrule%
0.96 \\ %
  0.958 \\ %
  0 \\ %
  -0.004*** \\ %
  0 \\ %
  0* \\ %
  0 \\ %
  0 \\ %
  0 \\ %
  0* \\ %
   \bottomrule%
\end{tabular}%
	\endgroup %
	\begingroup \fontTables %
	\begin{tabular}{l}%
  \toprule%
  UB \\ %
 \midrule%
0.841 \\ %
  0.831 \\ %
  0.002*** \\ %
  -0.007*** \\ %
  0.002 \\ %
  -0.004*** \\ %
  0.001* \\ %
  0.002 \\ %
  0.002*** \\ %
  0 \\ %
   \bottomrule%
\end{tabular}%
	\endgroup %
	\begingroup \fontTables %
	\begin{tabular}{l}%
  \toprule%
  PopGeoNN \\ %
 \midrule%
0.912 \\ %
  0.907 \\ %
  0.004*** \\ %
  -0.015*** \\ %
  0.007*** \\ %
  -0.014*** \\ %
  0.002. \\ %
  0 \\ %
  0.001. \\ %
  -0.003. \\ %
   \bottomrule%
\end{tabular}%
	\endgroup %
	\begingroup \fontTables %
	\begin{tabular}{l}%
  \toprule%
  GeoBPRMF \\ %
 \midrule%
0.731 \\ %
  0.715 \\ %
  0.004*** \\ %
  -0.012*** \\ %
  0.006* \\ %
  -0.01*** \\ %
  0.002. \\ %
  -0.002 \\ %
  0.002* \\ %
  -0.003. \\ %
   \bottomrule%
\end{tabular}%
	\endgroup %
	\begingroup \fontTables %
	\begin{tabular}{l}%
  \toprule%
  BPRMF \\ %
 \midrule%
0.758 \\ %
  0.744 \\ %
  0.003*** \\ %
  -0.01*** \\ %
  0.008*** \\ %
  -0.009*** \\ %
  0.002. \\ %
  -0.001 \\ %
  0.003** \\ %
  0 \\ %
   \bottomrule%
\end{tabular}%
	\endgroup
	\begingroup \fontTables %
	\begin{tabular}{l}%
  \toprule%
  IRENMF \\ %
 \midrule%
0.889 \\ %
  0.882 \\ %
  0.003*** \\ %
  -0.007*** \\ %
  0.005*** \\ %
  -0.009*** \\ %
  0.001 \\ %
  -0.001 \\ %
  0.001 \\ %
  -0.002** \\ %
   \bottomrule%
\end{tabular}%
	\endgroup
\end{table*}%

\begin{table*}[h]
	\caption{Coefficients of the Regression Model for ItemExposure@20}
	\begingroup \fontTables %
	\begin{tabular}{l}%
  \toprule%
name \\ %
  \midrule%
$R^2$ \\ %
  $R^2$ (adj) \\ %
  Shape\ \\ %
  Density\ \\ %
  $Gini_U$\ \\ %
  StPB\ \\ %
  KuPB\ \\ %
  StRG\ \\ %
  MedDA\ \\ %
  KuDA\ \\ %
\addlinespace[-0.04cm]
   \bottomrule%
\end{tabular}%
	\endgroup %
	\begingroup \fontTables %
	\begin{tabular}{l}%
  \toprule%
  Popularity \\ %
 \midrule%
0.933 \\ %
  0.929 \\ %
  -1.58*** \\ %
  0.44 \\ %
  1.885*** \\ %
  1.443*** \\ %
  -0.519* \\ %
  0.682 \\ %
  3.415*** \\ %
  0.446 \\ %
   \bottomrule%
\end{tabular}%
	\endgroup %
	\begingroup \fontTables %
	\begin{tabular}{l}%
  \toprule%
  IB \\ %
 \midrule%
0.717 \\ %
  0.7 \\ %
  -4.227*** \\ %
  -0.135 \\ %
  5.417*** \\ %
  1.389 \\ %
  0.604 \\ %
  -4.454** \\ %
  1.056. \\ %
  -7.898*** \\ %
   \bottomrule%
\end{tabular}%
	\endgroup %
	\begingroup \fontTables %
	\begin{tabular}{l}%
  \toprule%
  UB \\ %
 \midrule%
0.714 \\ %
  0.697 \\ %
  -4.517*** \\ %
  -0.118 \\ %
  3.738** \\ %
  2.253* \\ %
  0.1 \\ %
  -3.889** \\ %
  1.368* \\ %
  -6.694*** \\ %
   \bottomrule%
\end{tabular}%
	\endgroup %
	\begingroup \fontTables %
	\begin{tabular}{l}%
  \toprule%
  PopGeoNN \\ %
 \midrule%
0.715 \\ %
  0.698 \\ %
  -1.474*** \\ %
  0.635 \\ %
  1.362* \\ %
  0.127 \\ %
  -2.026*** \\ %
  1.088 \\ %
  2.639*** \\ %
  2.106*** \\ %
   \bottomrule%
\end{tabular}%
	\endgroup %
	\begingroup \fontTables %
	\begin{tabular}{l}%
  \toprule%
  GeoBPRMF \\ %
 \midrule%
0.69 \\ %
  0.672 \\ %
  -2.5*** \\ %
  -0.473 \\ %
  -0.73 \\ %
  1.726*** \\ %
  -0.496. \\ %
  1.648** \\ %
  0.405* \\ %
  2.294*** \\ %
   \bottomrule%
\end{tabular}%
	\endgroup %
	\begingroup \fontTables %
	\begin{tabular}{l}%
  \toprule%
  BPRMF \\ %
 \midrule%
0.29 \\ %
  0.248 \\ %
  -1.515*** \\ %
  0.112 \\ %
  -2.924*** \\ %
  2.178*** \\ %
  -1.795*** \\ %
  2.376** \\ %
  -0.061 \\ %
  3.135*** \\ %
   \bottomrule%
\end{tabular}%
	\endgroup
	\begingroup \fontTables %
	\begin{tabular}{l}%
  \toprule%
  IRENMF \\ %
 \midrule%
0.868 \\ %
  0.861 \\ %
  -2.915*** \\ %
  -0.899** \\ %
  2.023*** \\ %
  0.463. \\ %
  -0.144 \\ %
  0.754. \\ %
  0.91*** \\ %
  1.103** \\ %
   \bottomrule%
\end{tabular}%
	\endgroup
\end{table*}

\section{Pairwise Correlations of Explanatory Variables}

\begin{figure*}[htb]
	\includegraphics[width=\textwidth]{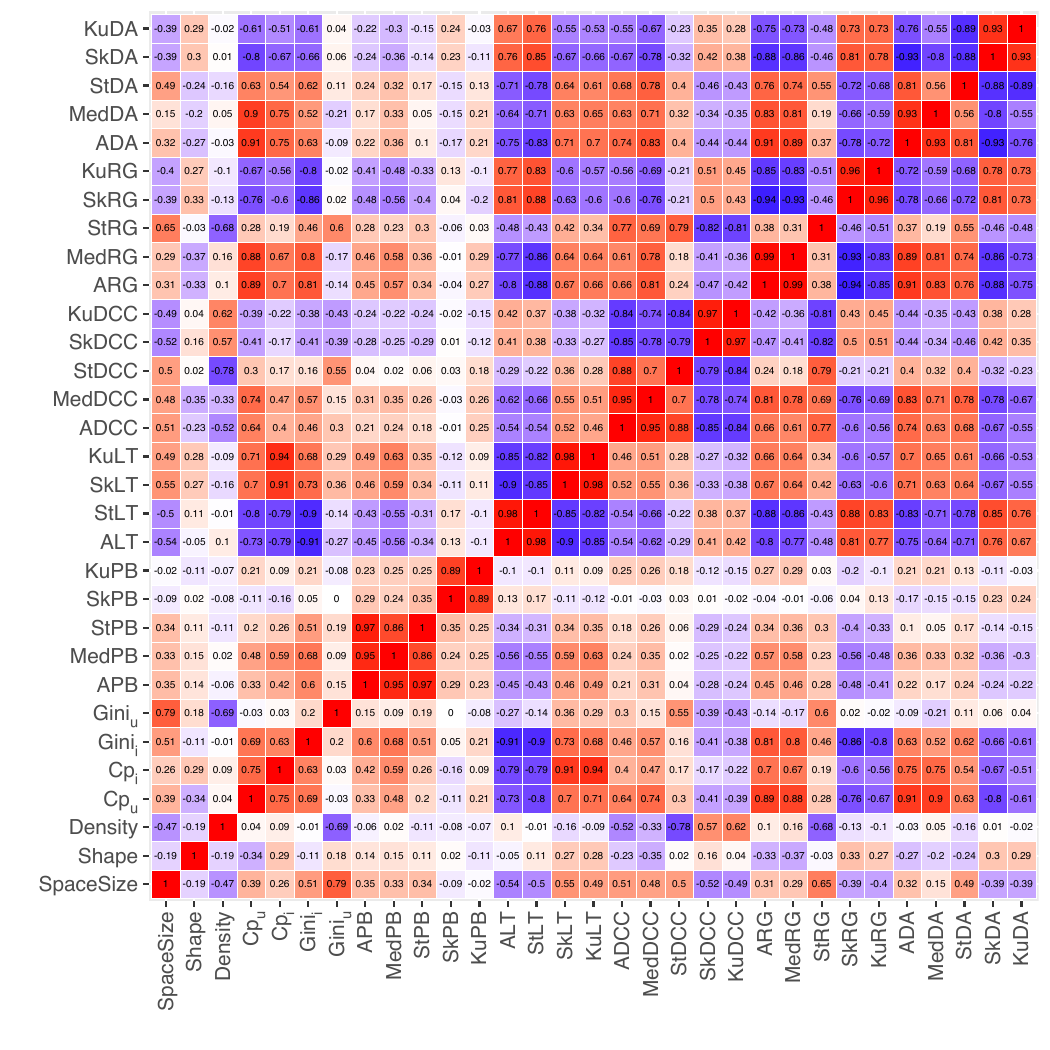}
	\includegraphics[width=0.8\textwidth]{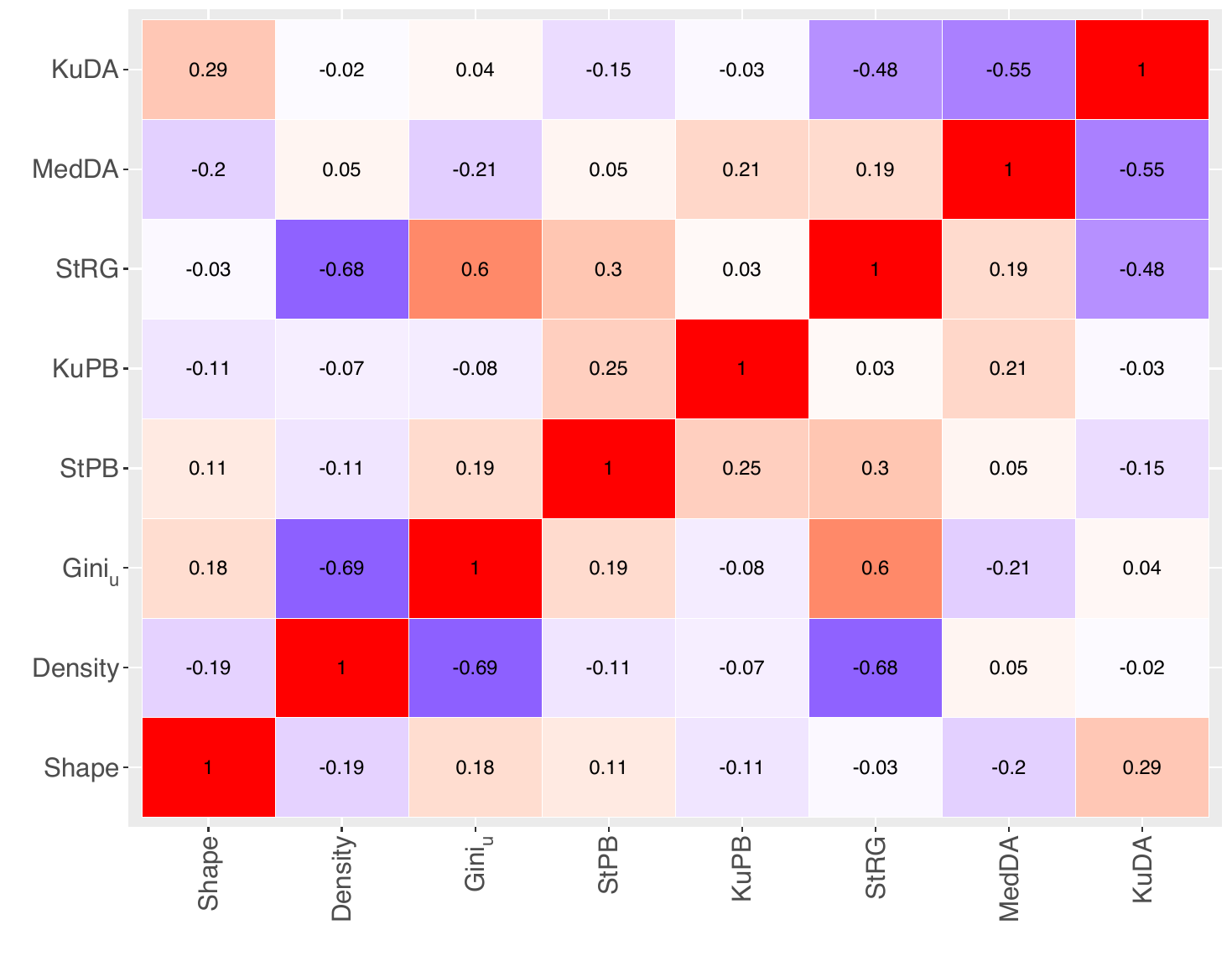}
	\caption{Above: Pairwise Pearson Correlations between all explanatory variables. Below: Pearson Correlation between explanatory variables after removing highly correlated variables using \autoref{alg:corr}.}
	\label{fig:correlation_vif_all}
\end{figure*}

The following \autoref{fig:correlation_vif_all} showcases the effect of the reduction of EVs using pairwise correlation plots.

\end{document}